\documentclass[final,5p,times,twocolumn]{elsarticle}
\journal{Physics Reports}

\usepackage{multicol}
\usepackage{amssymb}
\usepackage{amsmath}
\usepackage{subfigure}
\usepackage{pstricks}

\biboptions{sort&compress}

\def\openone{\leavevmode\hbox{\small1\kern-3.8pt\normalsize1}}

\newcommand{\et}{{\em et al.}}

\newcommand{\ba}{\begin{eqnarray*}}
\newcommand{\ea}{\end{eqnarray*}}
\newcommand{\beq}{\begin{equation}}
\newcommand{\eeq}{\end{equation}}
\newcommand{\be}{\begin{eqnarray}}
\newcommand{\ee}{\end{eqnarray}}
\newcommand{\magq}{|{\bf q}|}

\def\Llambdamu{L_{\lambda\mu}}
\def\Wlambdamu{W^{\lambda\mu}}
\def\lsim{\buildrel < \over {_{\sim}}}
\def\gsim{\buildrel > \over {_{\sim}}}

\newcommand{\m}{\mathbf}
\newcommand{\nn}{\nonumber}
\newcommand{\tetatt}{\mbox{$\theta_{23}$}}
\newcommand{\tetaot}{\mbox{$\theta_{13}$}}
\newcommand{\delot}{\mbox{$\Delta_{atm}$}}
\newcommand{\n}[1]{\ensuremath{|\mathbf{#1}|}}
\def\bea{\begin{eqnarray}}
\def\eea{\end{eqnarray}}
\usepackage{hyperref}
\hypersetup{
    colorlinks=true, 
    linktoc=all,     
    linkcolor=blue,  
    citecolor=blue
}

\begin{document}

\begin{frontmatter}



\title{Neutrino-nucleus interactions and the determination \\
of oscillation parameters}


\author[Roma1]{Omar Benhar}
\author[VT]{Patrick Huber}
\author[VT]{Camillo Mariani}
\author[Roma3]{Davide Meloni}

\address[Roma1]{INFN and Department of Physics, Sapienza University, I-00185 Roma, Italy}
\address[VT]{Center for Neutrino Physics, Virginia Tech, Blacksburg, Virginia 24061, USA}
\address[Roma3]{Dipartimento di Matematica e Fisica, Universit\`a Roma Tre, I-00146 Roma, Italy}

\begin{abstract}

We review the status and prospects of  theoretical studies of neutrino-nucleus interactions, 
and discuss the influence of the treatment of nuclear effects on the determination of oscillation 
parameters. The models developed to describe the variety of reaction mechanisms contributing to the
nuclear cross sections are analysed, with emphasis placed on their capability to explain the large
body of available electron scattering data. The impact of the uncertainties associated with the 
description of nuclear structure and dynamics on the determination of oscillation parameters is 
illustrated through examples, and possible avenues towards a better understanding of the signals 
detected by accelerator-based experiments are outlined.

\end{abstract}


\end{frontmatter}

\tableofcontents



\section{Introduction}
\label{sec:intro}
Neutrino physics is entering the age of precision measurements.
A number of experiments have firmly established the occurrence of neutrino oscillations and determined
the corresponding squared mass differences and mixing angles~\cite{SNO, KAMLAND, SUPERK, Aliu:2004sq, Abe:2014ugx, Adamson:2011qu}.
These measurements have provided unambiguous evidence  that
neutrinos\textemdash assumed to be massless in the standard model of particle physics\textemdash do have non-vanishing masses.
Reactor and accelerator-based experiments carried out over the past few years  
\cite{An:2012eh, Ahn:2012nd, Abe:2013hdq}
reported accurate measurements of the $\theta_{13}$ mixing angle,
whose value turned out to be $\sim$10 deg.
The large $\theta_{13}$ mixing angle will enable future experiments\textemdash such as the
Deep Underground Neutrino Experiment (DUNE) in the United States~\cite{DUNE}\textemdash to search
for leptonic CP violation in appearance mode, thus addressing one of the outstanding fundamental
problems of particle physics.
These searches will involve high precision determinations
of the oscillation parameters, which in turn require a deep understanding of neutrino interactions
with the atomic nuclei comprising the detectors. In view of the achieved and planned experimental accuracies, the treatment of nuclear
effects is indeed regarded as one of the main sources of systematic uncertainty~\cite{Abe:2013xua}.

Over the past decade, it has become more and more evident that the independent particle model of
nuclei\textemdash the ultimate implementation of which is the Relativistic Fermi Gas Model (RFGM) routinely
employed in simulations of neutrino-nucleus interactions\textemdash conspicuously fails to account for the complexity of nuclear dynamics and
the variety of reaction mechanisms contributing to the detected signals.

The large discrepancy between the results of Monte Carlo simulations and the double differential
cross section of charged current (CC) quasielastic (QE) interactions in carbon, measured
by the MiniBooNE Collaboration using a beam of average energy $\sim$0.8 (0.7) GeV in neutrino
(antineutrino) mode, is a striking manifestation of the above
problem~\cite{AguilarArevalo:2010zc,miniboone_ccqe_2,miniboone_antinu}. More recently, the analysis of the
inclusive $\nu_\mu$-nucleus
cross sections at beam energy in the range 2\textendash 20 GeV, measured by the MINER$\nu$A Collaboration using
a variety of targets, led to the striking conclusion that none of nuclear models implemented in Monte Carlo
simulations appears to be  capable of explaining the data ~\cite{minerva_2014}.

A great deal of effort is currently being devoted to the development of theoretical models providing a
fully quantitative description of the neutrino-nucleus cross section in the kinematical regime relevant
to most ongoing and future accelerator-based experiments, corresponding to beam energies ranging from a
 few hundred MeV to  a few GeV.  In this context, a key role is played by the availability of a wealth
of electron scattering data.

Electron scattering experiments have provided accurate
information on the electromagnetic response of a number of nuclei.
Static form factors and charge distributions have been extracted from
elastic scattering data, while the measurements of inelastic cross sections have allowed for
a systematic study of the dynamic response functions in a broad range
of energy and momentum transfer. Finally, with the advent of the last
generation of continuous beam accelerators, a number of exclusive channels
have been analysed to unprecedented precision.

The large body of measured electron scattering cross sections provides an indispensable benchmark for validation of
theoretical models. 
In fact, the ability to explain electron scattering data 
sholud be seen as an obvious requisite, to be met by any models of neutrino-nucleus interactions. In addition, new electron scattering experiments will be needed, 
to gain information on nuclei employed in neutrino detectors\textemdash most notably argon\textemdash for which the available data is scarce, or
nonexisting, \cite{argon}.

This review is organized as follows.
In Section ~\ref{enu:conf} we provide a comparative analysis between electron and neutrino-nucleus scattering, aimed at pointing out the 
difficulties involved in the interpretation of
the flux-integrated neutrino cross sections. The theoretical description of the reaction mechanisms contributing to the electron-nucleus cross section is outlined in Section~\ref{eA:xsec}, where
the ability of different approaches to explain the data is also illustrated. The generalisation of the formalism based on the impulse approximation to the case of neutrino interactions and the
problems associated with the interpretation of the measured CC QE cross sections are discussed in Sections~\ref{nuA:xsec} and \ref{interpretation}, respectively, while Section~\ref{implementation}
is devoted to a discussion of the issues involved in the implementation of the spectral function formalism in neutrino event generators. In Section~\ref{dop} we provide few examples showing
how the treatment of nuclear effects affects the determination of neutrino oscillation parameters. Finally, in Section~\ref{sec:concl} we summarise our assessment of both the present status and the future prospects of the field.


\section{Comparison between electron- and neutrino-nucleus scattering}
\label{enu:conf}

The description of electron-nucleus interactions involves a variety of
non-trivial problems, arising from the complexity of both nuclear and nucleon structure and
dynamics.

\begin{figure}[h!]
\vspace*{.1in}
\begin{center}
\includegraphics[scale= 0.45]{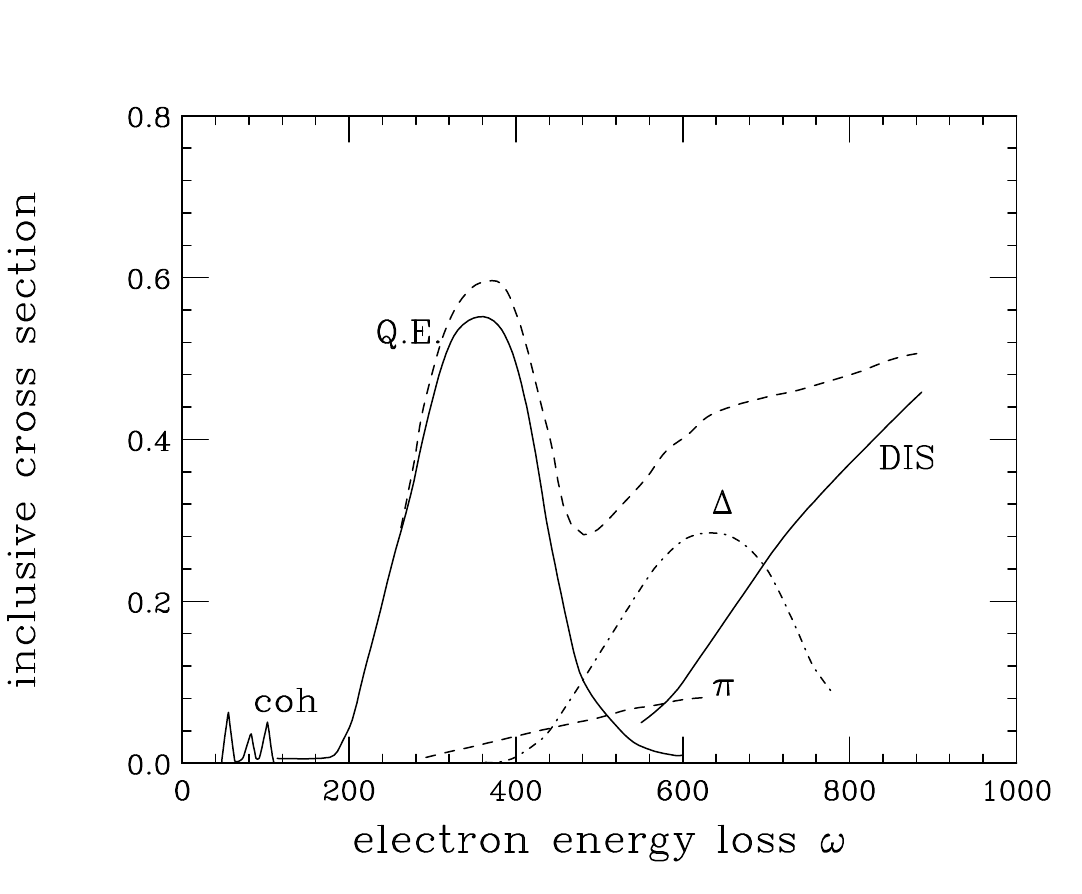}
\end{center}
\vspace*{-.2in}
\caption{Schematic representation of the inclusive electron-nucleus cross section at beam energy around 1 GeV and fixed electron scattering angle, displayed 
as a function of the energy transfer~\cite{Benhar:2006wy}.}
\label{regime}
\end{figure}

Figure \ref{regime} shows the typical behaviour of the double differential cross section of the inclusive process
 \beq
 \label{eeprime}
 e + A \rightarrow e^\prime + X \ ,
 \eeq
in which only the outgoing lepton is detected,  at beam energy around 1 GeV. Here, $A$ and $X$ denote the target nucleus
in its ground state and the undetected hadronic final state, respectively.

The data is shown for fixed electron scattering angle, $\theta_e$, as a function of the energy transfer $\omega= E_e - E_{e^\prime}$, the value of which 
is the main factor determining the dominant
reaction mechanism.  The bump centered at $\omega \approx \omega_{\rm QE} = Q^2/2m$, where $m$ is the nucleon mass and $Q^2=4E_e E_{e^\prime} \sin^2(\theta_e/2)$, corresponds to single nucleon knockout, while the structure visible at larger $\omega$ reflects the onset of more complex mechanisms, such as coupling to nucleons belonging to correlated pairs
or two-nucleon currents arising from meson exchange processes, excitation of nucleon resonances and deep inelastic scattering.

The analysis of electron scattering data has clearly exposed the limitations of the independent particle picture, providing the
conceptual framework of the nuclear shell model \cite{RMP_old}.

Accurate measurements of the coincidence $(e,e^\prime p)$ cross section in the region of low to moderate missing energy and missing momentum have unambiguously
demonstrated that, while the spectroscopic lines corresponding to knock out of nucleons occupying the shell model orbits are clearly visible in the missing energy spectra, the associated
spectroscopic factors\textemdash yielding the normalization of the single-nucleon states\textemdash are considerably less than unity, regardless of the nuclear mass number $A$ (for a recent overview of $(e,e^\prime p)$ data, see Ref.~\cite{Benhar:NPN}).
Complementary data collected at large missing energy and missing momentum strongly suggest that  this feature is a manifestation of
 dynamical nucleon-nucleon (NN) correlations, leading to the excitation of  nucleon pairs to continuum 
 states and to a corresponding depletion of the bound states belonging to the Fermi sea.  This interpretation,  strongly supported by the results of the pioneering  $^3{\rm He}(e,e^\prime p)$ 
experiment described in Ref.~\cite{Marchand},  has been recently confirmed by measurements carried out at Jefferson Lab  \cite{daniela,Rohe2006152}
using a carbon target.

Advanced nuclear models, developed using  the formalism of many-body theory,
provide an overall satisfactory description of the observed cross sections over a broad kinematical range.
In particular, in the region in which QE scattering dominates the data is generally reproduced with an
accuracy of few percent \cite{Benhar05,Ankowski2013}  (for a recent review of electron-nucleus scattering in the QE sector see also  Ref. \cite{Benhar:2006wy}).

Nuclear Many-Body Theory (NMBT) is based on the tenet that nucleons can be treated as point like non relativistic particles, the
dynamics of which are described by the Hamiltonian
\beq
H = \sum_{i=1}^{A} \frac{{\bf p}_i^2}{2m} + \sum_{j>i=1}^{A} v_{ij}
 + \sum_{k>j>i=1}^A V_{ijk} \ .
\label{H:A}
\eeq
In the above equation, ${\bf p}_i$ is the momentum of the $i$-th nucleon, while the potentials
$v_{ij}$ and $V_{ijk}$ describe two- and three-nucleon interactions, respectively.

Phenomenological two-nucleon potentials
are obtained from an accurate fit to the available data on the two-nucleon system, in both bound and scattering
states, and reduce to the Yukawa one-pion-exchange potential at large distances  \cite{Wiringa95}.
The inclusion of the additional three-body term, $V_{ijk}$, is needed to reproduce the binding 
of light nuclei~\cite{Pudliner95b}. 

Recently, chiral perturbation theory, or $\chi PT$,  (for a review, see Refs.~\cite{Epelbaum,chiral:review}) has been also employed to obtain a 
theoretically sound and consistent model of both two- and three-nucleon interactions, constrained by the symmetries of the fundamental theory of strong interactions.
This approach, originally proposed by Weinberg, exploits the Goldstone boson nature of the pion
\cite{WeinBook1}, 
implying that the interactions of low energy pions are weak, and can be treated in perturbation theory.

The nuclear electromagnetic current,  $J^\mu \equiv (J^0,{\bf J})$, is related to the Hamiltonian (\ref{H:A}) through the continuity equation \cite{Riska89}
\beq
\label{continuity}
{\boldmath \nabla} \cdot {\bf J} + i [H,J^0] = 0 \ .
\eeq
Because the NN potential $v_{ij}$ does not commute with the charge operator $J^0$, the above equation
implies that $J^\mu$ comprises two-nucleon contributions, arising from meson exchange processes. Therefore, it can be conveniently written in the form
\beq
\label{def:curr}
J^\mu = J_1^\mu + J_2^\mu = \sum_i j^\mu_i + \sum_{\rm j>i} j^\mu_{ij} \ .
\eeq

The main difficulty associated with the extension of the theoretical approaches developed for electron-nucleus scattering to
the case of neutrino scattering arises from the fact that,  since neutrino beams are always produced as secondary decay products,
their energy is not sharply defined, but broadly distributed according to a flux $\Phi$.

Consider, for example, charged-current neutrino interactions.  In this instance, detection of the energy of the outgoing lepton,
$T_\ell$,  {\em does not} provide the information on the energy transfer, $\omega$, and different reaction mechanisms contribute to the double differential cross section measured at
fixed $T_\ell$ and lepton scattering angle, $\theta_\ell$.

\begin{figure}[h!]
\vspace*{.1in}
\begin{center}
\includegraphics[scale= 0.5]{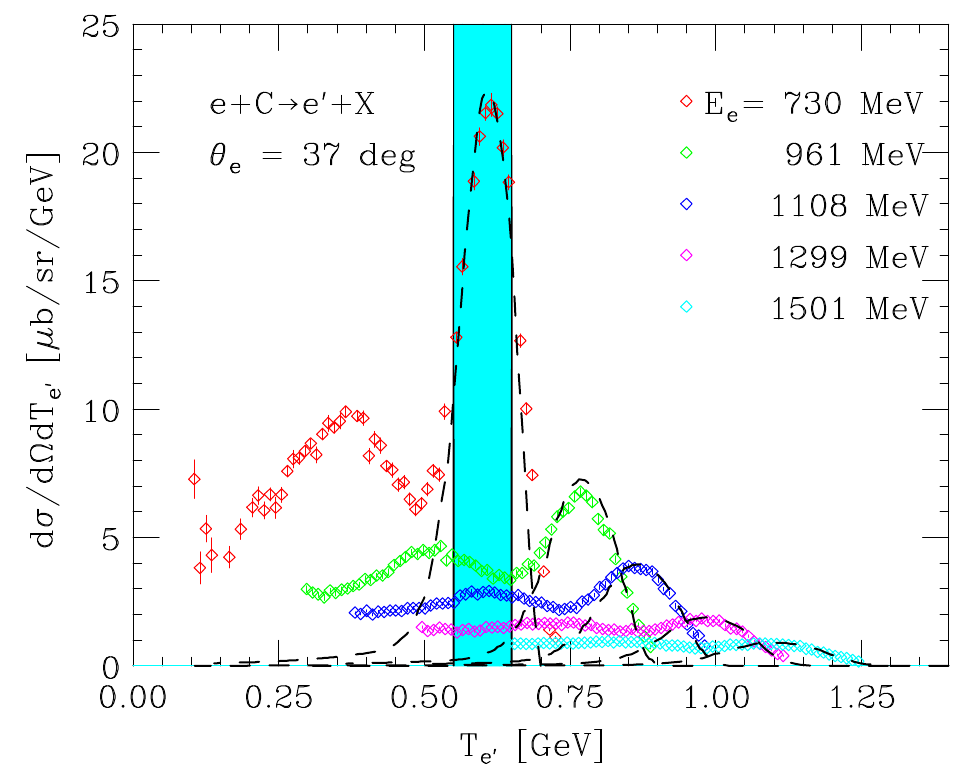}
\end{center}
\vspace*{-.25in}
\caption{Inclusive electron-carbon cross sections at $\theta_e = 37$ deg and beam energies
ranging between 0.730 and 1.501 GeV \cite{12C1,12C2}, plotted as a function of the energy of the outgoing electron.
The dashed lines represent the results of theoretical calculations, carried out within the spectral function approach (see Section~\ref{IA}) taking into account QE scattering only
\cite{Benhar:NUFACT11}.}
\label{diffmech}
\end{figure}

This feature is clearly illustrated  in Fig. \ref{diffmech}, showing the inclusive
electron-carbon cross sections at  $\theta_e~=~37$~deg and beam energies ranging between 0.730 and 1.501~GeV, as a function of the energy of the
outgoing electron \cite{12C1,12C2}. It clearly appears that the highlighted  electron energy bin ($550~<~T_{e^\prime}~< 650$~MeV), corresponding to QE
kinematics at $E_e~=~730 \ {\rm MeV}$, picks up contributions from scattering processes taking place at different beam energies, in which
reaction mechanisms other than single nucleon knockout dominate.

\begin{figure}[h!]
\vspace*{.1in}
\begin{center}
\includegraphics[scale= 0.475]{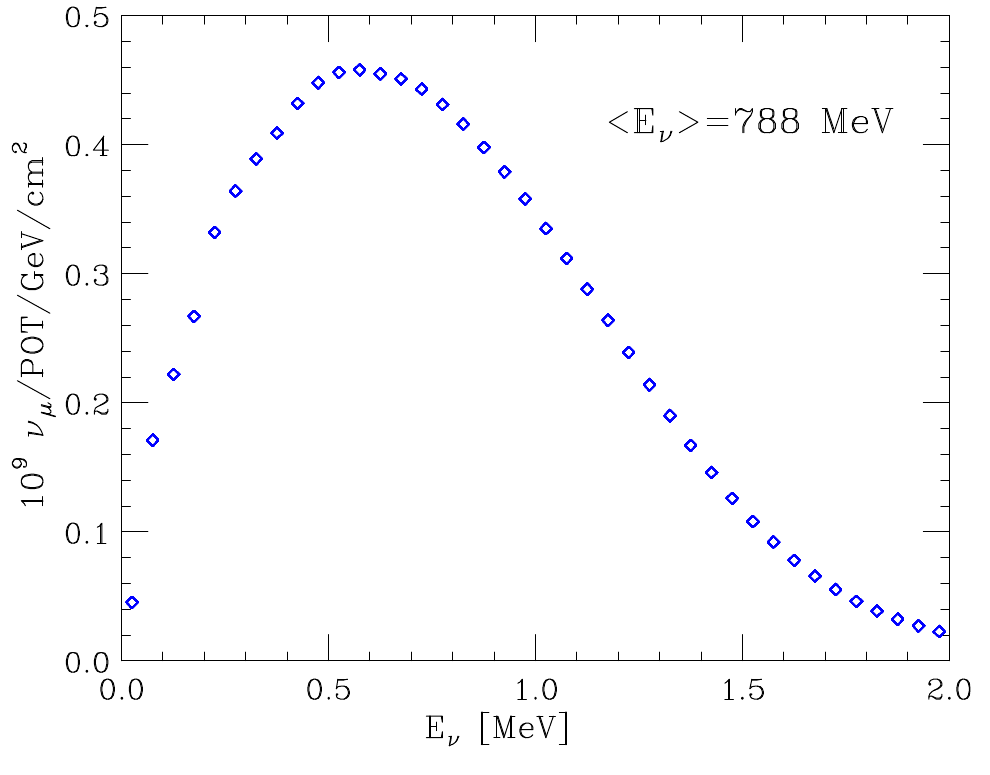}
\end{center}
\vspace*{-.25in}
\caption{Energy dependence of the MiniBooNE neutrino flux \cite{miniboone_ccqe_2}.}
\label{fluxav}
\end{figure}

To gauge the role played by different
contributions in a typical neutrino experiment, let us assume that the electron beam energy be distributed according to  the MiniBooNE
neutrino flux, displayed
in Fig. \ref{fluxav}. It turns out that the fluxes corresponding to energies $E_\nu =$ 730 and 961 MeV are within 
$\sim$20\% of one another. Hence, if we were to average the electron-carbon data of Fig.~\ref{diffmech} with the flux of Fig.~\ref{fluxav}, the
cross sections corresponding to beam energies 730 and 961 MeV would contribute to the measured cross section in the highlighted bin with about
the same weight.

The above discussion implies that the understanding of the flux-averaged neutrino cross section requires the development of
theoretical models providing a {\em consistent} treatment of {\em all} reaction mechanisms active in the broad kinematical range
corresponding to the relevant neutrino energies.


\section{The electron-nucleus cross section}
\label{eA:xsec}

The differential cross section of process (\ref{eeprime}),
in which an electron of initial four-momentum $k_e\equiv(E_e,{\bf k}_e)$ scatters off
a nuclear target to a state of four-momentum
$k^\prime_e~\equiv~(E_{e^\prime},{\bf k}_{e^\prime})$, the target final state being undetected,
can be written in Born approximation as (see, e.g., Ref.~\cite{Itzykson80})
\beq
\frac{d^2\sigma}{d\Omega_{e^\prime} dE_{e^\prime}} =
\frac{\alpha^2}{Q^4}\frac{E_{e^\prime}}{E_e}\ \Llambdamu\Wlambdamu \ ,
\label{eA_xsec}
\eeq
where $\alpha=1/137$ is the fine structure constant,
$d\Omega_{e^\prime}$ is the differential solid angle in the direction
specified by ${\bf k}_{e^\prime}$, $Q^2=-q^2$ and
$q=k_e-k_{e^\prime} \equiv (\omega,{\bf q})$ is the four momentum transfer.

The tensor $L^{\lambda \mu}$, that neglecting the electron mass reduces to
\beq
L^{\lambda \mu} = 2 \left[ k_e^\lambda k_{e^\prime}^\mu + k_e^\mu k_{e^\prime}^\lambda
 - g^{\lambda \mu} (k_e k_{e^\prime}) \right] \ ,
\label{lepten}
\eeq
where $g^{\lambda \mu} \equiv {\rm diag}(1,-1,-1,-1)$ and $(k_e k_{e^\prime})=E_e                                  
E_{e^\prime}-{\bf k}_e \cdot {\bf k}_{e^\prime}$,  is fully specified by the measured
electron kinematical
variables.

All information on target structure is contained
in the tensor $\Wlambdamu$, the definition of which involves the initial and final nuclear
states $|0\rangle$ and $|X\rangle$, carrying four-momenta $P_0$ and $P_X$,
as well as the nuclear current operator of Eq.~(\ref{def:curr}). It can be cast in the form
\beq
\Wlambdamu=\sum_{X}  \langle 0 |J^\lambda|X \rangle\langle X|J^\mu|0\rangle
\delta^{(4)}(P_0+q-P_X)\ ,
\label{nuclear:tensor}
\eeq
\noindent
where the sum includes all hadronic final states.

The most general expression of the target tensor of Eq.~(\ref{nuclear:tensor}),
fulfilling the
requirements of Lorentz covariance, conservation of parity and gauge invariance, can be
written in terms of two structure functions $W_1$ and $W_2$ as
\begin{align}
\notag
\Wlambdamu & =  W_1 \left( -g^{\lambda \mu} + \frac{q^\lambda q^\mu}{q^2} \right)  \\
  & + \frac{W_2}{M_A^2} \left(P_0^\lambda - \frac{(P_0 q)}{q^2}q^\lambda \right)
                   \left(P_0^\mu - \frac{(P_0 q)}{q^2}q^\mu \right) ,
\label{genw}
\end{align}
where $M_A$ is the target mass and the structure functions depend on the two scalar quantities  $Q^2$
and $(P_0 q)$. In the target rest frame $(P_0 q) = M_A\omega$,  and $W_1$ and $W_2$ become
functions of the measured momentum and energy transfer,  $|{\bf q}|$ and ${\omega}$.

Substitution of Eq.~(\ref{genw}) into Eq.~(\ref{eA_xsec}) leads to
\begin{align}
\frac{d^2\sigma}{d\Omega_{e^\prime} dE_{e^\prime}} = 
\left( \frac{d\sigma}{d\Omega_{e^\prime}}\right)_{\rm M} 
  \left[ W_2(|{\bf q}|,\omega)
            + 2 W_1(|{\bf q}|,\omega) \tan^2\frac{\theta_e}{2} \right] \ ,
\label{eA:xsec12}
\end{align}
where $(d\sigma/d\Omega_{e^\prime})_{\rm M}= \alpha^2 \cos^2(\theta_e/2)/4E_e^2\sin^4(\theta_e/2)$
is the  Mott cross section.

The right-hand side of Eq.~(\ref{eA:xsec12}) can be conveniently rewritten in terms of
the contributions arising from scattering processes involving longitudinally (L) and
transversely (T) polarized virtual photons. The resulting expression is
\begin{align}
\label{eA:xsecLT}
\frac{d^2\sigma}{d\Omega_{e^\prime} dE_{e^\prime}}  = 
\left( \frac{d\sigma}{d\Omega_{e^\prime}}\right)_{\rm M}  & \ \left[
\left( \frac{Q^2}{\magq^2} \right)^2 \ R_L(|{\bf q}|,\omega)  \right. \\
\notag
&  +  \left. \left( \frac{1}{2} \frac{Q^2}{\magq^2}
+ \tan^2\frac{\theta}{2} \right)  R_T(|{\bf q}|,\omega) \right] \ ,
\end{align}
where the longitudinal and transverse structure functions, $R_L$ and $R_T$, are trivially related
to $W_1$ and $W_2$ through
\beq
R_T = 2 W_1
\eeq
and
\beq
\left( \frac{Q^2}{\magq^2} \right)^2 R_L = W_2 -
\frac{Q^2}{\magq^2} W_1 \ .
\eeq

The initial state of the target nucleus appearing in Eq.~\eqref{nuclear:tensor} can be safely treated using the non relativistic approximation, regardless 
of the kinematical regime. At large momentum transfer, however,  this approximation can not
be used to describe either the nuclear final state, comprising at least one particle carrying
momentum $\sim {\bf q}$, or the current operator,  which depends explicitly on momentum transfer.


At low and moderate momentum transfer, typically ${\bf |q|}~<~500\, {\rm MeV}$,  accurate NMBT calculations of the tensor $\Wlambdamu$ of Eq.~(\ref{nuclear:tensor})
have been carried out for the few-nucleon systems, using nuclear
wave functions derived from the Hamiltonian of Eq.~(\ref{H:A})
to describe the initial and final states and expanding the current operator in powers of ${\bf |q|}/m$  \cite{Golak}. Valuable information have been also obtained 
exploiting integral transform techniques \cite{Efros94,Efros97,Efros04,CS92,Carlson98}.

On the other hand,  additional assumptions are unavoidably required for the treatment of nuclear interactions in the region of large
momentum transfer, the understanding of which is relevant to accelerator based neutrino experiment. To importance of relativistic effects can be easily gauged considering that 
the mean momentum transfer of CC QE events obtained by averaging over the MiniBooNE \cite{miniboone_ccqe_2} and Miner$\nu$a \cite{minerva_2014} neutrino fluxes
turn out to be $\sim 640$ and $\sim 880$ MeV, respectively.


\subsection{The impulse approximation}
\label{IA}

\begin{figure}[h!]
\begin{center}
\includegraphics[scale= 0.8]{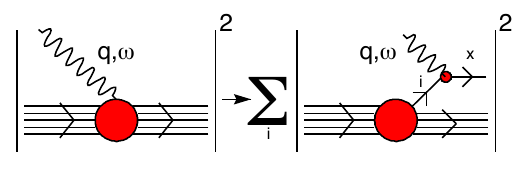}
\end{center}
\caption{Schematic representation of the IA regime, in which
the nuclear cross section is replaced by the incoherent sum of
cross sections describing scattering off individual nucleons, with the
recoiling $({\rm A}-1)$-nucleon system acting as a spectator.}
\label{IA:cartoon}
\end{figure}

The Impulse Approximation (IA) scheme, extensively employed to analyze electron-nucleus scattering data in the region in which the non relativistic approximation breaks down  \cite{Benhar:2006wy}, is based on the premise that
at momentum transfer $\m{q}$ such that $|{\bf q}|^{-1}~\ll~d/\pi$, $d$ being the average distance between nucleons in the target nucleus,
the nuclear scattering process reduces to an incoherent sum of collisions involving individual nucleons\textemdash as schematically illustrated by the diagram 
of Fig. \ref{IA:cartoon}\textemdash the remaining ${\rm A}-1$ particles acting as spectators.
Moreover, as a first approximation, final state interactions (FSI) between the outgoing hadrons and the spectator nucleons are assumed to be negligible.
Within this picture, the relativistic particles in the final state are
completely decoupled from the recoiling nucleus, and the description of their
motion becomes a trivial kinematical problem\footnote{In the literature the approximation scheme in which FSI are neglected is sometimes referred 
to as Plane Wave Impulse Approximation (PWIA).}.

Under the assumptions underlying the IA, the nuclear current of Eq.\eqref{def:curr} simplifies to a sum of one-body terms,
while the final state factorizes into the product of the hadronic state $x$, produced at the interaction vertex
with momentum ${\bf p}_x$, and the state describing the recoiling $(A-1)$-nucleon system, carrying
momentum ${\bf p}_{\rm R}$\footnote{Note that the discussion of this Section is not restricted to the QE sector. To the extent to which a description of the elementary interaction process 
is available, it can be readily applied to resonance production and deep-inelastic scattering.}. As a consequence, in Eq.~\eqref{nuclear:tensor} we can replace
\begin{align}
\label{factorization}
|X\rangle \longrightarrow |x,{\bf p}_x \rangle \otimes | {\rm R}, {\bf p}_{\rm R}\rangle \ ,
\end{align}
implying
\begin{align}
\label{sum:final}
\sum_X |X\rangle\langle X|\rightarrow & \sum_x\int \,{d^3p_x}|x, \m{p}_x\rangle\langle \m{p}_x,x|  \\
\nonumber
& \times \sum_{\rm R}   \ \int d^3p_{\rm R} |{\rm R},\m{p}_{\rm R}\rangle\langle \m{p}_{\rm R}, {\rm R}| \ ,
\end{align}
where the integrations include a sum over all discrete quantum numbers.
The resulting nuclear transition matrix elements take the form
\begin{align}
\label{fact:matel}
\langle 0|J^{\mu}|X\rangle  =  \sqrt{  \frac{m}{E_{p_{\rm R}}} } M_{\rm R}({\bf p}_{\rm R})   \sum_{i} \langle -\m{p}_{\rm R},N|j^{\mu}_i|x,\m{p}_x\rangle \ ,
\end{align}
where the state $|N,\m{k}\rangle$ decribes a non interacting nucleon ($N=p,\ n$ denotes protons and neutrons) carrying momentum ${\bf k}$, and
\begin{align}
\label{def:ampl1}
M_{\rm R}({\bf p}_{\rm R}) =  \langle 0| \{ | {\rm R},  \m{p}_{\rm R} \rangle \otimes | N,-\m{p}_{\rm R} \rangle \} \ , 
\end{align}
with $E_{p_{\rm R}} = \sqrt{ |{\bf p}_{\rm R}|^2 + m^2 }$.

Being independent of ${\bf q}$, the nuclear amplitude $M_{\rm R}({\bf p}_{\rm R})$ can
be safely obtained from NMBT. On the other hand,  the calculation of the matrix element of the current operator between free nucleon states
can be carried out for  all values of the momentum transfer without employing any approximations, assumin that the nucleon form factors are precisely known.

Substituting Eq.\eqref{fact:matel}  into Eq. \eqref{nuclear:tensor} and using Eq.\eqref{sum:final}, one can rewrite the target tensor in the concise and
transparent form
\begin{align}
\label{tens:IA}
W^{\lambda \mu}  =   \int {d^3k \ dE} \  \frac{M}{E_k} P(\m{k},E)  
  \left[ Z \mathcal{W}_p^{\lambda \mu}                          + (A-Z)  \mathcal{W}_n^{\lambda \mu}\right] \ , 
\end{align}
where $Z$ is the target charge, 
and the spectral function $P(\m{k},E)$,
yielding the probability of removing a nucleon with momentum $\m{k}$ from the target ground state leaving the residual system with excitation
energy $E$,  is defined as\footnote{In deriving Eq. \eqref{tens:IA} we have made the assumption, largely justified in isoscalar nuclei, that the proton and neutron spectral functions be identical.}
\cite{PKE,LDA}
\begin{align}
\label{def:pke}
P(\m{k},E) = \sum_{\rm R} |M_{\rm R}({\bf k})|^2 \delta( E + M_A - m - E_{\rm R}) \ ,
\end{align}
$E_{\rm R}$ being the energy of the state $| {\rm R}, \m{k} \rangle$.

The tensor
\begin{align}
\label{nucleon:tensor}
\nonumber
 \mathcal{W}_N^{\lambda \mu} = \sum_x  & \int d^3 p_x   \langle \m{k}, N|j^{\lambda}_N | x, {\bf p}_x \rangle \langle {\bf p}_x, x |j^{\mu}_N | N , \m{k}\rangle \\
&  \times  \delta( \tilde{\omega} + E_k - E_x) \delta^{(3)}({\bf k} + {\bf q} - {\bf p}_x) \ ,
\end{align}
where $E_x$ is the energy of the hadronic final state carrying momentum ${\bf p}_x = {\bf k} + {\bf q}$, describes the electromagnetic interactions of a nucleon with four
momentum $k \equiv (E_k,{\bf k})$ in free space. Note, however,  that the four momentum transfer $q$ is replaced by $\tilde{q} \equiv ( \tilde{\omega}, {\bf q})$, with
\begin{align}
\label{def:omegatilde}
\tilde{\omega} & = E_x - E_k  = \omega + m - E - E_k \ .
\end{align}
The substitution $\omega \to \tilde{\omega}$ is needed to take into account the fact that a fraction $\delta\omega$ of the
energy transfer to the target goes into excitation energy of the spectator system. Equation \eqref{nucleon:tensor} shows that the elementary scattering process is described as if it
took place in free space with energy transfer $\tilde{\omega}~=~\omega-\delta\omega$.

It has to be pointed out that,  while sensible on physics grounds, the use of $\tilde{q}$ in the nucleon tensor \eqref{nucleon:tensor} poses a non trivial
conceptual problem, in that it leads to a violation of current conservation. This problem is inherent in the IA scheme, which does not allow
energy and current to be simultaneously conserved. A very effective prescription to restore gauge invariance,  extensively employed in the analysis of $(e,e^\prime p)$ experiments,
is based on the use of off-shell extrapolations of the electron-nucleon cross section,  referred to as $cc1$ and $cc2$,  developed by de Forest in the early 80s \cite{Forest83}.
Note, however, that in QE kinematics, because the struck nucleon is nearly on the mass shell, the effect of using de Forest's prescription turns out to be quite small.

Collecting the above results, the nuclear cross section can be finally written in the form
\begin{align}
\label{sigma1}
\frac{ d^2 \sigma_{IA} }{ d\Omega_{e^\prime} d E_{e^\prime} }
 =  \int    \ d^3k \ dE  \  P({\bf k},E)  
 \left[  Z \frac{d^2\sigma_{ep}}{d\Omega_{e^\prime}dE_{e^\prime} }  +  N \frac{d^2\sigma_{en}}{d\Omega_{e^\prime}dE_{e^\prime} }  \right], 
\end{align}
with
\begin{align}
\label{sigma2}
\frac{d^2\sigma_{\rm e N}}{d\Omega_{e^\prime} dE_{e^\prime}} =
\frac{\alpha^2}{Q^4}\frac{E_{e^\prime}}{E_e}\ \Llambdamu   \mathcal{W}^{\lambda \mu}_N   \ .          
\end{align}

Equation \eqref{sigma1}, first derived by the authors of Ref.~\cite{Dieperink76}, has been widely used in the course of four decades. It shows that, within the IA scheme,  the electron-nucleus cross section can be obtained folding the cross sections of the processes involving individual nucleons\textemdash which can be, at least in principle, measured using proton and deuteron targets\textemdash with the energy and momentum distribution of the participating nucleon,  described by the spectral function.

Note that in Eq.\eqref{sigma1} the effect of Pauli blocking on the phase space available to the struck nucleon in the final state\textemdash which becomes vanishingly small in the limit of large momentum transfer\textemdash is
disregarded altogether. 
A simple and reasonable procedure to take it into account is based on the replacement
\begin{align}
P(\m{k},E) \to P(\m{k},E) \theta( |{\bf k} + {\bf q}| - {\bar k}_F ) \ , 
\end{align}
where $\theta(x)$ is the Heaviside step function and ${\bar k}_F$ is an average nuclear Fermi momentum, derived within the
local Fermi gas model  \cite{Benhar05,Ankowski2013}.

\subsection{The nuclear spectral function}
\label{PKE}


Within the mean field approximation underlying the independent particle model (IPM) of the nucleus, the sum over the sates
of the residual $({\rm A}-1)$-nucleon system appearing in Eq. \eqref{def:pke} is  restricted
to {\em bound} one-hole states. The corresponding spectral function can be written in the form
\begin{align}
\label{def:pkeMF}
        P_{\rm MF}(\m{k},E) = \sum_{\alpha \in \{ F \} } |\phi_\alpha(\m{k})|^2 \delta(E-\epsilon_\alpha) \ ,  
\end{align}
where the sum includes all single particle states belonging to the Fermi sea $\{ F \}$, labeled by the index $\alpha$,  with $\phi_\alpha(\m{k})$ and $\epsilon_\alpha$ being
the corresponding momentum-space wave function and energy, respectively. Note that  $|\phi_\alpha(\m{k})|^2$ yields the probability of finding a nucleon with
momentum ${\bf k}$ in the state $\alpha$.

The mean field approximation provides a fairly good  description of the spectral functions at
$|{\bf k}| \lsim 250 \ {\rm MeV}$, and $E$ lower than the energies required to remove a nucleon from the shell model states.

A wealth of experimental information on the nuclear spectral functions in the kinematical regime in which mean field dynamics is dominant
has been extracted from the cross sections of the $(e,e^\prime p)$ process, measured using a variety of targets (for extensive reviews, see 
Refs.~\cite{eep1,eep2}).

Within the RFGM,  the single-particle states, labeled by
the momentum ${\bf k}$, are occupied with unit probability for $|{\bf k}| < k_F$, $k_F$ being the Fermi momentum, while all levels corresponding to  $|{\bf k}| > k_F$ are empty. As a consequence,
the spectral function reduces to
\begin{align}
\label{oke:FG}
P_{\rm FG}(\m{k},E) = \frac{3}{4 \pi k_F^3} \ \theta(k_F - |{\bf k}|)  \ \delta(E - E_k - {\bar \epsilon}) \ ,
\end{align}
where $E_k$ is the energy of a non interacting nucleon of momentum ${\bf k}$,  and ${\bar \epsilon}$ is an average binding energy.

The values of the two
parameters of the RFGM, $k_F$ and ${\bar \epsilon}$, are inferred from the width and position of the peak exhibited by the measured electron-nucleus
cross sections in the QE channel, respectively \cite{Moniz}.

\begin{figure}[h!]
\begin{center}
\includegraphics[scale= 0.4]{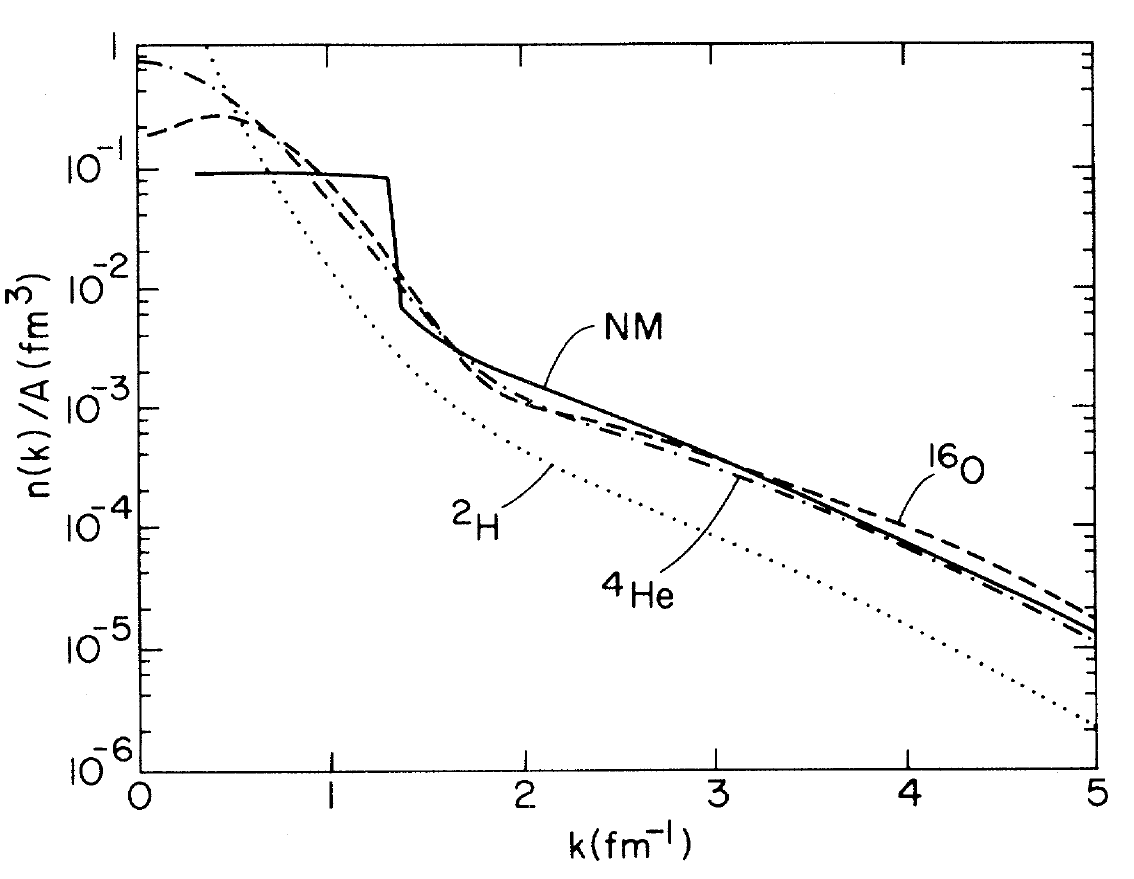}
\end{center}
\vspace*{-.2in}
\caption{Momentum distribution per nucleon in $^2$H, $^4$He,
$^{16}$O and uniform isospin-symmetric nuclear matter (NM), obtained from NMBT using realistic nuclear 
Hamiltonians \cite{Schiavilla86,RMP_old}.}
\label{momdis}
\end{figure}

It is very important to realise that, when using  a realistic nuclear model, in which the effects of NN correlations are taken into account,
more complex states, with at least one of the spectator nucleons excited to the continuum, give non vanishing contributions to the spectral function. Accurate
calculations carried out for a variety of nuclear systems suggest that these contributions, arising form short-range dynamics, are largely
unaffected by surface and shell effects, and are therefore nearly independent of $A$ for $A>2$. This feature is illustrated in Fig.~\ref{momdis}, showing the $A$-dependence of the
momentum distribution, defined as
\begin{align}
\label{def:nk}
n({\bf k}) = \int dE P(\m{k},E) \ .
\end{align}
It clearly appears that in the region of $|{\bf k}| \gsim 1.5 \ {\rm fm}^{-1}$, or $|{\bf k}|  \gsim 300 \ {\rm MeV}$,  in which short-range
interactions dominate, the curves corresponding to systems other than the deuteron come very close to one another. Note that in this region the 
momentum distributions predicted by the IPM are order of magnitudes lower than those displayed in Fig.~\ref{momdis}, or
vanish altogether.

Highly accurate theoretical calculations of the spectral function can be carried out for uniform nuclear matter, exploiting the simplifications
granted by translation invariance  \cite{PKE}. The results
of these calculations have
been combined with the information obtained from $(e,e^\prime p)$ experiments
to obtain spectral functions of a variety of nuclei within the Local
Density Approximation (LDA) \cite{LDA,Benhar05}.

Being trivially related to the two-point Green's function, the spectral function can be split into two parts exhibiting distinct analytical structures, as prescribed by the K\"all\'en-Lehman representation \cite{Itzykson80}. The resulting decomposition, which combined with the universality of correlation effects provides the 
basis of the LDA scheme, turns out to be \cite{Benhar:KL}
\beq
P(\m{k}, E)= P_{\rm MF}(\m{k}, E) + P_{\rm corr}(\m{k}, E) \ .
\eeq
The first term in the right-hand side of the above equation, describing the contribution arising from the nuclear mean field, exhibits a collection of
peaks corresponding to the energies of the shell-model states belonging to the Fermi sea, while correlations provide a smooth background extending to 
large energy and momentum. 

The mean field term is usually written in the factorized form [compare to Eq. \eqref{def:pkeMF}]
\beq
P_{\rm MF}(\m{k}, E)= \sum_{\alpha \in \{  F \} } Z_\alpha|\phi_\alpha(\m{k})|^2F_\alpha(E - \epsilon_\alpha) \ .
\eeq
The spectroscopic factors $Z_\alpha < 1$ and the functions $F_\alpha(E - \epsilon_\alpha)$, accounting for the finite width of the $\alpha$-th shell-model state, describe the effects of
residual interactions not included in the mean-field picture. In the absence of these interactions,  $Z_\alpha~\rightarrow~1$,
$F_\alpha(E-\epsilon_\alpha)~\rightarrow~\delta(E~-~\epsilon_\alpha)$, and Eq. \eqref{def:pkeMF} is recovered.

Within LDA, the correlation contribution is obtained from 
\beq
P^{\rm LDA}_{\rm corr}(\m{k}, E)= \int \,{d^3r} \ \varrho_A(\m{r}) \ P^{\rm NM}_{\rm corr}(\m{k}, E;\varrho= \varrho_A(\m{r})) \ ,
\eeq
where $\varrho_A(\m{r})$ is the nuclear density distribution and $P^{\rm NM}_{\rm corr}(\m{k},E;\varrho)$ is the continuum part of the spectral function of  nuclear
matter at uniform density $\varrho$. Note that the spectroscopic factors $Z_\alpha$ are constrained by the requirement
\beq
\int \,{d^3k \ dE}\ P_{LDA}(\m{k}, E)= 1 \ , 
\eeq
with 
\beq
P_{\rm LDA}(\m{k}, E)= P_{\rm MF}(\m{k}, E) + P^{\rm LDA}_{\rm corr}(\m{k}, E) \ .
\eeq
Typically, the mean-field contribution accounts for $\sim 80\%$
of the normalisation. The $\sim20\%$ correlation contribution, residing at large $|{\bf k}|$ and $E$, has been
recently measured at Jefferson Lab using a carbon target. The results of this analysis are consistent with the data at low missing energy and missing momentum,
as well as with the results of theoretical calculations carried out within NMBT \cite{daniela,Rohe2006152}.

\begin{figure}[h!]
\begin{center}
\includegraphics[scale= 0.725]{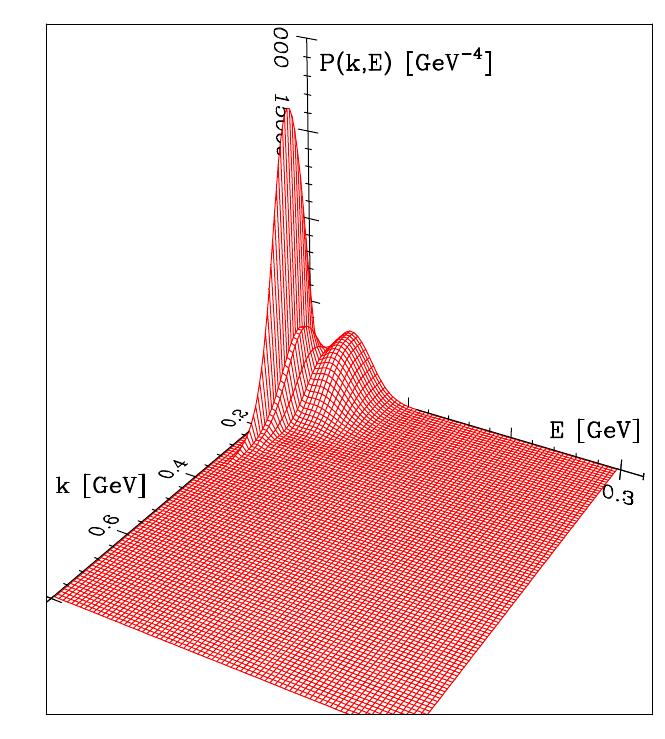}
\end{center}
\vspace*{-.2in}
\caption{Three-dimensional plot
of the oxygen spectral function of  Ref.~\cite{Benhar05}, obtained from the LDA
approach.}
\label{pke16O}
\end{figure}

The oxygen spectral function of Ref.~\cite{Benhar05}, obtained within the LDA
approximation using the results of nuclear matter calculations performed in Correlated Basis Function (CBF) perturbation theory, is shown in Fig.~\ref{pke16O}. 
The peaks corresponding to the shell model states are clearly visible, as is the broad background contribution
arising from removal of a nucleon belonging to a correlated pair.


\begin{figure}[h!]
\begin{center}
\includegraphics[scale= 0.45]{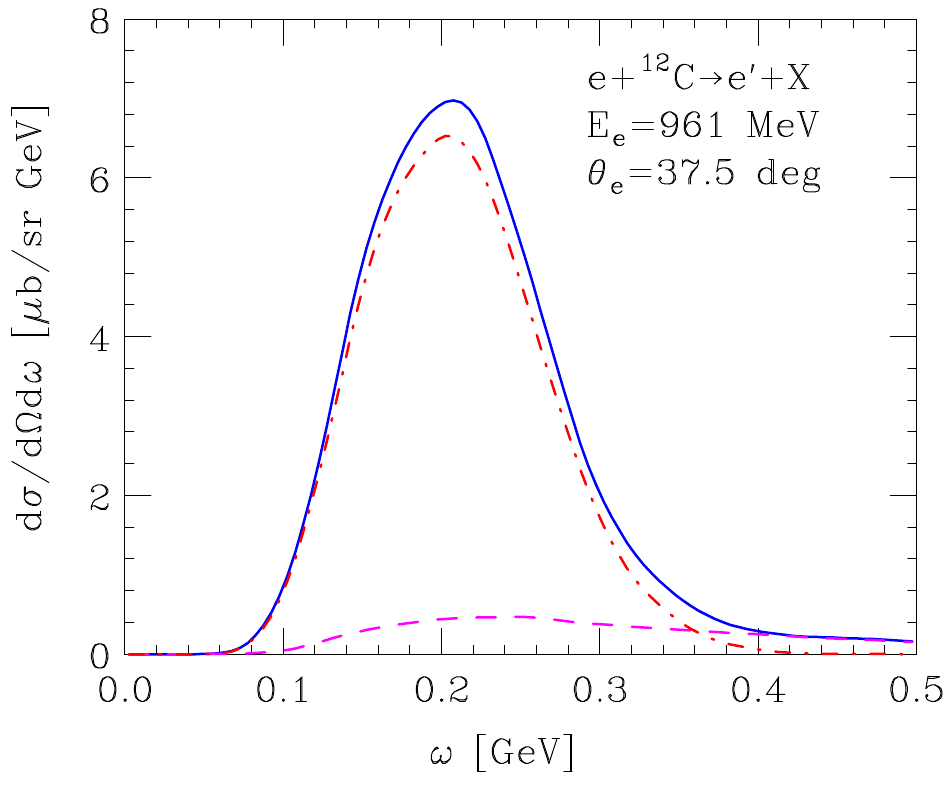}
\end{center}
\vspace*{-.2in}
\caption{Inclusive electron-carbon cross section at $E_e~=~961 \ {\rm MeV}$ and  $\theta_e~=~37.5 \ {\rm deg}$, plotted as a function of the energy loss $\omega$. 
The solid line corresponds to the result of the full IA calculation, while the dot-dash and dashed lines have been obtained, respectively,  including only 1p1h and 2p2h final 
states \cite{Benhar2015a}.}
\label{2bbreak}
\end{figure}

A very important consequence of the presence of the continuum component of the spectral function is that  Eq.~\eqref{sigma1}, in addition to single-nucleon knock
out processes\textemdash in which the target nucleus is left in a bound one-particle\textendash one-hole (1p1h) state\textemdash also describes
interactions leading to the excitation of two-particle\textendash two-hole (2p2h) final states.
As an example, Fig.~\ref{2bbreak}, shows the inclusive electron-carbon cross section in the QE channel, at beam energy $E_e = 961 \ {\rm MeV}$ and
scattering angle $\theta_e = 37.5 \ {\rm deg}$. The dot-dash and dashed lines, obtained using Eq. \eqref{sigma1} and the carbon spectral function of 
Ref.~\cite{LDA}, correspond to the cross sections of processes involving 1p1h and 2p2h final states, respectively, while the solid line shows the results of the 
full calculation. 

Note that the 2p2h contribution displayed in Fig.~\ref{2bbreak} originates from initial state
dynamics only, and would be vanishing in the absence of 
ground-state correlations. The appearance of additional 2p2h contributions arising form FSI will be discussed in Section~\ref{fsi}.

\subsection{$y$-scaling and superscaling}
\label{yscaling}

Scaling in the variable $y$ in the QE sector is a manifestation of the dominance of single nucleon knock out, which allows
to write the equation expressing conservation of energy in a simplified form.

As a consequence, in the limit of large
momentum transfer the function
\begin{align}
\label{def:Fscal}
F(|{\bf q}|,\omega)  = \frac{1}{ Z  \sigma_{ep} + (A-Z) \sigma_{en} } \ \left( \frac{d \omega}{ d k_\parallel} \right)_{k=k_{\rm min}} 
\frac{d^2 \sigma}{d\Omega_{e^\prime} dE_{e^\prime}} \ , 
\end{align}
which in general depends on {\em both} $|{\bf q}|$ and $\omega$, becomes a function of the  single variable
$y=y(|{\bf q}|, \omega)$, defined by the equation  \cite{West75,Sick80}
\begin{align}
\label{def:y}
\omega + M_A  & =  \sqrt{ (y + |{\bf q}|)^2 + m^2 } + \sqrt{y^2 + (M_A - m + E_{\rm min})^2 } \ .
\end{align}
In Eq. \eqref{def:Fscal}, $k_\parallel = |{\bf k} \cdot {\bf q}|$, and $k_{\rm min} \equiv(E_{\rm min},{\bf k}_{\rm min}$), with $E_{\rm min}$
and  $|{\bf k}_{\rm min}|$ being the lowest values of the energy and momentum of the struck nucleon allowed by the kinematical setup. The
quantities  $\sigma_{ep}$ and $\sigma_{en}$ are the elementary electron-proton and electron-neutron cross sections in the QE channel, evaluated at
$k = k_{min}$ and stripped of the
energy conserving $\delta$-function.

\begin{figure}[h!]
 \includegraphics[scale= 0.5]{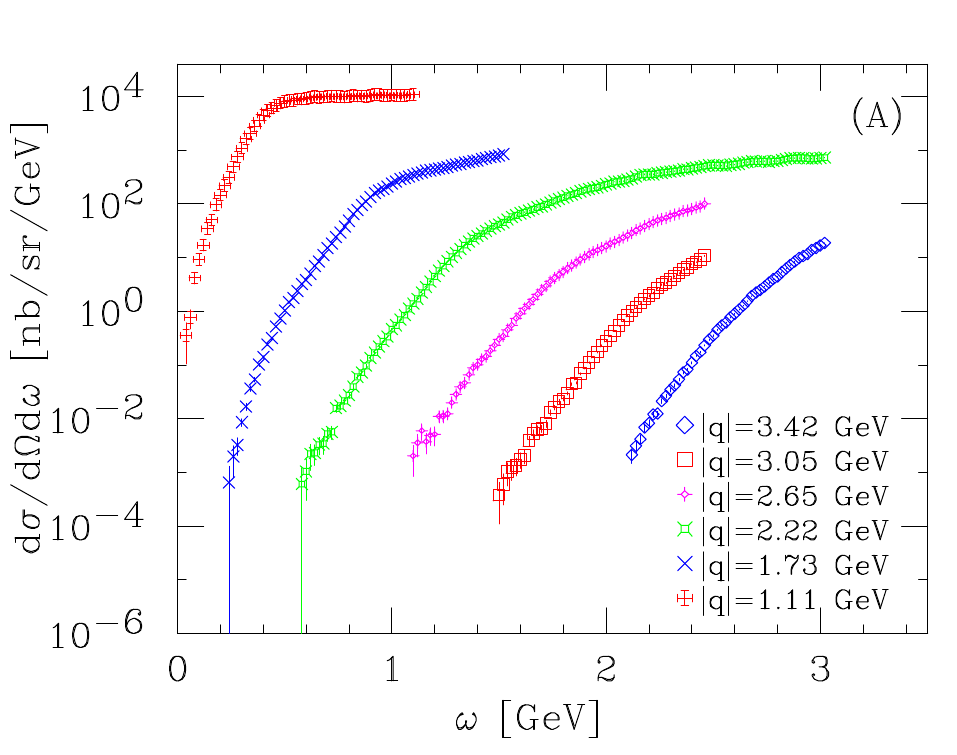}
  \includegraphics[scale= 0.5]{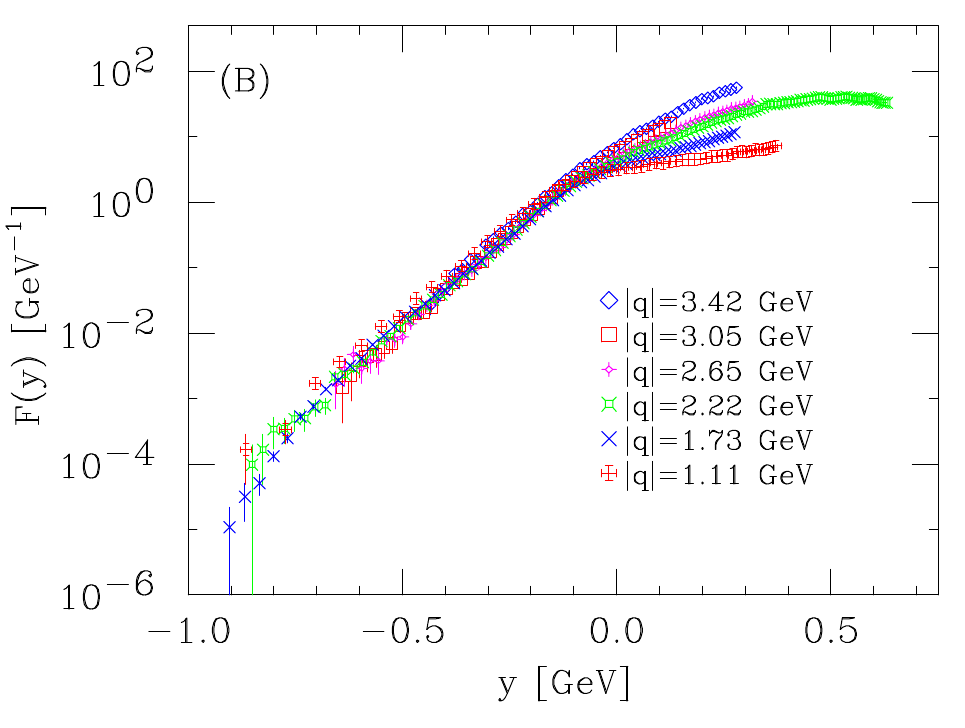}
\vspace*{-.05in}
\caption{Panel (A): inclusive electron-iron cross sections measured at Jefferson Lab \cite{E89-008_1}. Panel (B): $y$-scaling function, defined by  Eqs. \eqref{def:Fscal}
and \eqref{def:y}, obtained from the data shown in panel (A). The values of $|{\bf q}|$ correspond to QE kinematics.}
\label{scaling_AB}
\end{figure}

The 
onset of $y$-scaling is clearly illustrated in Fig. \ref{scaling_AB}, showing the iron data collected at Jefferson Lab by the E89-008 Collaboration \cite{E89-008_1}.
The values of $|{\bf q}|$ listed in the figures correspond to QE kinematics, i.e. to $\omega=\omega_{\rm QE} = Q^2/2m$.
It clearly appears that the inclusive cross sections measured over a broad range of momentum transfer, displayed in panel  (A), collapse to the scaling function $F(y)$ of panel $(B)$ in the
region $y<0$, corresponding to energy transfer $\omega < \omega_{\rm QE}$. On the other hand, large scaling violations, arising  from the presence of reaction mechanisms other than
single nucleon knock out, are visible at $y>0$, or $\omega > \omega_{\rm QE}$  (see Fig.~\ref{regime}).

\begin{figure}[h!]
\begin{center}
 \includegraphics[scale= 0.65]{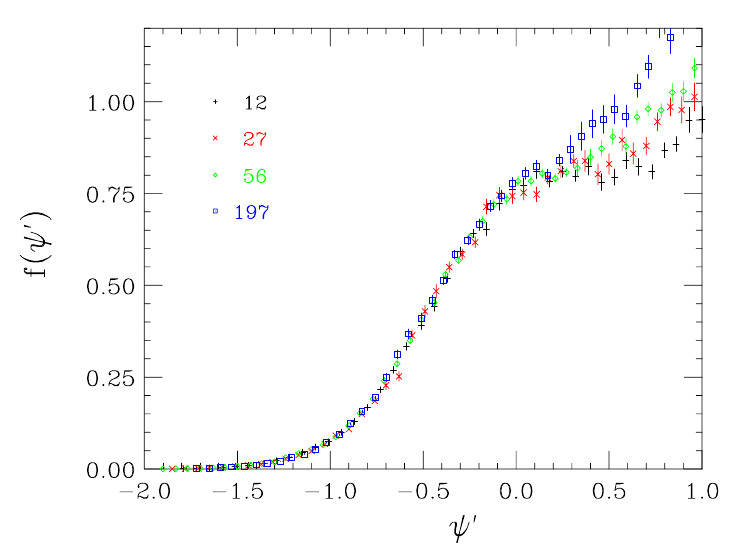}
 \end{center}
\vspace*{-.2in}
\caption{Illustration of scaling of second kind. The scaling functions for nuclei with mass number $12 \leq {\rm A} \leq 197$,
obtained from the data of Ref.~\cite{SLAC} at beam energy $E_e = 3.6 \ {\rm GeV}$ and  electron scattering angle $\theta_e = 16 \ {\rm deg}$, corresponding to $|{\bf q}| \sim 1 \ {\rm GeV}$, are
shown as a function of the variable $\psi^\prime = y/k_F$ \cite{Donnelly_superscaling}.}
\label{superscaling}
\end{figure}

As pointed out above, scaling in the variable $y$, also referred to as scaling of first kind,  reflects the $|{\bf q}|$-independence of the nuclear response at large
 momentum transfer.
A more general form of scaling, dubbed scaling of second kind, allows to eliminate the dependence of the scaling function $F(y)$ on the nuclear
target \cite{Donnelly_superscaling}.  In Fig. \ref{superscaling}, the secon-kind  scaling functions of nuclei with mass number $12 \leq {\rm A} \leq 197$ are shown  as a function of the variable $\psi^\prime= y /k_F$, where
$y$ is defined by Eq. \eqref{def:y} and $k_F$ is the nuclear Fermi momentum.  The functions $f(\psi^\prime)$
have been obtained from the data of Ref.~\cite{SLAC} at beam energy $E_e = 3.6 \ {\rm GeV}$ and  electron scattering angle $\theta_e = 16 \ {\rm deg}$,
corresponding to $|{\bf q}| \sim 1 \ {\rm GeV}$. Simultaneous occurrence of scaling of first and second kind is referred to as superscaling.

The better quality of scaling of second kind, clearly appearing from a 
comparison between Fig.~\ref{scaling_AB} (B) and Fig.~\ref{superscaling} can be easily 
explained considering that---unlike those of Fig.~\ref{scaling_AB}---the data of  
Fig.~\ref{superscaling} correspond to a fixed kinematical setup. Therefore, scaling in  
the variable $\psi^\prime$ merely reflects the universality of the high 
momentum tail of the nuclear momentum distributions, illustrated in 
Fig.~\ref{momdis}.

The superscaling hypotesis has been recently extended to the kinematical region corresponding to $\omega > \omega_{\rm QE}$, in which single-nucleon knockout 
predominantly leads to sresonance production. Within this approach, the contribution of inelastic channels is obtained by subtracting from the data the effective 
QE cross section resulting from the superscaling analysis \cite{Maieron}.

Besides allowing to identify the dominant reaction mechanism,  the occurrence of superscaling can be exploited to predict the nuclear cross section for
kinematical regions and targets not covered by the available data. However,  the inclusion of contributions arising from mechanisms leading to large 
scaling violations, such as meson-exchange currents (MEC), necessarily requires the use of specific models.
The universal scaling function extracted from electron scattering data has been extensively used to obtain
both charged- and neutral-current neutrino-nucleus cross sections \cite{Amaro,Martinez}.

\subsection{Final state interactions}
\label{fsi}

The occurrence of strong 
FSI  in electron-nucleus scattering processes\textemdash not taken into account within the IA
scheme\textemdash has long been experimentally established. The results of a number of $(e,e^\prime p)$ measurements,
covering the kinematical domain corresponding to $0.5 \lsim Q^2 \lsim 8.0$ (GeV/c)$^2$
\cite{Garino92,O'Neill95,Abbott98,Garrow02,Rohe05}, clearly show that the flux
of outgoing protons is strongly suppressed with respect to the IA predictions.
The observed attenuation, parametrized by the nuclear transparency $T_A$,  
ranges from 20-40\% in Carbon to 50-70\% in Gold, and becomes independent 
of momentum transfer for $Q^2 \gtrsim 1$ GeV. This behaviour is        
consistent with the results of theoretical studies of nuclear matter, 
providing clear cut evidence of the persistence of FSI at large momentum 
transfer~\cite{BenharPRL99}.

Being only sensitive to interactions taking place within a distance
$\sim 1/|{\bf q}|$ of the primary vertex, the inclusive cross section at
high momentum transfer is, in general, largely unaffected by FSI.
However, the role of FSI turns out to be appreciable, or even dominant, in the low $\omega$
region, where the cross section obtained within the IA becomes very small.

Let us consider nuclear interactions at fixed beam energy and electron scattering angle in the purely nucleonic sector.
As long as the sum over final states comprises a complete set, FSI {\em do not} affect the $\omega$-integrated 
cross section. Therefore, they can only give rise
to two effects: (i) a shift in $\omega$,
arising from the interaction of the struck nucleon with the mean field of the residual nucleus, and
(ii) a redistribution of the strength---leading to a quenching
of the quasi elastic peak and a corresponding enhancement of the tails---arising from
NN scattering processes coupling 1p1h states to more complex final states.

Within the IPM, FSI can be described replacing the plane wave describing the struck nucleon in the final state with
a wave function obtained from the solution of the Schr\"odinger equation involving a complex optical potential.
This approach, referred to as Distorted Wave Impulse Approximation, or DWIA  (see, e.g.,  Ref.~\cite{BoffiPR}), can be generalised to allow for a consistent treatment
of relativistic effects, and has been widely applied to both inclusive and exclusive processes \cite{Udias,Meucci_etal,Meucci,Giusti_etal}. In inclusive processes, however, there
    is no absorption, and FSI are usually described using a real optical potential.
    An alternative relativistic formalism employed for the description of
    inclusive processes is based on the expansion of the Green's function entering
    the definition of the target response tensor in eigenfunctions of a non-hermitian
    optical potential \cite{Capuzzi1991681}.

In the widely employed convolution approach \cite{Sosnik91,Benhar91} the nuclear cross section is written in terms of the
IA result according to
\begin{align}
\label{convolution}
\frac{d \sigma}{d\Omega_{e^\prime} dE_{e^\prime}} = \int d\omega^\prime \ \frac{d \sigma_{IA}}{d\Omega_{e^\prime} dE_{e^\prime}} \ 
F_{\bf q}(\omega - \omega^\prime) \ ,
\end{align}
where the folding function, defined as
\begin{align}
\label{def:ff}
F_{\bf q}(\omega) = \sqrt{T_A}\delta(\omega)  + (1 - \sqrt{T_A}) f_{\bf q}(\omega) \ , 
\end{align}
embodies all FSI effects.

Equation \eqref{def:ff} shows that the description of FSI involves (i) the nuclear transparency and
(ii) the finite-width folding function $f_{\bf q}(\omega)$. Note that these quantities are both strongly affected by
short-range correlations, since the repulsive core of the NN potential reduces the probability that the struck nucleon may interact with one of the spectator particles
within a distance $\lsim~1 \ {\rm fm}$ of the primary interaction vertex \cite{BenharCT}. In the absence of FSI, $T_A \to 1$, implying that
the residual nucleus is fully transparent to the struck nucleon, $f_{\bf q}(\omega) \to \delta(\omega)$, and the IA cross section of Eq.~\eqref{sigma1}  is recovered.

As pointed out in Ref.~ \cite{Benhar2013}, the convolution approach can be regarded as a generalisation of the spectral function formalism described
in Secion~\ref{PKE}, since the function $F_{\bf q}(\omega)$ turns out to be simply related to the spectral function describing
the propagation of a nucleon in a continuum state. However, for large momentum transfer this quantity cannot be obtained  using 
the non relativistic formalism.

Within the approach developed by the authors of Ref.~\cite{Benhar91}, the folding function is derived within the eikonal approximation, which basically amounts
to assuming that (i) the struck nucleon moves along a straight line with constant velocity, and (ii) the spectator nucleons are seen by the struck nucleon as a collection of fixed
scattering centres. Under these assumptions, the elements entering the calculation of $F_{\bf q}(\omega)$ are the NN scattering amplitude, extracted from the
measured cross sections, and the distribution of the spectator nucleons in coordinate space, that can be consistently obtained within NMBT using the
same dynamical model employed for the description of the initial state.

\begin{figure}[h!]
\begin{center}
\includegraphics[scale= 0.45]{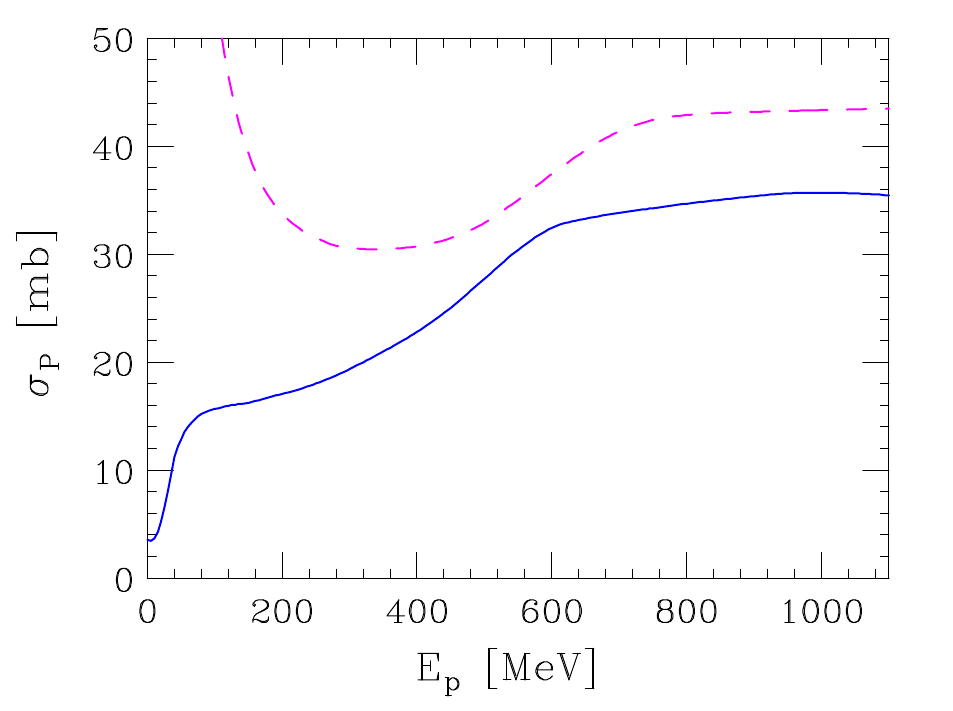}
\end{center}
\vspace*{-.2in}
\caption{Total proton-neutron cross section as a function of the
projectile kinetic energy in the Lab frame \cite{Benhar2013}. The dashed line shows the free space cross section,
while the solid line has been obtained including medium modifications according to the
procedure developed in Refs.~\cite{panpiep,higherord}.}
\label{inmediumxsec}
\end{figure}

The real part of the NN scattering amplitude is the source of the shift of the folded cross section of Eq. \eqref{convolution}, with respect to the IA result. As shown in Ref.~\cite{Ankowski2013}, this effect can be accurately parametrized in terms of  a phenomenological real optical potential.  On the other hand, the imaginary part\textemdash related to the total NN scattering cross section through the optical theorem\textemdash determines the shape of the function $f_{\bf q}(\omega)$.

It has to be pointed out that the NN cross section, driving the rescattering processes, is strongly influenced by the presence of the nuclear medium, which affects both
the incoming flux and the available phase space. The medium modifications of the total cross section in the proton-neutron channel is illustrated by the results of Ref.~\cite{panpiep}, shown in Fig.~\ref{inmediumxsec}.



\subsection{Two-nucleon currents and 2p2h final states}
\label{2bodycurr}

\begin{figure}[h!]
\vspace*{.1in}
\begin{center}
\includegraphics[scale= 0.45]{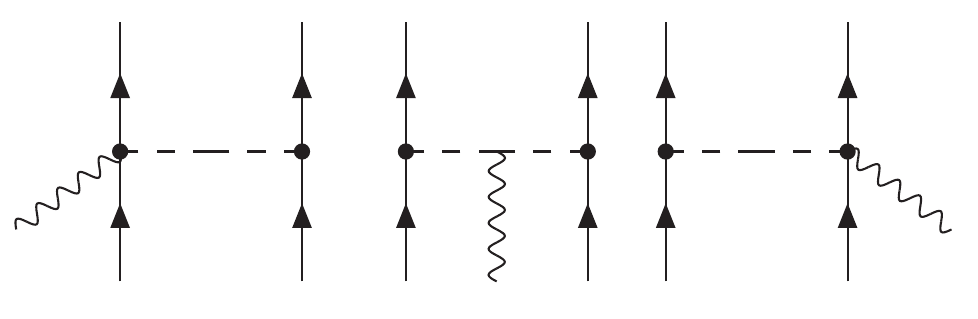}
\end{center}
\vspace*{-.2in}
\caption{Diagrams depicting processes contributing to the electromagnetic  two-nucleon current.
Oriented lines correspond to nucleons, while the wavy and dashed lines are associated with the photons and the exchanged mesons, respectively. }
\label{MEC_1}
\end{figure}

In addition to NN correlations in the initial and final states, interactions involving electromagnetic two-nucleon currents, arising from processes in which the photon couples to a meson exchanged between two interacting 
nucleons, also lead to the excitation of 2p2h final states. As an example, the simplest such processes contributing to the
electron scattering cross section are depicted in Fig. \ref{MEC_1}.

The two-body currents are linked
to the potential describing NN interactions through the continuity equation \eqref{continuity}, establishing
a relation between the nuclear Hamiltonian $H$
and the longitudinal component of the current $J^\mu$.
As a consequence, the operator $J^\mu$ can be separated into model-dependent and
model-independent contributions, the latter being constrained by the NN potential \cite{Riska89}.

As pointed out above, in the regime of low to moderate momentum transfer the
nuclear matrix element of the two-nucleon current can be evaluated using realistic nuclear wave functions, obtained within the
framework of NMBT,  and a non relativistic reduction of the current operator, based on its expansion in powers
of $|{\bf q}| /m$ \cite{Carlson98}. The model-de\-pen\-dent component of the current, being transverse in nature, is not determined by the NN potential.
Existing calculations typically take into account  the isoscalar
$\rho \pi \gamma$ and isovector $\omega \pi \gamma$ transition currents, as well as the isovector
current associated with excitation of intermediate $\Delta$-isobar resonances.
 The two-body charge operators include the $\pi$-, $\rho$-, and
$\omega$-meson exchange charge operators,  the (isoscalar) $\rho \pi \gamma$ and (isovector)
$\omega \pi \gamma$  couplings and the single-nucleon Darwin-Foldy and spin-orbit relativistic corrections  \cite{Schiavilla90}.

The role of the two nucleon current in electron scattering is best illustrated by comparing the longitudinal and
transverse contributions to the scaling function $F(y)$, defined by Eqs.~\eqref{def:Fscal} and \eqref{def:y}.

It is important to recall that the occurrence of scaling of first kind provides a strong handle on the
identification of the reaction mechanism, while the observation of scaling violations reveals the role played
by processes beyond the IA. In this context, valuable information is provided by the scaling analysis of the
longitudinal (L) and transverse (T) contributions to the measured cross sections [see Eq.~\eqref{eA:xsecLT}].

\begin{figure}[h!]
\begin{center}
\includegraphics[scale= 0.5]{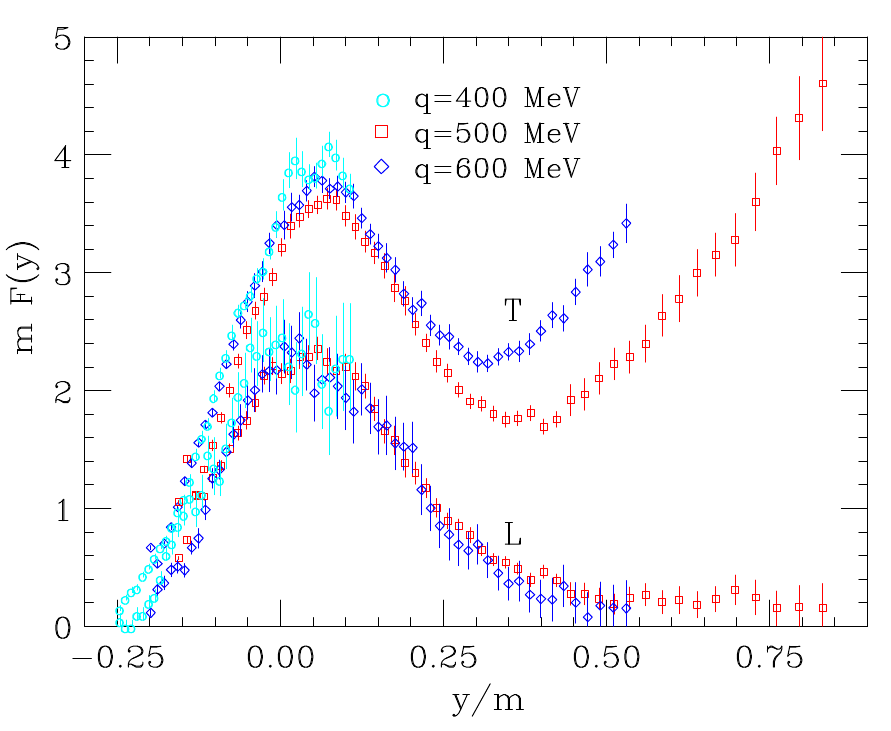}
\end{center}
\vspace*{-.2in}
\caption{$y$-dependence of the longitudinal (L) and transverse (T) scaling functions of Carbon at $|{\bf q}| =$ 400, 500, and 600 MeV \cite{Finn84}, obtained  from
the analysis of the data of  Ref.~\cite{Barreau83}. Note that $y$ is given in units of the nucleon mass, $m$, and that the scaling function is multiplied by $m$,
to obtain a dimensionless quantity.}
\label{scaling}
\end{figure}

Figure \ref{scaling} shows the $y$-dependence of the L and T scaling functions obtained by the authors of Ref.~\cite{Finn84} using the
corresponding carbon responses, extracted from the cross sections measured at Saclay \cite{Barreau83}.
The onset of scaling is manifest in the region of the quasi free peak, corresponding to $y \sim 0$,
where the data points at different momentum transfer tend to sit on top of one another as $|{\bf q}|$ increases. On the other hand, large scaling violations,
mainly arising  from non QE processes, such as resonance production, are clearly visible in the transverse channel at $y>0$, corresponding to $\omega > \omega_{\rm QE}$.
In addition, the T scaling function turns out to be significantly 
larger than the L one, while within the IA picture\textemdash neglecting the small
convection terms in the nucleon current\textemdash the L and T scaling functions are predicted to be identical (see, e.g., Ref.~\cite{Cennietal}).

The results of {\em ab initio} Green's Function Monte Carlo (GFMC)
calculations of the longitudinal and transverse responses of nuclei
with ${\rm A} \leq 12$ \cite{Carlson02,Inversion} provide convincing evidence that the   
pattern observed in Fig.~\ref{scaling} is driven by processes involving 
two-nucleon currents, whose contributions, while being negligible 
in the longitudinal channel, give rise to a significant enhancement of 
the transverse response.



The role of the two-body currents in determining the sum rules of the L and T
responses, defined as
\begin{align}
\label{SL}
S_L(|{\bf q}|) = \frac{1}{Z} \int_{\omega_{\rm th}}^\infty d\omega \ R_L(|{\bf q}|,\omega) \ ,
\end{align}
and
\begin{align}
\label{ST}
S_T(|{\bf q}|) = \frac{2}{Z \mu_p + N \mu_n} \frac{m^2}{|{\bf q}|^2} \int_{\omega_{\rm th}}^\infty d\omega \ R_T(|{\bf q}|,\omega) , 
\end{align}
has been thoroughly analysed by  the authors of Ref.~\cite{Carlson02} using the GFMC approach. In the above equations, $R_L$ and $R_T$ are
the response functions defined in Eq.\eqref{eA:xsecLT}, $\mu_p$ and $\mu_n$ are the proton and neutron magnetic moments, respectively, and the lower
integration limit,  $\omega_{\rm th}$, corresponds to the threshold of inelastic scattering.

The numerical results of Ref.~\cite{Carlson02}, including the L and T sum rules of
$^3$He, $^4$He and $^6$Li at momentum transfer $300 \leq |{\bf q}| \leq 700 \  {\rm MeV}$, indicate that two-nucleon currents are responsible for
a  $\sim 20-40$\% enhancement of  of the T sum rule, while the typical contribution to $S_L$ is a $\sim 5$\% decrease. A similar pattern emerges from the analysis of Ref.~\cite{Inversion}, whose authors have computed the longitudinal and transverse responses of $^{12}{\rm C}$.

\begin{figure}[h!]
\begin{center}
 \includegraphics[scale= 0.45]{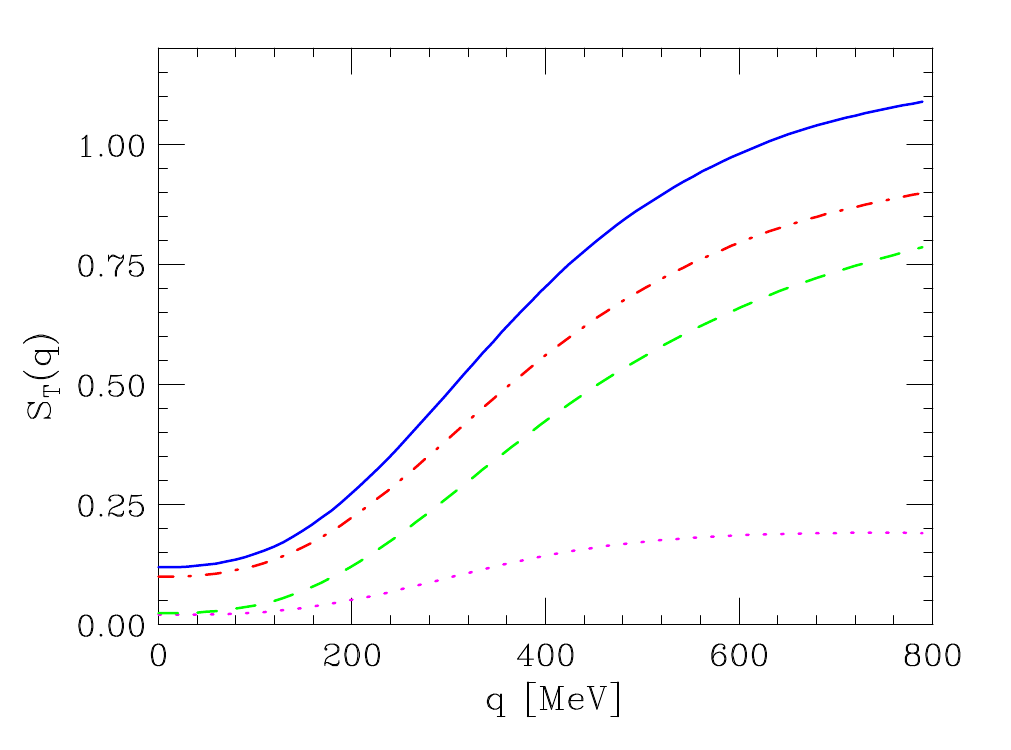}
\end{center}
\vspace*{-.2in}
\caption{Sum rule of the electromagnetic response of carbon in the transverse channel, defined by Eq.~\eqref{ST}. The dashed line
shows the results obtained including the one-nucleon current only, while the solid line corresponds to the
 full calculation. The dot-dash line represents the sum rule computed neglecting interference
terms, the contribution of which is displayed by the dotted line. The results are normalised so that the dashed line approaches
unity as $|{\bf q}| \to \infty$ \cite{Benhar2015a}. }
\label{sumrule}
\end{figure}

As pointed out above, owing to the presence of NN correlations 2p2h final states can be excited in processes involving
both one- and the two-body currents. Within the IA scheme, the contribution of the one-body current can be taken into account
using spectral functions derived from realistic nuclear models, in which the ground state has non vanishing overlaps with the two hole-one particle
states of the residual system \cite{PKE}. On the other hand, the discussion of Section \ref{PKE} implies that all models based on the mean
field approximation fail to meet this requirement.

A consistent treatment of the one- and  two-nucleon current contributions to the nuclear cross section in the 2p2h sector requires that interference between
the corresponding amplitudes\textemdash including the one associated with the excitation of 2p2h final states in the aftermath of a rescattering of the knocked
out particle, to be discussed below\textemdash be carefully taken into account.

The role of interference terms in determining the transverse electromagnetic response of $^{12}$C has been recently analysed within the GFMC approach.
The results of this study, displayed in Fig.~\ref{sumrule}, clearly show that interference is the source of a sizeable fraction of the sum rule.
At momentum transfer $|{\bf q}| \gsim 300$ MeV, its contribution turns out to be comparable to\textemdash in fact even larger than\textemdash the one arising from the squared
matrix element of the two-nucleon current \cite{Benhar2015a}.

A fully consistent description of one- and two-body current contributions to the nuclear cross sections in the region in which the non relativistic
approximation is expected to break down involves serious difficulties. Existing calculations have been mainly carried out using diagrammatic approaches, based on simplified descriptions of the the
nuclear initial and final states, obtained from either the RFGM or more advanced implementations of the mean field approximation \cite{MEC_TO,MEC_PV}.

A novel approach, recently proposed in Refs.~\cite{Benhar:2013bwa,Benhar2015a}, is based on a generalisation of the factorisation {\em ansatz} described in Section~\ref{IA}.
The 2p2h final state is written in the form [compare to Eq.~\eqref{factorization}]
\beq
\label{fact2}
| X \rangle = | {\bf p} {\bf p}^\prime \rangle \otimes | {\rm R}_{A-2} , {\bf p}_{\rm R} \rangle \ ,
\eeq
where the states $| {\bf p} {\bf p}^\prime \rangle$ and $| {\rm R}_{A-2} , {\bf p}_{\rm R} \rangle$ describe two non interacting nucleons, carrying  momenta ${\bf p}$ and ${\bf p}^\prime$,  and
the recoiling $({\rm A}-2)$-particle spectator system, respectively.

Using Eq. \eqref{fact2} and following the procedure described in Section \ref{IA}, the nuclear matrix element of the two-nucleon current can be written in terms of
two-body matrix elements according to
\begin{align}
\label{matel:2}
\langle 0 | J_2^{\mu} | X \rangle  = \frac{m}{ \sqrt{E_p E_{p^\prime}} }
  \int d^3k d^3 k^\prime   {\mathcal M}_{\rm R} ({\bf k},{\bf k}^\prime)    
   \sum_{j>i}\langle {\bf p} {\bf p}^\prime | j_{ij}^\mu | {\bf k} {\bf k}^\prime \rangle  \ ,
\end{align}
with the amplitude ${\mathcal M}_{\rm R}({\bf k},{\bf k}^\prime)$ given by
\begin{align}
\label{def:Mn}
{\mathcal M}_{\rm R}({\bf k},{\bf k}^\prime) = \left\{ \langle {\rm R}_{(A-2)} , {\bf p}_{\rm R} | \otimes \langle  {\bf k}  {\bf k}^\prime | \right\} | 0 \rangle \ .
\end{align}

Within the scheme outlined in Eqs. \eqref{fact2}-\eqref{def:Mn}, the nuclear amplitude ${\mathcal M}_{\rm R}({\bf k},{\bf k}^\prime)$ turns out to be independent of ${\bf q}$,
and can therefore be obtained within NMBT. On the other hand, the two-nucleon matrix element between free
nucleon states can be evaluated using the fully relativistic expression of the current.

The connection with the spectral function formalism
becomes apparent noting that the two-nucleon spectral function $P({\bf k},{\bf k}^\prime,E)$,
yielding the probability of removing {\em two} nucleons with momenta ${\bf k}$ and ${\bf k}^\prime$ from the nuclear ground state, leaving the residual system with excitation energy $E$, is
defined as~\cite{spec2}
\begin{align}
\label{def:pke2}
P({\bf k},{\bf k}^\prime,E) = \sum_{\rm R} |{\mathcal M}_{\rm R}({\bf k},{\bf k}^\prime)|^2 \delta(E + E_0 - E_{\rm R}) \ ,
\end{align}
with ${\mathcal M}_{\rm R}({\bf k},{\bf k}^\prime)$  given by Eq. \eqref{def:Mn}.

\begin{figure}[h!]
\begin{center}
\includegraphics[scale=0.45]{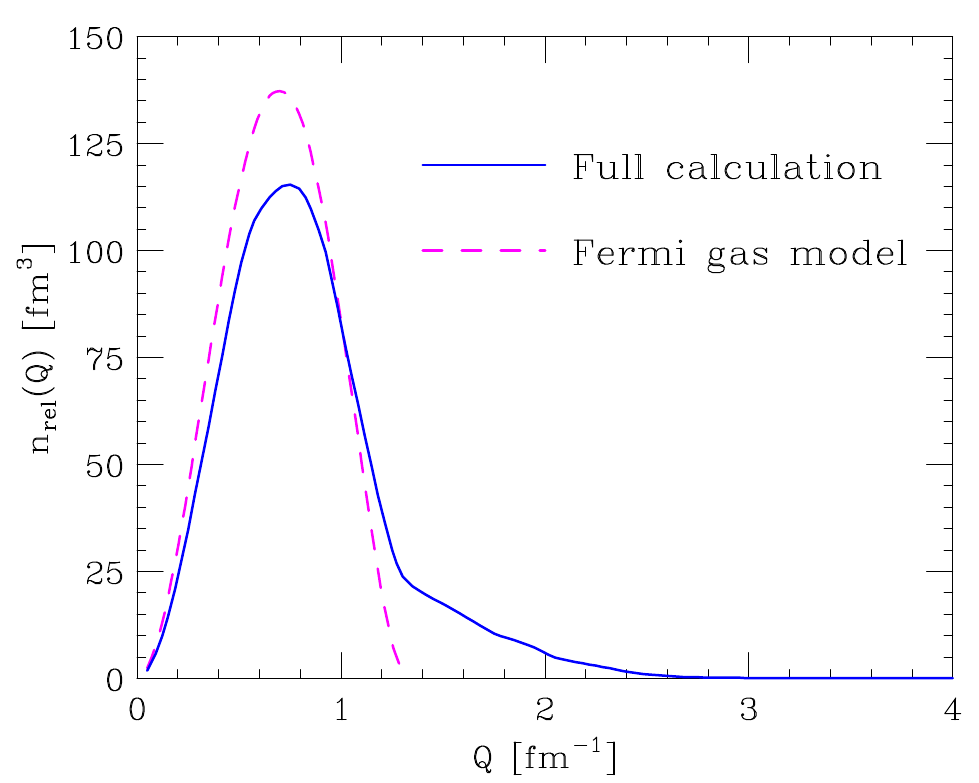}
\end{center}
\vspace*{-.1in}
\caption{Relative momentum distribution of a nucleon pair in isospin symmetric nuclear matter at equilibrium density \cite{spec2}.}
\label{SF2}
\end{figure}

The two-nucleon spectral function of uniform and isospin symmetric nuclear matter at equilibrium density has been calculated
 by the authors of Ref.~\cite{spec2} using CBF perturbation theory and a realistic Hamiltonian. The resulting relative momentum distribution, defined as
\begin{align}
\label{rel:dist}
n_{\rm rel}({\bf Q}) = 4 \pi |{\bf Q}|^2 \int d^3 K \ n\left( \frac{ {\bf K} }{2} + {\bf Q}, \frac{ {\bf K} }{2} - {\bf Q} \right)
\end{align}
where ${\bf K} = {\bf k}+{\bf k}^\prime$, ${\bf Q} = ({\bf k}-{\bf k}^\prime)/2$, and
\begin{align}
n({\bf k},{\bf k}^\prime) = \int dE  \ P({\bf k},{\bf k}^\prime,E) \ ,
\end{align}
is shown by the solid line of Fig. \ref{SF2}. Comparison with the prediction of the Fermi Gas model, represented by the dashed line,
clearly illustrates the significance of correlation effects.

\subsection{Collective excitations}
\label{RPA}

At low momentum transfer, the interaction with the beam particle may involve more than one nucleon, and give rise to long-range correlations
leading to the appearance of collective excitations of the target nucleus.

The contribution of collective excitations can be approximated writing the final state appearing
in Eq.~\eqref{nuclear:tensor} as a superposition of 1p1h states. The resulting expression of the nuclear response involves the
propagator of the particle-hole pair  excited at the the interaction vertex, $\Pi({\bf q},\omega)$, carrying momentum and energy ${\bf q}$ and $\omega$.

Within the commonly used Random Phase Approximation (RPA),  $\Pi({\bf q},\omega)$ is obtained from an integral equation,  which allows to take
into account the contributions of the so called ring diagrams to all orders. Most existing calculations have been performed within a theoretical framework
based on the IPM, in which the particle-hole interaction is described in terms of perturbative meson exchange, augmented by
a set of phenomenological parameters. This scheme has been also extended to include contributions involving 
$\Delta$-resonance  production~\cite{Alberico,Gil1997543}.   

In recent years, a significant effort has been devoted to the development of a treatment of RPA correlations based on more realistic dynamical models~\cite{Cowell,Farina}. 
In Refs.~\cite{LLB,LBGL}, the weak responses of nuclear matter at low momentum transfer have been computed using an effective particle-hole 
interaction obtained from a phenomenological  Hamiltonian including two- and three-nucleon potentials. The results of these studies indicate that\textemdash as 
it was to be expected on the basis of very general quantum mechanical considerations\textemdash the effects of collective excitations, while being large at $|{\bf q}| < 100 \ {\rm MeV}$, 
become less and less important with decreasing momentum transfer, and turn out to be vanishingly small
at $|{\bf q}| \gsim 400 \ {\rm MeV}$ \cite{Farina}.

\subsection{Comparison to data}
\label{compare}

The ability to account for electron scattering data is the obvious prerequisite to be fulfilled by any model of neutrino interactions.
Below, we provide some representative comparisons between the results of the approaches outlined in the
previous Sections and the available empirical information, obtained from the measured cross sections.

Non relativistic {\em ab initio} calculations of the electromagnetic responses of few-nucleon systems with $A \leq 4$, performed 
using realistic nuclear Hamiltonians, have reached a remarkable degree of accuracy, and provide a good description of a large body 
of data (see, e.g., Refs.~\cite{Golak,CS92,Efros94,Efros97,Efros04}). 
In this review, however, we will focus on the heavier nuclei employed to detect neutrino interactions, such as carbon, oxygen and argon.


The electromagnetic longitudinal  and transverse responses of carbon, defined as in Eq~ \eqref{eA:xsecLT}, have been analysed using the GFMC formalism.
The available results include the Euclidean responses, related to the corresponding $\omega$-space responses by Laplace transformation \cite{Euclidean}, as well as the sum 
rules defined by Eqs.\eqref{SL} and \eqref{ST} \cite{LovatoSR,Inversion}. 

\begin{figure}[bth]
\begin{center}
\includegraphics[scale=0.3]{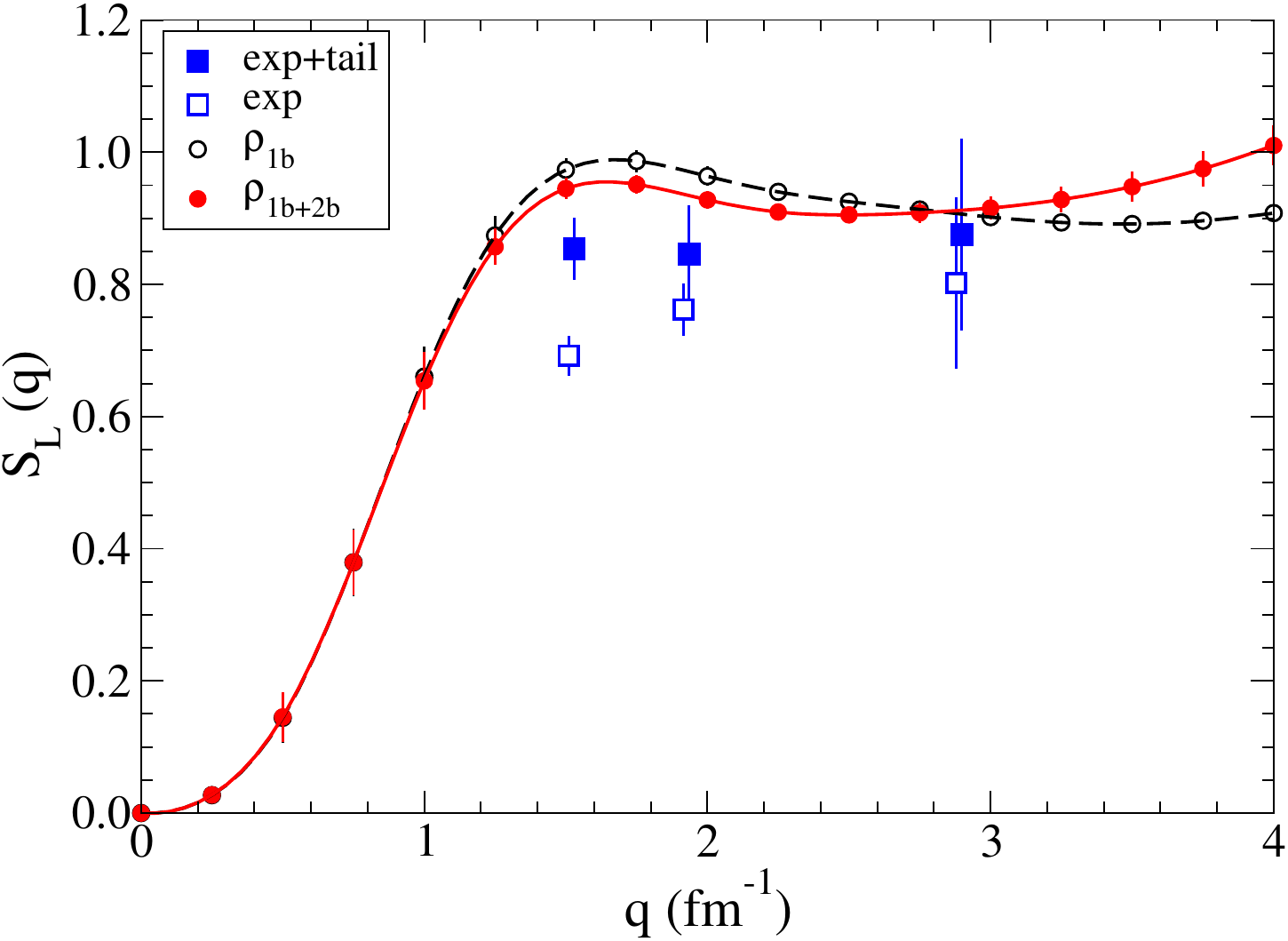}
\end{center}
\vspace*{-.15in}
\caption{Longitudinal sum rule of $^{12}$C, computed by the authors of Ref.~\cite{LovatoSR} using the
GFMC formalism. 
The solid line has  been obtained using the full current operator of Eq.\eqref{def:curr},
while the dashed line does not take into account the contribution of two-nucleon terms.
The experimental data, with (full squares) and without (open squares) tail correction, correspond to the 
response functions resulting from the analysis of Ref.~\cite{Jourdan96}.
}
\label{SL_carbon}
\end{figure}
%
\begin{figure}[bth]
\vspace*{.25in}
\begin{center}
\includegraphics[scale=0.3]{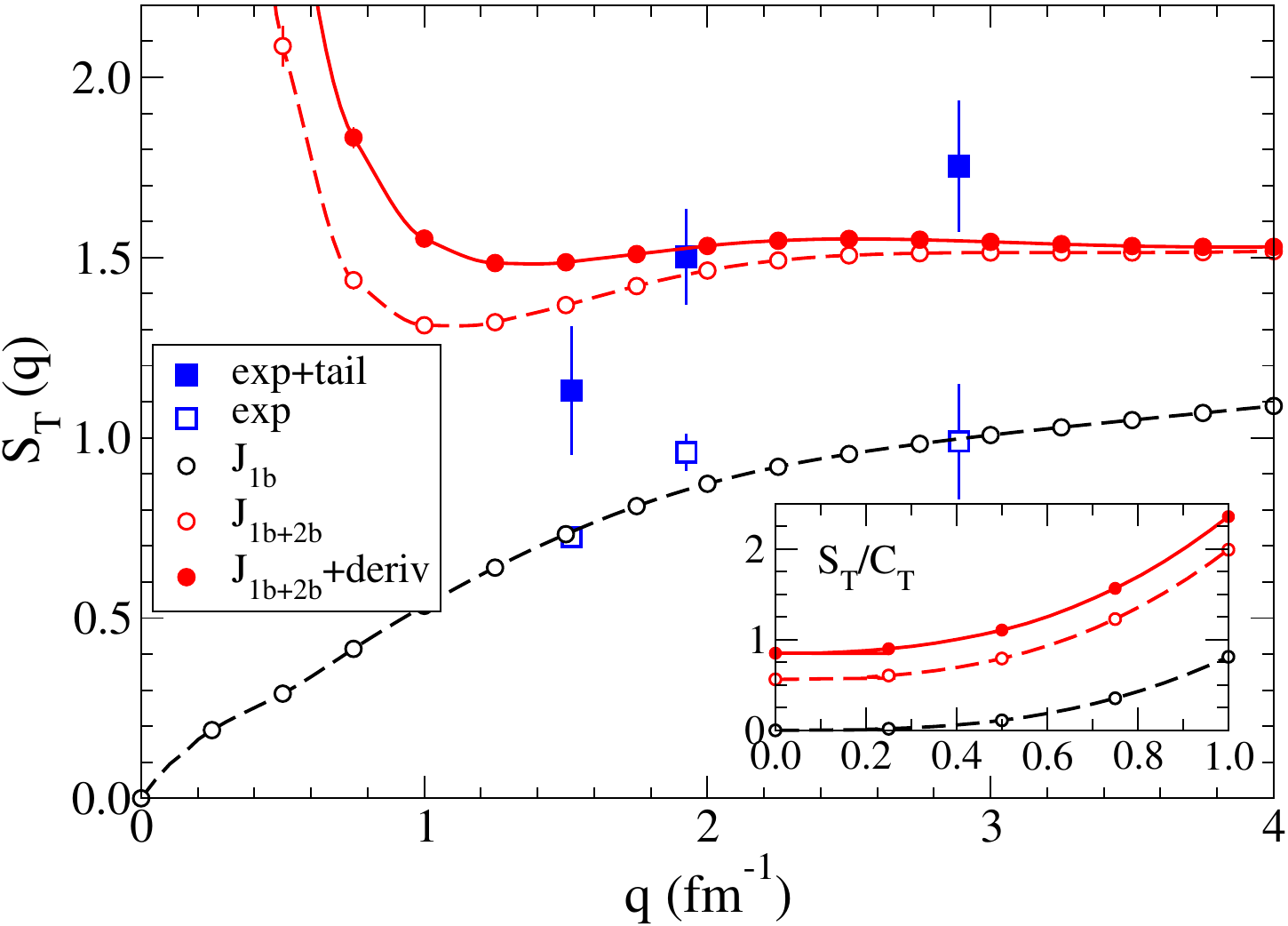}
\end{center}
\vspace*{-.15in}
\caption{Same as in Fig.~\ref{SL_carbon}, but for the
transverse sum rule.  The inset shows  the
low-$|{\bf q}|$ behaviour obtained removing from the definition of $S_T$, Eq. \eqref{ST}, the $\sim 1/|{\bf q}|^2$ divergent coefficient
    \cite{LovatoSR}.}
\label{ST_carbon}
\end{figure}

Figures \ref{SL_carbon} and \ref{ST_carbon} show the longitudinal and transverse sum rules of carbon computed by the authors of  Ref.~\cite{LovatoSR}. Theoretical results turn out to be in satisfactory agreement with the data, corrected to take into account the contribution of the
region of large $\omega$ not covered by the experiments 
(see Ref.~\cite{LovatoSR} and references therein). In addition, a comparison between the solid and dashed lines confirms that
the two-nucleon currents, while playing a nearly negligible role in the longitudinal channel, provide a large contribution to the transverse sum rule.
A recent, more accurate, analysis has shown that the agreement between the GFMC longitudinal sum rule of carbon and the data can be further improved taking 
into account the contributions of the low-lying $J^\pi = 2^+,  \ 0_2^+ \ {\rm (Hoyle)},  \ {\rm and} \  4^+$ states~\cite{Inversion}.

The main limitation of the sum rules is the lack of information on the energy distribution of strength.
As a consequence, the analysis based on sum rules does not answer the question of whether the excess of transverse strength arising fom interactions involving the two-nucleon
currents occurs mostly at large $\omega$\textemdash well beyond the quasi-elastic peak\textemdash or it is also found at $\omega \approx \omega_{\rm QE}$.

The inversion of the Euclidean response, needed to retrieve the energy dependence, is
long known to involve severe difficulties. A groundbreaking result has been recently reported in Refs.~\cite{Euclidean,Inversion}, whose authors
exploited the maximum entropy technique to obtain the
electromagnetic longitudinal and transverse responses of $^4$He and $^{12}$C.

\begin{figure}[bth]
\begin{center}
\includegraphics[scale=1.1]{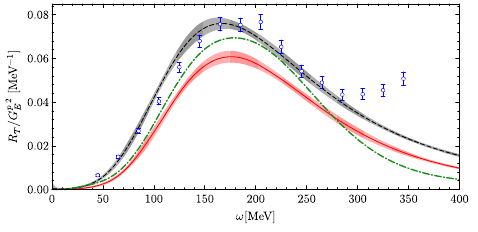}
\end{center}
\vspace*{-.2in}
\caption{Transverse electromagnetic
response functions of $^{12}$C at  $|{\bf q}|=570$ MeV, computed within the GFMC approach \cite{Inversion}. The dashed and solid lines have been 
obtained using the full nuclear current operator and neglecting the two-body terms, respectively, while the dot-dash line
shows the results of a PWIA calculation. The shaded areas represent the uncertainty associated with the inversion of the euclidean response.  
The data is taken from Ref.~\cite{Jourdan96}}
\label{fig:RT_inverted}
\end{figure}

\begin{figure}[bth]
\begin{center}
\includegraphics[scale=1.1]{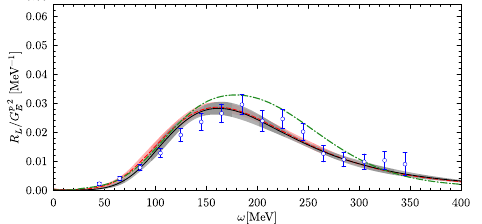}
\end{center}
\vspace*{-.2in}
\caption{Longitudinal electromagnetic
response functions of $^{12}$C at  $|{\bf q}|=570$ MeV, computed within the GFMC approach \cite{Inversion}. The solid and dashed lines have been 
obtained using the full nuclear current operator and neglecting the two-body terms, respectively, while the dot-dash line
shows the results of a PWIA calculation. The shaded areas represent the uncertainty associated with the inversion of the euclidean response.  
The data is taken from Ref.~\cite{Jourdan96}}
\label{fig:RL_inverted}
\end{figure}

Figures~\ref{fig:RT_inverted} and \ref{fig:RL_inverted}  show the $\omega$ dependence of  
the transverse and longitudinal responses of $^{12}$C at $|\mathbf{q}~|=~570 \ {\rm MeV}$,  obtained by inverting the corresponding euclidean response \cite{Inversion}.
It appears that the GFMC approach provides a quantitative account  of the data in the region in which the description 
in terms of purely nucleonic degrees of freedom is applicable. The contribution arising from processes involving two-nucleon currents is sizeable  
in the transverse channel, and extends over the whole $\omega$ range. On the other hand, it turns out to be nearly negligible in the 
longitudinal channel.
 

\begin{figure*}[t!]
\centering
    \subfigure{\label{fig:electrons_a}}
    \subfigure{\label{fig:electrons_b}}
    \subfigure{\label{fig:electrons_c}}
    \subfigure{\label{fig:electrons_d}}
    \subfigure{\label{fig:electrons_e}}
    \subfigure{\label{fig:electrons_f}}
    \subfigure{\label{fig:electrons_g}}
    \subfigure{\label{fig:electrons_h}}
    \subfigure{\label{fig:electrons_i}}
    \includegraphics[width=0.90\textwidth]{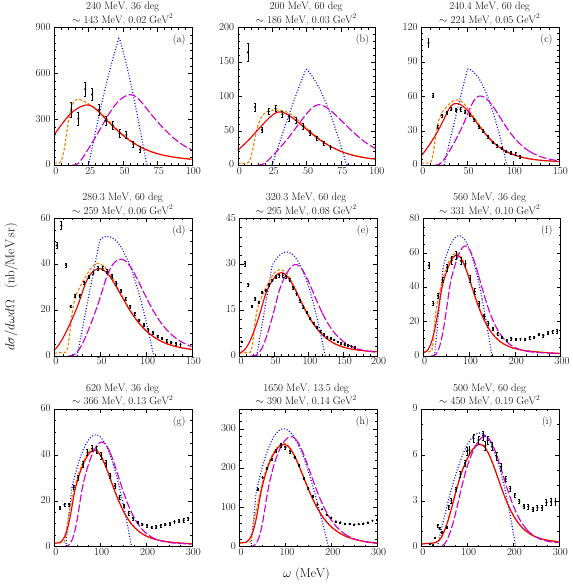}
\caption{\label{fig:electrons}Double differential electron-carbon cross sections in the QE channel, computed by the authors of Ref.~\cite{Ankowski2013}
within the spectral function approach, compared to the data of 
Ref.~\cite{Barreau83,Baran88,Whitney74}. The solid lines correspond to the result of the full calculation, whereas the long-dashed lines have been obtained neglecting FSI.
The difference between the solid and short-dashed lines illustrates the effect of using alternative treatments of Pauli blocking. For comparison, the prediction of the RFGM are
also shown, by the dotted lines. The panels are labeled according to beam energy, scattering angle, and values of $|{\bf  q}|$ and $Q^2$ at the quasielastic peak.}
\end{figure*}

The results of the approach based on the IA and the spectral function formalism, corrected to take into account the effects of FSI
within the convolution scheme described in Section \ref{fsi} \cite{Ankowski2013}, are displayed by the solid lines of Fig. \ref{fig:electrons}.
It clearly appears that the description of nuclear dynamics based on a realistic Hamiltonian and nuclear many-body theory, which does not include
any adjustable parameters,  provides a quantitative description of the data in the region in which QE scattering is dominant.
On the other hand, the dotted lines show that the RFGM, while yielding an acceptable account of few measured cross sections,
conspicuously fails to explain the data over the entire ranges of beam energy and scattering angle.

\begin{figure}[bth]
\begin{center}
\includegraphics[scale=1.1]{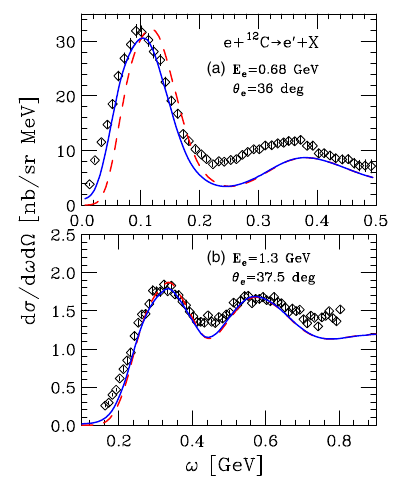}
\end{center}
\vspace*{-.2in}
\caption{(a): double differential electron-carbon cross section at beam energy $E_e=680 $ MeV and scattering angle $\theta_e= 36$ deg \cite{NoemiPRL}. The solid and dashed lines have been 
obtained from the spectral function approach, with and without inclusion of FSI.  The experimental data are taken from Ref.~\cite{Barreau83}. (b): same as (a) but for $E_e= 1300$ MeV and $\theta_e= 37.5$ deg. The experimental data are taken from Ref.~\cite{12C2}.}
\label{Noemi}
\end{figure}

Although the results of Ref.~\cite{Ankowski2013} are limited to the QE sector,  the spectral function formalism provides a framework for the consistent description of both elastic and inelastic single-nucleon knock out processes. Moreover,  as pointed out above, it can be generalised to include both FSI and the contributions of processes involving two-nucleon currents. Figure~\ref{Noemi} shows a comparison between the measured electron-carbon cross sections of Refs.~\cite{Barreau83,12C2} and the results of the calculations discussed in 
Ref.~\cite{NoemiPRL}, in which MEC and FSI are both taken into account.

Besides the {\em ab initio} approaches based on the Hamiltonian of Eq. \eqref{H:A},  a number of schemes based on different---and somewhat simplified---descriptions  of nuclear structure and dynamics have been employed to study electron-nucleus scattering.

\begin{figure}[h!]
\begin{center}
 \includegraphics[scale=0.85]{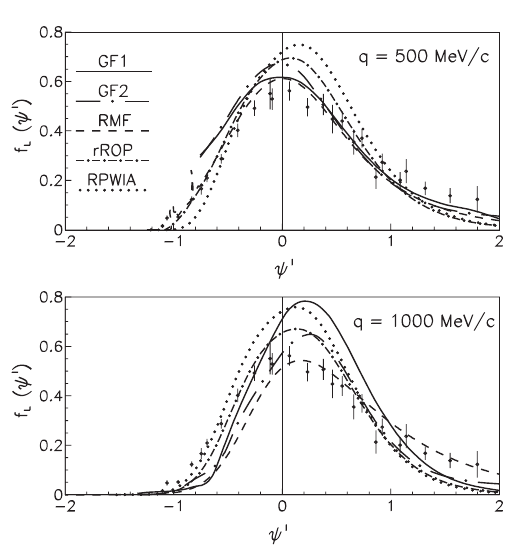}
 \end{center}
  \vspace*{-.2in}
\caption{Superscaling functions computed by the authors of Ref.~\cite{Meucci_etal_2}. The meaning of the labels is explained throughout the text.}
\label{meanfield}
\end{figure}

As an example, Fig.  \ref{meanfield} shows the superscaling functions at $|{\bf q}| =$ 500 and 1000 MeV, obtained using the mean field approximation for the initial state and two alternative
approaches for the treatment of FSI. The curves labeled GF1 and GF2 have been obtained from the Green's function formalism  (see Section \ref{fsi}) and two different complex optical potentials,
while the curve labeled RMF corresponds to calculations carried out within the
   relativistic mean field scheme, using the same 
   real (scalar and vector) potentials to describe  both bound and scattering
   nucleon states. On the other hand, the curve labelled rROP shows the results obtained using the real part of the relativistic
   optical potential to determine the outgoing nucleon wave function.
For comparison, the result obtained neglecting FSI altogether are also shown, by the line labelled RPWIA  \cite{Meucci_etal_2}.

\begin{figure}[h!]
\begin{center}
 \includegraphics[scale=1.45]{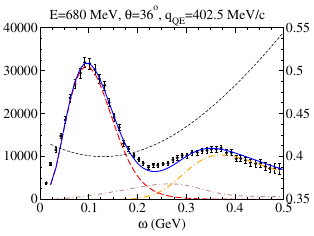} \\
 \includegraphics[scale=1.45]{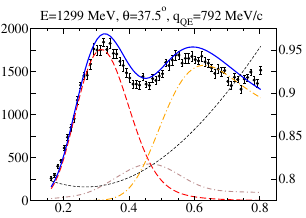}
 \end{center}
\caption{Comparison between the measured electron-carbon cross sections of Refs.~\cite{Barreau83} (upper panel) and \cite{12C2} (lower panel) and the results of the 
SuSav2-MEC approach of Ref.~\cite{Megias2016}, represented by the solid lines. The vertical axis on the left corresponds to the double-differential cross section  
in units of nb/sr/GeV.}
\label{megias}
\end{figure}

The phenomenological superscaling analysis has been recently extended with the inclusion of additional information, obtained from studies carried out 
within the RMF approach,  which allow to 
pin down the nuclear responses in the isoscalar and isovector channels \cite{SuSav2}. The model described in Ref.~\cite{SuSav2}, referred to as SuSav2, has been further 
developed to take into account inelastic channels, and augmented considering the contributions of processes involving MEC, described according to the RFGM~\cite{Megias2016}. The results displayed in Fig.~\ref{megias} show that the 
hybrid procedure derived in Ref.~\cite{Megias2016}, dubbed SuSav2-MEC, describes the measured electron-carbon cross sections with accuracy comparable 
to that obtained using the spectral function formalism. 

\begin{figure}[bth]
\begin{center}
\includegraphics[scale=0.85]{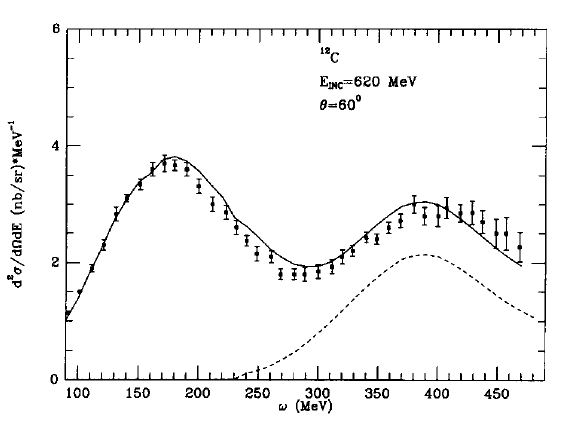}\\ \includegraphics[scale=0.85]{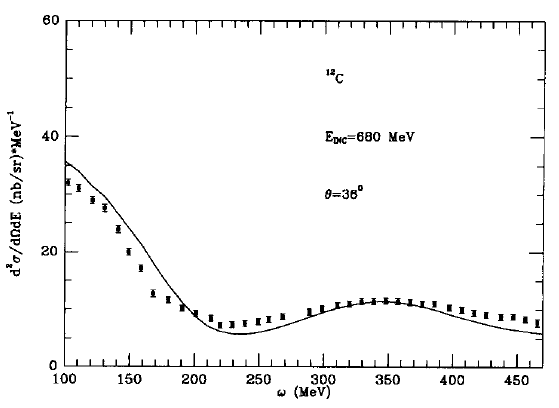}
\caption{Electron-carbon cross sections at $E_e = 620$ MeV and $\theta_e = 60$ deg (upper panel) and $E_e = 680$ MeV and $\theta_e = 36$ deg (lower panel)
obtained from the model of Ref.~\cite{Gil1997543}. The data points are taken from Ref.~\cite{Barreau83}.}
\end{center}
\label{oset}
\end{figure}

The authors of Refs.~\cite{Alberico,Gil1997543} developed diagrammatic approaches to study the inclusive electron nucleus cross section,  in 
which the effects of long range RPA correlations is explicitly  taken into account, and found to be significant. 
Within these models, in which the description of the target ground state is based on the RFGM, interaction effects are included at the level of perturbative 
meson exchange, and short-range correlations are taken into account through phenomenological modifications of the NN amplitudes. 
As an example, Fig.~\ref{oset} shows the electron-carbon 
cross sections reported by the authors of Ref.~\cite{Gil1997543} for two kinematical setups, corresponding to $|{\bf q}| \sim 600 \ {\rm MeV}$ (upper panel) and $\sim 400 \ {\rm MeV}$ (lower panel).
The model of Ref.~\cite{Gil1997543} has been also applied to the description of a variety of semi-inclusive electron scattering processes, including $(e,e^\prime N)$, $(e,e^\prime NN)$,
and $(e,e^\prime \pi)$ \cite{Gil1997599}.


\section{The flux-integrated neutrino-nucleus cross section}
\label{nuA:xsec}

For definiteness, we will consider charged-current neutrino-nucleus interactions at fixed neutrino energy.
The formalism discussed in this Section can be readily generalized
to the case of neutral current interactions (see, e.g., Ref.~\cite{veneziano}).

The double differential cross section of the
process [compare to Eq. \eqref{eeprime}]
\beq
\nu_\ell + A \to \ell^- + X \ ,
\eeq
can be written in the form (see, e.g., Ref.~\cite{Benhar:2006nr})
\begin{align}
\label{nucl:xsec}
\frac{d^2\sigma}{d\Omega_{\ell} dE_\ell}& =\frac{G_F^2\,V^2_{ud}}{16\,\pi^2}\,
\frac{| {\bf k}_\ell |}{ |{\bf k}_\nu |}\,L_{\lambda \mu}\, W^{\lambda \mu} \ .
\end{align}
Here $k_\nu \equiv (E_\nu,{\bf k}_\nu)$ and $k_\ell \equiv (E_\ell,{\bf k}_\ell)$
are the four momenta carried by the incoming neutrino and the outgoing charged lepton, respectively,
$G_F$ is the Fermi constant and $V_{ud}$ is the CKM matrix element coupling $u$
and $d$ quarks. Neglecting the term proportional to $m_\ell^2$,
where $m_\ell$ is the mass of the charged lepton, the tensor $L^{\lambda\mu}$ can be cast in the form
\begin{align}
\label{leptensor}
L^{\lambda \mu} =  8 \,\left[ k^\lambda_\ell \,k^\mu_\nu  + k_\nu^{\lambda}\,k_\ell^\mu- 
g^{\lambda \mu}(k_\ell \cdot k_\nu) -i\,\varepsilon^{\lambda\mu\alpha\beta}\,{k_\ell}_\beta\,{k_\nu}_\alpha \right] \ , 
\end{align}
where $\varepsilon^{\lambda\mu\alpha\beta}$ is the fully antisymmetric Levi-Civita tensor.

The target tensor is written as in Eq.~\eqref{nuclear:tensor}, replacing the electromagnetic current with the
weak charged current. Within the IA scheme, it reduces to
\begin{align}
\label{tens:IAnu}
W^{\lambda\mu}  =  N  \int {d^3k \ dE} \  \frac{M}{E_k} P(\m{k},E)   \mathcal{W}_n^{\lambda\mu} \ , 
\end{align}
where the tensor $ \mathcal{W}_n^{\lambda\mu}$ describes the interaction of a free neutron of four momentum $k$ at
four momentum transfer $\tilde{q} \equiv ( \tilde{\omega}, {\bf q})$, with $\tilde{\omega}$ given by Eq.~\eqref{def:omegatilde}.
Its most general expression
can be written in terms of five structure functions according to
\begin{align}
\label{W2}
{\mathcal W}^{\lambda\mu} =
-g^{\lambda\mu}W_1 + \frac{k^{\lambda}k^{\mu}}{m^2}W_2   & -i\epsilon^{\lambda\mu\varrho\sigma}\frac{ k_\varrho k_\sigma} {m^2}W_3 \\
\notag
 & + \frac{q^{\lambda}q^{\mu}}{m^2}W_4
+ \frac{k^{\lambda}q^{\mu} + k^{\lambda}q^{\mu}}{m^2}W_5\ .
\end{align}
Note that in scattering processes involving isolated nucleons, after contraction of ${\mathcal W}^{\lambda\mu} $ with the lepton tensor $L_{\lambda \mu}$, the structure functions $W_4$ and $W_5$ give vanishing contributions to the cross section. Owing to the replacement $q \to {\tilde q}$ in the
arguments of $\mathcal{W}^{\lambda \mu}$,  in neutrino-nucleus scattering this is no longer the case. However, the
results of numerical calculations suggest that the contributions of the terms involving $W_4$ and $W_5$ are small,
and can be safely neglected \cite{Benhar:2006nr}.

From Eqs.~\eqref{leptensor} and \eqref{W2} one obtains
\beq
L^{\lambda\mu}{\mathcal W}_{\lambda\mu}= 16 \ \sum_i W_i \ \Big( \frac{A_i}{m^ 2}\Big) \ ,
\eeq
the kinematical factors $A_i$ being given by
\begin{align}
\label{def:A}
\nonumber
A_1 & = m^2 \ (k \cdot k^\prime)  \ , \\
\nonumber
A_2 & = (k \cdot  p)\,(k^\prime \cdot p)-\frac{A_1}{2} \ , \\ 
A_3 & =(k\cdot  p)\,(k^\prime \cdot  \tilde q)-(k \cdot \tilde q)\,(k^\prime \cdot  p) \ , \\
\nonumber
A_4 & =(k\cdot \tilde q)\,(k^\prime \cdot \tilde q)-\frac{\tilde q^2}{2}\,\frac{A_1}{m^2} \ , \\
\nonumber
A_5 &  = (k\cdot  p)\,(k^\prime \cdot \tilde q)+(k^\prime \cdot  p)\,(k\cdot \tilde q)-(\tilde q \cdot p)\,\frac{A_1}{m^2} \ .
\end{align}

The flux-integrated double differential neutrino-nucleus cross section, defined as
\begin{align}
\label{fluxint:sigma}
\frac{d \sigma}{ dT_\ell d\cos \theta_\ell}  = \frac{1}{N_\Phi}
 \int dE_\nu \Phi(E_\nu) \frac{ d \sigma}{dE_\nu dT_\ell d\cos \theta_\ell} \ , 
\end{align}
where $T_\ell = E_\ell - m_\ell$ is the kinetic energy of the outgoing charged lepton and
\begin{align}
N_\Phi =  \int dE_\nu \Phi(E_\nu) \ , 
\end{align}
can be readily obtained from the above equations, yielding the expression of the double differential cross section at fixed neutrino energy derived within the
IA.

Equations \eqref{nucl:xsec}-\eqref{tens:IAnu} show that, within the scheme based on the IA, the calculation of the neutrino-nucleus cross section requires two ingredients: (i) the structure functions
describing the elementary neutrino-nucleon interactions and (ii) the  spectral function, describing the properties of the nuclear initial state. It follows that, to the extent to which the structure functions $W_i$ are known, the  spectral function formalism provides a unified  framework, suitable to describe neutrino-nucleus scattering in different kinematical regimes, in which different reaction
mechanisms dominate.


In Section \ref{W:QE} we report the explicit expression of the structure functions relevant to CC QE interactions, which are the ones
exploited for oscillation analyses. For completeness, the corresponding structure functions  for the resonance production and deep inelastic scattering channels
     are briefly discussed in Sections \ref{res} and \ref{dis}.

Owing to the moderate value of the mean energy of the neutrino flux of Fig. \ref{fluxav},  $\langle E_\nu \rangle \sim 800$ MeV, the MiniBooNE CC QE sample accounts for $\sim$ 60\% of the total cross section, the remaining $\sim$ 40\% being associated with inelastic processes. On the other hand, the DUNE neutrino flux has its maximum at $ E_\nu  \sim 2.5$ GeV and exhibits a long high-energy tail \cite{Akiri:2011dv}. As a consequence, the fractions of CC QE, resonance production and deep inelastic events turn out to  be $\sim$ 5\%, $\sim$ 35\% and $\sim$ 60\%, respectively, and the identification of the relevant signal will require a fully quantitative
understanding of the background of non CC QE interactions \cite{PhysRevD.89.093003}.


It has to be pointed out that, while in this review we mainly focus on the scheme based on the IA and the spectral function formalism\footnote{Note that this scheme provides the 
basis of a variety of different models, including the RFGM.},
a unified description of  nuclear effects has been also developed within a completely different approach, based on transport theory.
This approach,  thoroughly described in Ref.~\cite{Buss:2011mx}, provides the conceptual framework underlying the Giessen Boltzmann-Uehling-Uhlenbeck (GiBUU) event generator, extensively applied to a variety of processes, ranging from pion-induced nuclear reactions to heavy-ion collisions and photon- and lepton-nucleus scattering.

\subsection{Quasielastic scattering}
\label{W:QE}

In the CC QE channel, the structure functions involve the energy conserving $\delta$-function enforcing the condition that  the
scattering process be elastic. Therefore, they can be conveniently written in the form
\beq
\label{def:Wtilde}
W_i = {\widetilde W}_i \ \delta\Big(\tilde{\omega} + \frac{\tilde{q}^2}{2M}\Big) \ ,
\eeq
with the ${\widetilde W}_i$ determined by the matrix elements of the weak nucleon current. Exploiting the
CVC and PCAC hypoteses, the resulting structure functions, can be written in terms of the vector form factors,
$F_1$ and $F_2$, and the axial-vector form factor, $F_A$, according to
\begin{align}
\label{strfunc}
\nonumber
{\widetilde W}_1 & = 2[F_A^2(1 + \tau) + \tau(F_1 + F_2)^2] \  ,  \\
\nonumber
{\widetilde W}_2 & = 2[F_A^2 + F_1^2 + \tau F_2^2] \ , \\
{\widetilde W}_3 & = 2F_A(F_1 + F_2)  \ , \\ 
\nonumber
{\widetilde W}_4 &= [F_2^2(1 + \tau) - 2F_2(F_1 + F_2)]/2  \ , \\
\nonumber
{\widetilde W}_5 & = W_2/2 \ ,
\end{align}
with $\tau = - {\tilde q}^2/4m^2$.

The form factors appearing in the vector current, $F_1(q^2)$ and $F_2(q^2)$, are obtained from the measured proton and neutron electric and magnetic
form factors, $G^N_E$ and $G^N_M$ $(N = p, n$), through the relations
\begin{align}
F_i(q^2) = F_i^p(q^2) - F_i^n(q^2)
\end{align}
with $i = 1,2$ and
\begin{align}
F_1^N(q^2) & = \frac{1}{(1 -\tau)} [ G_E^N(q^2) - \tau G_M^N(q^2) ] \  ,  \\
F_2^N(q^2) & = \frac{1}{(1 -\tau)}[ -G_E^N(q^2) + G_M^N(q^2)] \ .
\end{align}

While more refined parametrisations of the large body of electron scattering data are available \cite{Kelly2004,BBBA}, the form factors $G_E^N$ and $G_M^N$ are often written in the simple dipole approximation
\begin{align}
\nonumber
G_E^p(q^2) & = \Bigg( 1 - \frac{q^2}{m_V^2}\Bigg)^{-2} \  \ \ ,  \ \ \ G_E^n(q^2) = 0 \ , \\
G_M^N(q^2) & = \mu_N \Bigg( 1 - \frac{q^2}{m_V^2}\Bigg)^{-2} \ ,
\end{align}
with $m_V^2= 0.71\ \rm{GeV}^2$. The axial form factor, $F_A$, is also written in the same form
\beq
\label{FA}
F_A(q^2)= g_A\Bigg( 1 - \frac{q^2}{m_A^2}\Bigg)^{-2} \ .
\eeq
The  axial coupling constant, $g_A=~-1.2761^{+14}_{-17}$, is obtained from neutron $\beta$-decay \cite{g_A}, while the axial mass
determined from elastic neutrino- and antineutrino-nucleon scattering, charged pion
electro-production off nucleons and muon capture on the proton is $m_A~=~1.03 \ {\rm GeV}$ \cite{bernard,bodek2}.

Note that in Eq.~\eqref{strfunc} the contributions  involving the pseudoscalar form factor, $F_P$, have been neglected. This
approximation is largely justified, except  for the case of $\nu_\tau$ scattering.

\subsection{Resonance production}
\label{res}

The generalisation of the formalism summarised in the previous section to describe resonance production, driven by elementary processes such as
\begin{align}
\label{delta++}
\nu_\mu + p \to \mu^- + \Delta^{++}  \to \mu^- + p + \pi^+ \ , 
\end{align}
only requires minor changes \cite{Benhar:2006nr}.
In this case, the nucleon tensor $\mathcal{W}^{\lambda\mu}$ involves  the matrix elements of the weak current describing the nucleon-resonance transitions.
As a consequence, the structure functions\textemdash which can still be written in terms of phenomenological vector and axial-vector form factors\textemdash  depend on both $q^2$ and $W^2$, the squared invariant mass of the hadronic final state. In addition, the energy conserving $\delta$-function in
Eq.\eqref{def:Wtilde} is replaced by a Breit-Wigner factor according to
\beq
 \delta(W^2 - m^2)  \to \frac{M_R\Gamma_R}{\pi}\frac{1}{(W^2-M_R^2)^2 + M_R^2\Gamma_R^2} \ ,
\eeq
where $M_R$ and $\Gamma_R$ denote the resonance mass and decay width, respectively \cite{Benhar:2006nr,LP} .

\begin{figure}[h!]
\begin{center}
 \includegraphics[scale= 0.5]{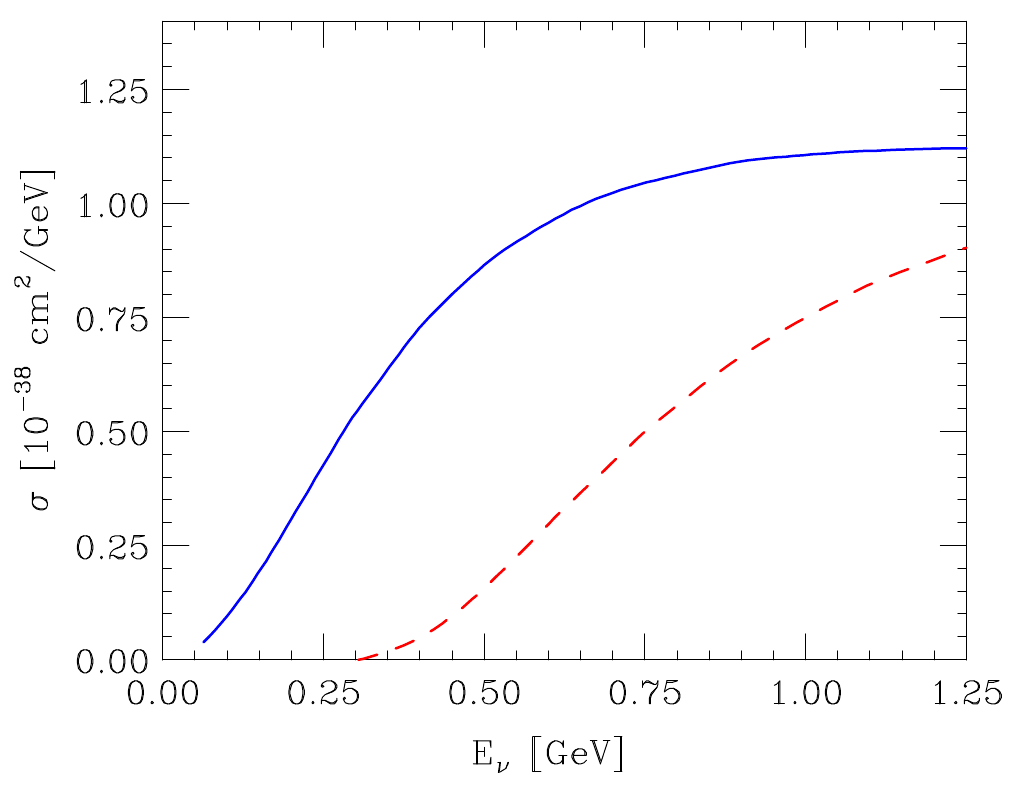}
\end{center}
\vspace*{-.2in}
\caption{QE (solid line) and resonance production (dashed line) contributions to the total cross section of charged-current
neutrino-nucleon interactions (adapted from Ref.~\cite{Benhar:2006nr}).} 
\label{sigtot}
\end{figure}

As an example, illustrating the relative weight of QE scattering and resonance producion, the CC QE and resonance contributions to the total neutrino-nucleon cross section reported in Ref.~\cite{Benhar:2006nr} are shown in Fig. \ref{sigtot} as a function of neutrino energy. The resonance-production cross section has been obtained taking into account both
the $P_{33}(1232)$ $\Delta$ resonance and the three isospin $1/2$ states lying in the so-called second resonance region: $P_{11}(1440)$,  $D_{13}(1520)$
and $S_{11}(1535)$ \cite{Benhar:2006nr}. It clearly appears that at beam energies $\gsim 1$ GeV, QE scattering and resonance production 
turn out to be provide comparable contributions.

It must be kept in mind that he decay of the $\Delta$ resonance [see Eq.~\eqref{delta++}] is a prominent mechanism leading to the appearance of pions, 
and that events in which the final state pion is absorbed in the target nucleus are a major background to QE processes. 
A detailed discussion of both coherent and incoherent pion production can be found in Refs.~\cite{sajjad} and \cite{Leitner:2008ue}.

\subsection{Deep inelastic scattering}
\label{dis}

From the observational point of view, the Deep Inelastic Scattering (DIS) regime corresponds to hadronic final states with more than one pion.

In principle, the three nucleon structure functions entering the definition of the IA nuclear cross section, Eqs.~\eqref{sigma1} and \eqref{sigma2},
may be obtained combining neutrino and antineutrino scattering cross sections. However, as the available structure functions have been
extracted from nuclear cross sections (see, e.g., Ref.~\cite{CDHS}),
their use in {\em ab initio}  theoretical studies aimed at identifying nuclear effects entails obvious conceptual difficulties.

An alternative approach, allowing to obtain the structure functions describing DIS on isolated nucleons, can be developed within the conceptual 
framework of the quark-parton model, exploiting the large database of accurate DIS data collected using charged lepton beams and  hydrogen and deuteron targets (see, e.g., Ref.~\cite{roberts}).
Within this scheme,  the function $F_2^{\nu N} = \omega W_2$, where $\omega$ is the energy transfer and  $W_2$ is the weak structure function of  an isoscalar nucleon,
defined as in Eq. \eqref{W2}, can be simply related to the corresponding structure function extracted from electron scattering data, $F_2^{e N}$, through\footnote{For simplicity, here and in
what follows we will neglect the contributions of $s$- and $c$-quarks.} 
\begin{align}
\notag
F_2^{\nu N}(Q^2,x) & = x \ [ \ u(Q^2,x) + \overline{u}(Q^2,x)  + d(Q^2,x) + \overline{d}(Q^2,x) \ ] \\
 & = \frac{18}{5} \  F_2^{e N}(Q^2,x)  \ , 
 \label{DIS1}
\end{align}
where $x$ is the Bjorken scaling variable, 
while $q(Q^2,x)$ and $\overline{q}(Q^2,x)$, with $q = u, d$, denote the quark and antiquark distributions, respectively. 
In addition, the relation
\begin{align}
\label{DIS2}
x \omega W_3(Q^2,x)  = x F_3^{\nu N}  & = x \ [ \ u_{\rm v}(Q^2,x)  + d_{\rm v}(Q^2,x)  \ ] \ , 
\end{align}
with $u_{\rm v}(Q^2,x)$  and $d_{\rm v}(Q^2,x)$ being the valence quark distributions, implies
\begin{align}
\label{DIS3}
x F_3^{\nu N}  = F_2^{eN} - 2 x \  [ \overline{u}(Q^2,x) + \overline{d}(Q^2,x)]   \ . 
\end{align}

Using Eqs.~\eqref{DIS1}-\eqref{DIS3} and the Callan-Gross relation,  linking  $F_1^{\nu N} =  m W_1$ to $F_2^{\nu N}$,  one can readily obtain all the weak
structure functions from the existing parametrisations of the measured electromagnetic structure functions and of the antiquark distribution (see, e.g., Ref~.\cite{GRV98}).  
Alternatively, the quark and antiquark distributions can be also used to
obtain the structure function $F_2^{e N}(Q^2,x)$ from Eq.\eqref{DIS1}.

The above procedure rests on the tenet, underlying the IA scheme, that the elementary neutrino-nucleon interaction is {\em not} affected by the
presence of the nuclear medium. While this assumption is strongly supported by electron-nucleus scattering data in the quasi elastic channel, showing no medium modifications of the nucleon vector form factors, analyses of neutrino DIS data are often
carried out allowing for a medium modification of the nucleon structure functions \cite{petti,haider}, or of the parton distributions entering their definitions \cite{kumano}.

The approach of  Ref.~\cite{petti,haider} makes use of a model of the nuclear spectral function, and includes a variety of medium effects, such as the
$\pi$- and $\rho$-meson cloud contributions and nuclear shadowing. On the other hand,  Ref.~\cite{kumano} provides a parametrization of the {\em nuclear}
parton distributions to order $\alpha_s$\textemdash $\alpha_s$ being the coupling constant of strong interactions\textemdash obtained from a fit to the measured nuclear cross sections.


\section{Interpretation of CC QE events}
\label{interpretation}

The data set of CC QE events collected by the MiniBooNE collaboration \cite{miniboone_ccqe_2} provides an unprecedented opportunity to carry out a systematic
study of the double differential cross section of the process,
\beq
\nu_\mu + ^{12}\mkern -5mu C \rightarrow \mu^- + X \ ,
\eeq
averaged over the neutrino flux shown in Fig. \ref{fluxav}.

As pointed out in the previous section, the CC QE neutrino-nucleon process is described in terms of three
form factors. The proton and neutron electromagnetic form factors, which have been precisely measured
up to large values of $Q^2$ in electron-proton and electron-deuteron scattering experiments,
and the nucleon axial form factor $F_A$, parametrized in terms of the axial mass $m_A$ as in Eq. \eqref{FA}.
The data analysis performed using the RFGM yields an axial mass $m_A \approx 1.35$ GeV, significantly larger than that obtained from deuteron
data \cite{bernard,bodek2}. A large value of the axial mass, $m_A \approx 1.2$ GeV, has been also obtained from the analysis of the CC QE neutrino-oxygen
cross section carried out by the K2K collaboration \cite{Gran:2006jn}, while the NOMAD collaboration reported the value $m_A = 1.05$ GeV\textemdash
compatible with the world average of deuteron data\textemdash resulting from the analysis of CC QE neutrino- and antineutrino-carbon interactions at much larger beam energies
($\langle E_\nu \rangle \sim 26$ GeV) \cite{Lyubushkin:2008pe}.

\begin{figure}[h!]
\begin{center}
\includegraphics[scale= 0.45]{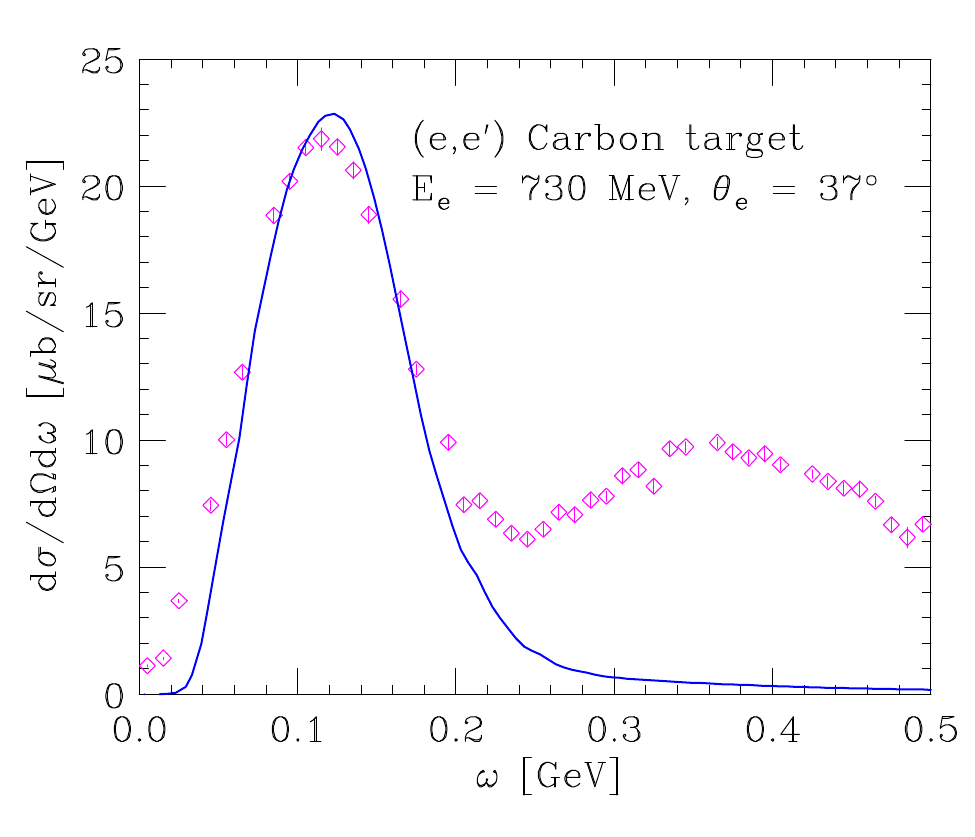}
\end{center}
\vspace*{-.2in}
\caption{ Inclusive electron-carbon cross section at beam energy $E_e=$ 730 MeV and electron scattering
angle $\theta_e=37^\circ$, plotted as a function of the energy loss $\omega$  \cite{Benhar:2010nx}. The solid line shows the results 
obtained using the spectral function formalism. The data points are taken from Ref.~\cite{12C1}.}
\label{compare1}
\end{figure}

\begin{figure}[h!]
\begin{center}
\includegraphics[scale= 0.60]{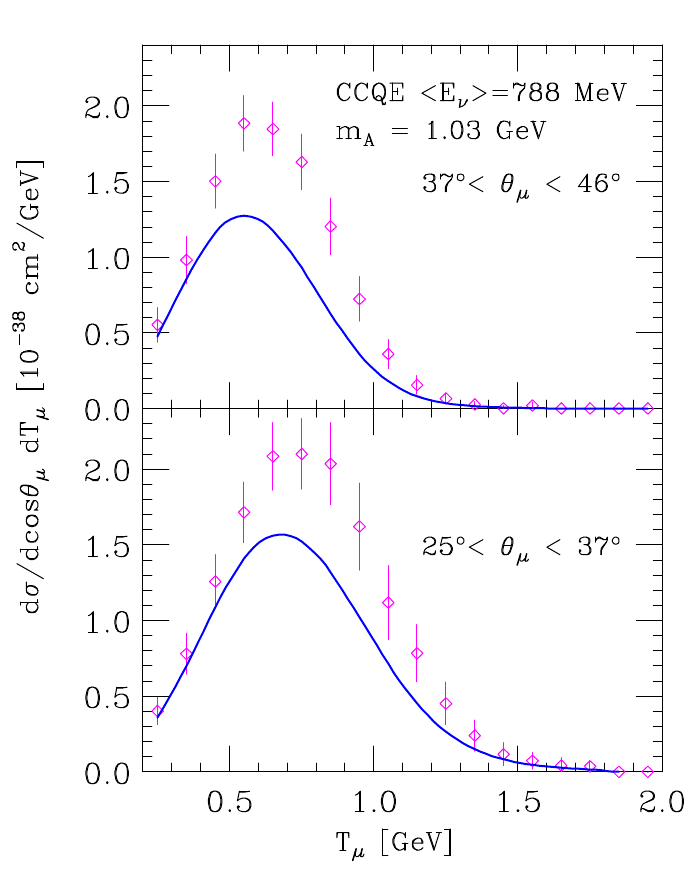}
\end{center}
\vspace*{-.25in}
\caption{Flux averaged double differential CC QE cross section measured by the MiniBooNE collaboration
\cite{miniboone_ccqe_2},  shown as a function of the kinetic energy of the outgoing muon  \cite{Benhar:2010nx}. The upper and lower panels correspond 
to different values of the muon scattering angle. Theoretical results have been obtained using the same spectral
functions and  vector form factors employed in the calculation of the electron scattering cross section of  Fig.~\ref{compare1},
and a dipole parametrization of the axial form factor with $m_A=1.03$ GeV.}
\label{compare2}
\end{figure}

 It would be tempting  to interpret the value of $m_A$ reported by MiniBoonNE as an {\em effective} axial mass, modified by nuclear
effects not included in the RFGM. However, theoretical studies carried out within the IA scheme with a realistic carbon spectral function---an approach providing a satisfactory account of the
electron scattering cross section in the quasi elastic sector---fail to describe the flux averaged double differential cross section of 
Ref.~\cite{miniboone_ccqe_2}.
This striking feature is illustrated in Figs. \ref{compare1} and \ref{compare2}.
Figure \ref{compare1} shows a comparison between the electron scattering data data of Ref.~\cite{12C1} and the results obtained using the spectral function of  Ref.~\cite{LDA},
while in Fig. \ref{compare2} the results obtained within the same scheme and setting $m_A=1.03$ MeV are compared to the flux averaged double differential CC QE cross
section measured by the MiniBooNE collaboration,  shown as a function of kinetic energy of the outgoing muon \cite{Benhar:2010nx}.
It is apparent that height, position and width of the QE peak measured in electron scattering, driven by the energy and momentum dependence of the
spectral function, respectively, are well reproduced. On the other hand,  the peaks exhibited by the neutrino cross sections are largely underestimated.

The authors of Ref.~\cite{Benhar:2010nx} argued that the differences observed comparing Fig. \ref{compare1} to Fig. \ref{compare2} are to be largely ascribed to the flux average involved in the determination of the neutrino
cross section, leading to the appearance of contributions of reaction mechanisms not taken into account in the IA picture.

In MiniBooNE data analysis, an event is labeled as CC QE if no final state pions are detected in addition to the outgoing muon.
The adjective {\em elastic} is therefore intended as alternative to {\em inelastic}, as it should, and event selection is performed in
a model independent fashion.

The simplest reaction mechanism compatible with the above qualification is single nucleon knockout, induced by the one-nucleon
contributions to the nuclear current [see Eq. \eqref{def:curr}]. As pointed out above, in the absence of NN correlations the spectator
$(A-1)$-particle system is left in a bound state, and the final state, consisting of the knocked out nucleon and the recoiling
residual nucleus, is a 1p1h state.

It has been suggested that the observed excess of CC QE cross section may be traced back to the occurrence of events
with 2p2h final states, discussed in Section \ref{2bodycurr} \cite{martini,Nieves:2011yp}. According to the above classification, these events
cannot be distinguished from those with 1p1h final states . Therefore, they are often referred to as CC QE-like.
The excitation of 2p2h final states at higher energies, up to 10 GeV, has been also recently discussed by the authors of Ref.~\cite{Gran:2013kda}.

The role of neutrino interactions leading to  the appearance of multi-particle\textendash multi-hole final states was first
pointed out by Marteau and collaborators \cite{marteau2},  who also took into account the effect of long range correlations within the RPA scheme.
The underlying nuclear model,
     originally developed in the early 1990s to describe ($^3$He,$^3$H)
     charge exchange reactions \cite{delorme}, included perturbative
     $\pi$- and $\rho$-exchange interactions, supplemented
     with contact terms providing an approximate description of short-range
    dynamics \cite{marteau1}.

\begin{figure*}
\begin{center}
\includegraphics[scale= 0.45]{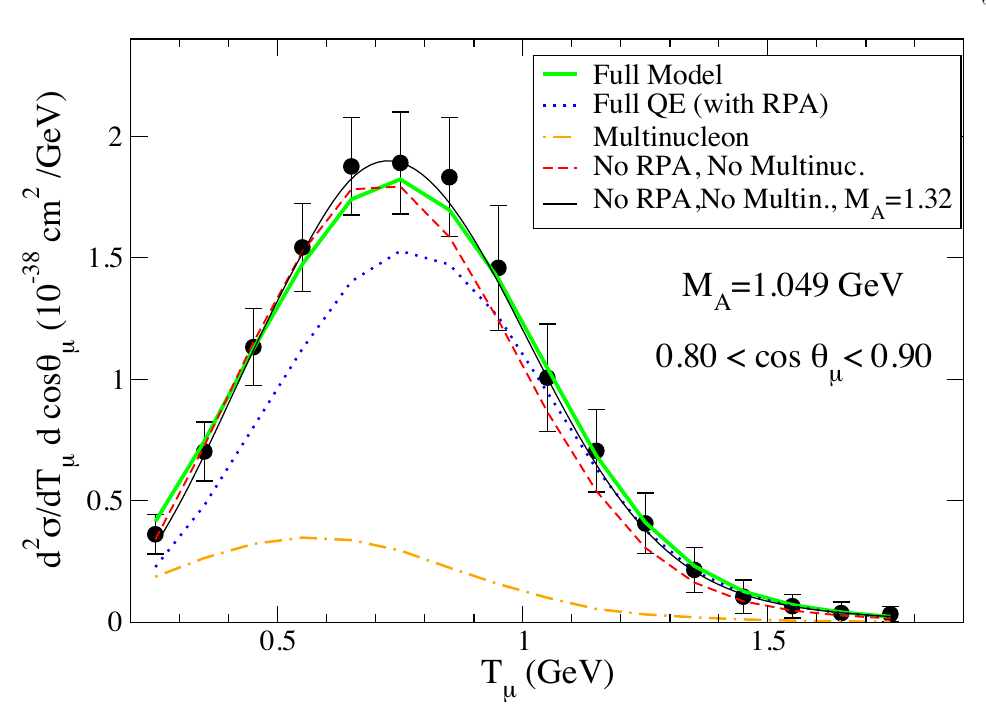} \hspace*{.2in}  \includegraphics[scale= 0.850]{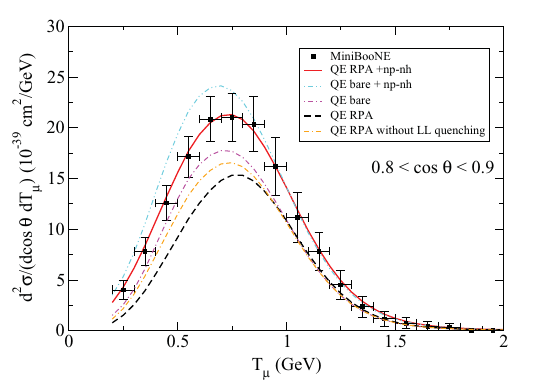}
\end{center}
\vspace*{-.1in}
\caption{Comparison between the flux averaged muon energy spectrum
at muon scattering angle $\theta_\mu$ corresponding to $0.8 \leq \cos \theta_\mu  \leq 0.9$, measured by the MiniBooNE collaboration \cite{miniboone_ccqe_2}, and the theoretical
results of  Ref.~\cite{martini} (thick solid line of the right panel) and Ref.~\cite{Nieves:2011yp} (thick solid line of the left panel). Both theoretical calculations
have been carried out using a value of the axial mass consistent with those reported in Refs.\cite{bernard,bodek2}. Note that the data in
left panel have been rescaled by a factor 0.9.}
\label{numec1}
\end{figure*}

The authors of Refs.~\cite{martini} and \cite{Nieves:2011yp} carried out extensive calculations of the CC QE neutrino-carbon cross section, averaged over the MiniBooNE flux, taking into account the effects of the two-nucleon current as well as collective nuclear excitations, which are expected to play a role at
low momentum transfer.  As an example, in Fig. \ref{numec1} the results of these approaches, obtained using a value of the axial mass consistent with
the one extracted from deuteron data, are compared to the MiniBooNE muon energy spectrum
at muon scattering angle $\theta_\mu$ such that $0.8 \leq \cos \theta_\mu \leq 0.9$. After inclusion of the reaction mechanisms beyond
single-nucleon knock out, both schemes turn out to provide a significantly improved description of the measured cross section. Note that in the left panel the data 
have been rescaled by a factor 0.9, to allow for a 10\% normalisation uncertainty \cite{Nieves:2011yp}.

\begin{figure}
\begin{center}
\includegraphics[scale= 1.4]{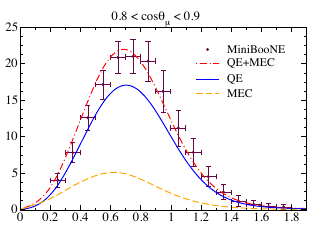} 
\end{center}
\vspace*{-.2in}
\caption{Comparison between the flux averaged muon energy spectra measured by the MiniBooNE collaboration
at muon scattering angle $\theta_\mu$ corresponding to  $0.8 \leq \cos \theta_\mu \leq 0.9$  \cite{miniboone_ccqe_2}
and the results of  the SuSav2-MEC approach \cite{Megias:nu}.}
\label{SuSav2-MEC_nu}
\end{figure}

The contribution of  interactions involving the two-nucleon current has been also recently investigated by the authors of Ref.~\cite{Megias:nu},  
 who generalised the SuSav2-MEC approach\textemdash  briefly outlined in Section~\ref{eA:xsec}\textemdash to the case of neutrino interactions.
 The results of this study are illustrated in Fig.~\ref{SuSav2-MEC_nu}. 

The authors of Ref.~\cite{Megias:nu} carried out a detailed study of the cross    
section associated with processes involving MEC. The results of this
analysis show that in neutrino interactions the contribution             
of the longitudinal channel is more significant than in electromagnetic 
interactions, and arises mainly from the the axial-vector current.

The emerging picture strongly suggests that the inclusion of 
contributions other than single-nucleon knock out is needed to bring theoretical calculations into agreement with 
the measured cross sections. However, it must be pointed out that long range RPA correlations, while providing significant contributions to the 
cross sections reported in Refs.~\cite{martini} and \cite{Nieves:2011yp}, do not appear to be needed  within the  SuSav2-MEC approach of Ref.~\cite{Megias:nu}.


\section{Implementation of nuclear dynamics in Monte Carlo simulations}
\label{implementation}
As pointed out in the previous Sections, the generalisation of the theoretical description of electron-nucleus scattering to the case of neutrino interactions does not involve severe conceptual difficulties. However, while significant progress has been made in the understanding of the different reaction mechanisms contributing to the signals detected by neutrino experiments, the implementation of state-of-the-art models in the existing Monte Carlo generators has been lagging behind. 
Among the many reasons of this state of affairs, one of the most prominent is that in neutrino oscillation experiments event generators are used to predict how the signal and background events will appear in the neutrino detector. Therefore, each generator is expected to simulate all relevant interactions, and each simulation has to cover all possible kinematical regions. Additional complications arise from the requirement of being able to describe the variety of nuclei used for detection.

Most available neutrino event generators rely on the RFGM for the treatment of the nuclear ground state. Recently, an improved implementation of the spectral function approach \cite{LDA,Benhar05,Ankowski:2007uy}, based on the formalism described in Section~\ref{PKE}, has been included in the GENIE event generator~\cite{Andreopoulos:2009rq,Jen:2014aja}. The event generators NUWRO and NEUT also
feature a spectral function implementation, aimed at improving the description of the nucleon energy and momentum distribution \cite{NUWRO,NEUT}.

\begin{figure}[ht!]
\begin{center}
\includegraphics[scale= 0.45]{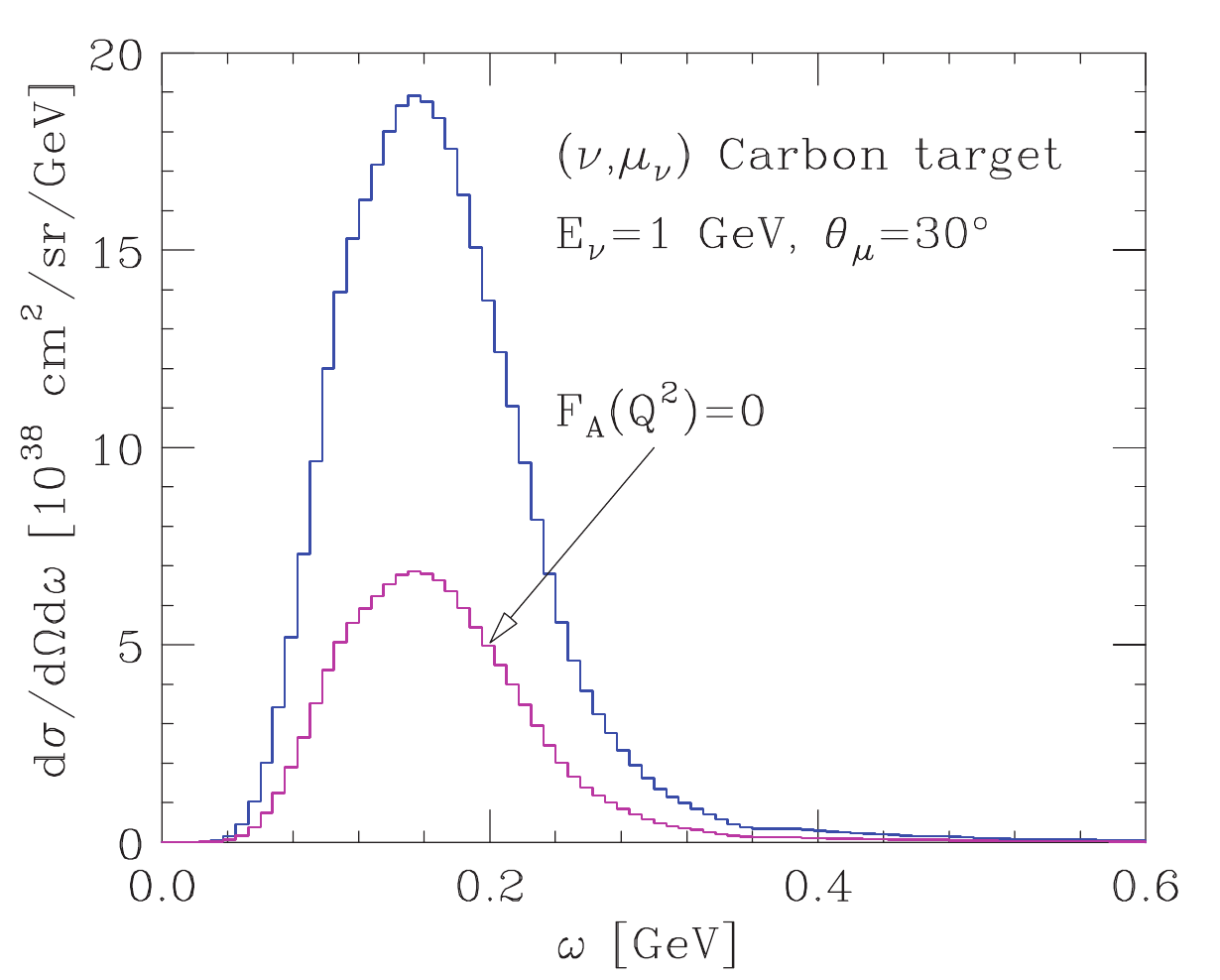}
\end{center}
\vspace*{-.25in}
\caption{Double differential cross section of the process $\nu_\mu+{^{12}\textrm{C}} \to \mu^- + X$ in the QE channel, obtained using the spectral function of Ref.~\cite{LDA}. The two histograms show the  results of the full calculation and those obtained setting $F_A(Q^2)=0$  \cite{Jen:2014aja}. \label{F3}}
\end{figure}

As a first step, the authors of Ref.~\cite{Jen:2014aja} focused on the CC QE channel, which accounts for a large fraction of the detected signal in many neutrino oscillation experiments.
As an example, Fig.~\ref{F3} presents the double differential cross section of the process
\begin{equation}
\nu_\mu + {^{12}\textrm{C}} \to \mu^- + X \ ,
\end{equation}
in the QE channel, at neutrino energy $E_\nu = 1$ GeV and muon scattering angle $\theta_\mu = 30$ deg, plotted as a function of the lepton energy loss $\omega$. The calculation has been carried out  using the carbon spectral
function of Ref.~\cite{LDA}. In order to illustrate the size of the axial-vector contributions, the result of the full calculation is compared to that obtained setting $F_A(Q^2)=0$.

To carry out the simulation following the scheme outlined above, few new modules were developed, and few modules from the official GENIE release $2.8.0$ were modified~\cite{Jen:2014aja}. From now on, we will refer
to them as GENIE $2.8.0+\nu T$. By analogy with Eq.~\eqref{sigma1}, the QE neutrino-nucleus cross section at beam energy $E_\nu$ can be written in the target rest frame as
\begin{align}
 \label{cross:section}
 & \left( \frac{d^2 \sigma_{\nu A}}{dE_\mu d \Omega_{\mu}} \right)  = \int d^3k dE
  \left(\frac{d^2\sigma_{\nu N}} {dE_\mu d\Omega_{\mu}} \right) P(|{\bf k}|,E) \\
  \notag
 ~& \ \ \ \ \ \ \times \delta(\omega + M_A -  \sqrt{ |{\bf k} + {\bf q}|^2 + m^2 } - E_{A-1} )  \ .
\end{align}
The integrations can be carried out using the Monte Carlo method, yielding
\begin{align}
\label{MC1}
\left( \frac{d^2 \sigma_{\nu A}}{dE_\mu d \Omega_{\mu}} \right)   
\approx  \frac{1}{N} \sum_{n=1}^N G( E_\nu, E_\mu, \cos \theta_\mu ; \{ k ,E , \cos \theta_N \}_n ) \ ,
\end{align}
where $E_\mu$ and $\theta_\mu$ denote the muon energy and scattering angle, respectively,
$\theta_N$ is the polar angle specifying the direction of the nucleon momentum, ${\bf k}$, and
\begin{align}
\label{MC2}
&G(E_\nu,  E_\mu ,\cos \theta_\mu ; k,E, \cos \theta_N)   = \left( \frac{d^2 \sigma_{\nu N}}{dE_\mu d \Omega_{\mu}} \right) \\
\notag
& \ \ \ \ \ \ \ \ \ \times \delta( \omega + M_A - \sqrt{ |{\bf k} + {\bf q}|^2 + m^2 } - E_{A-1} )  \ .
\end{align}
%
%
In Eq. \eqref{MC1},  $\{ k ,E , \cos \theta_N \}_n$ denotes the set of kinematical variables of the struck nucleon. Momentum and energy,
$k$ and $E$, are sampled from the probability distribution
$F(k,E) = 4 \pi |{\bf k}|^2 P(|{\bf k}|,E)$, while $\cos \theta_N$ is assumed to be uniformly distributed in the range $[-1,1]$.

In its release $2.8.0+\nu T$, the GENIE event generator  provides  a simulation of CCQE neutrino interactions based on two different description of the nuclear ground state: the RFGM and the spectral function approach. In addition to the CC QE channel, both nuclear models can be used to simulate interactions leading to different hadronic final states, such as resonance production and decay, pion production and deep-inelastic scattering.
A detailed description of the treatment of these processes can be found in the literature \cite{Andreopoulos:2009rq,Dytman:2011zza,Jen:2014aja}.

\begin{figure*}
\begin{center}
\subfigure[~$e+\textrm{C} \to e' + X$, $E_{e}=0.961$~GeV, $\theta_{e}=37.5$~deg]{\includegraphics[width=0.8\columnwidth]
{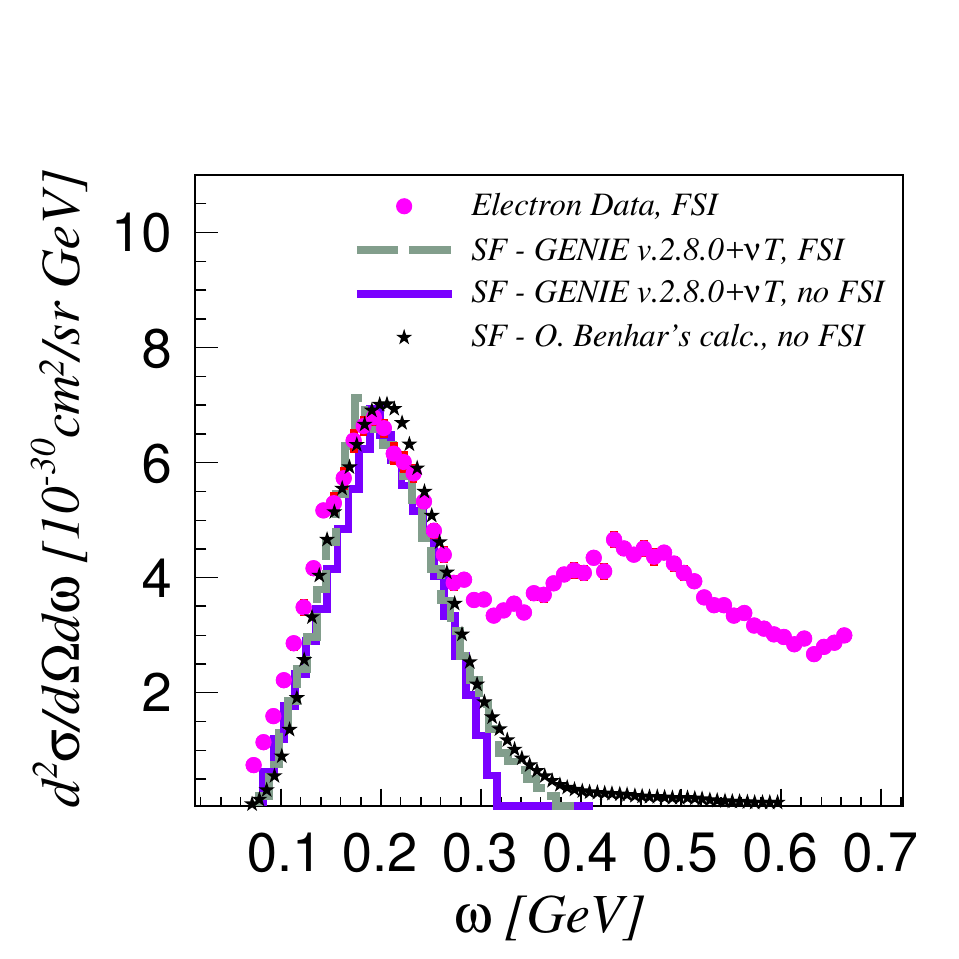}}
\subfigure[~$e+\textrm{C}\to e' + X$, $E_{e}=1.299$~GeV, $\theta_{e}=37.5$~deg]{\includegraphics[width=0.8\columnwidth]
{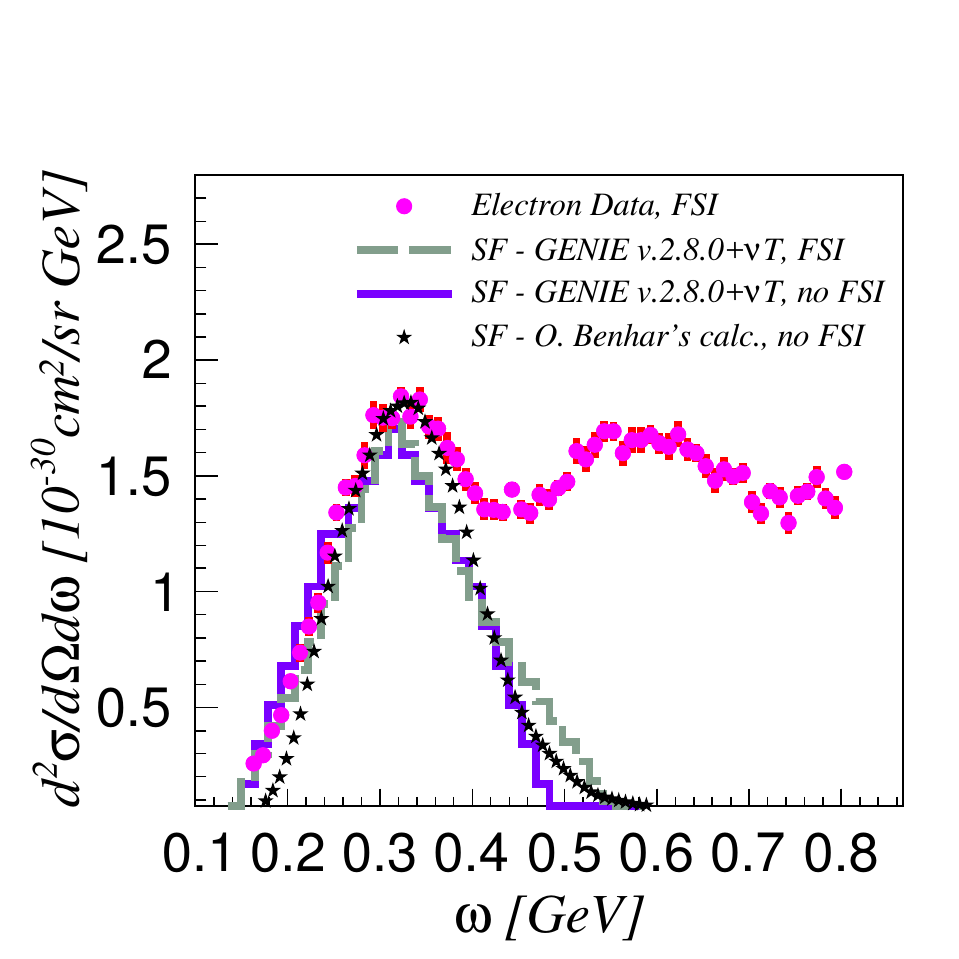}}
\subfigure[~$e+\textrm{Ca} \to e' + X$, $E_{e}=0.841$~GeV, $\theta_{e}=45.5$~deg]{\includegraphics[width=0.8\columnwidth]
{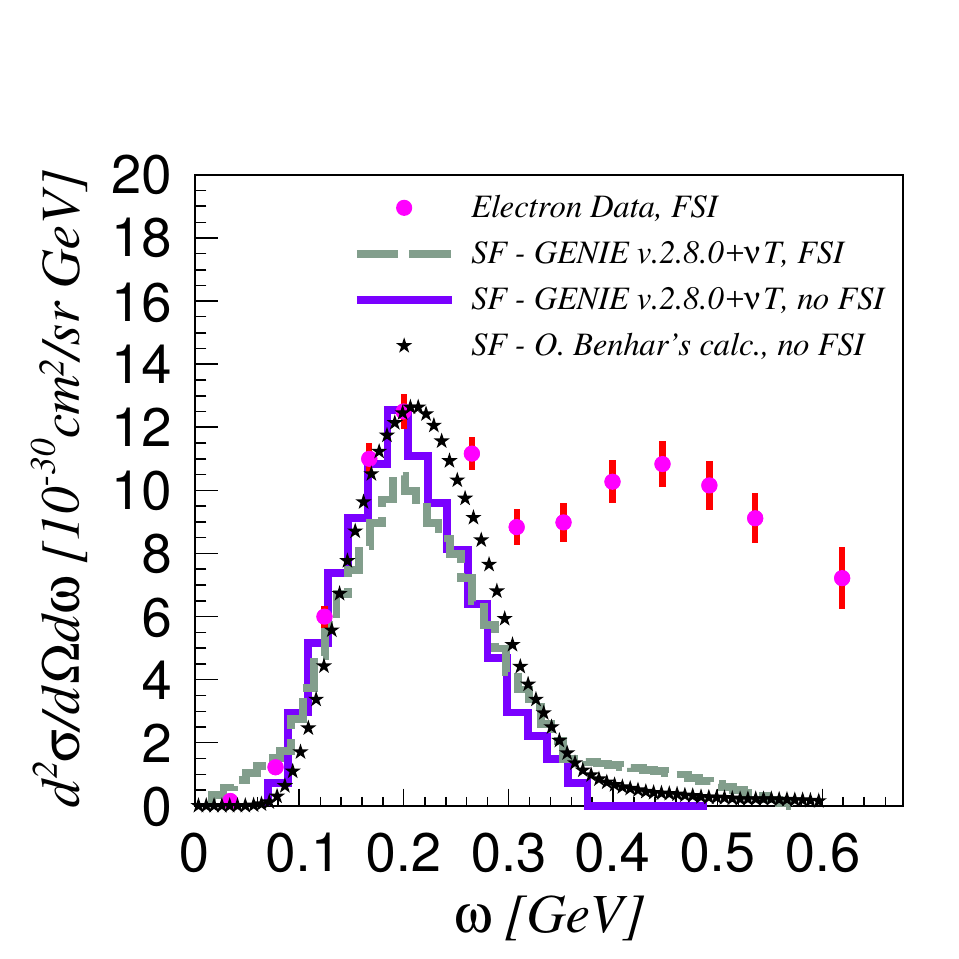}}
\subfigure[~$e+\textrm{Ar} \to e' + X$, $E_{e}=0.700$~GeV, $\theta_{e}=32$~deg]{\includegraphics[width=0.8\columnwidth]
{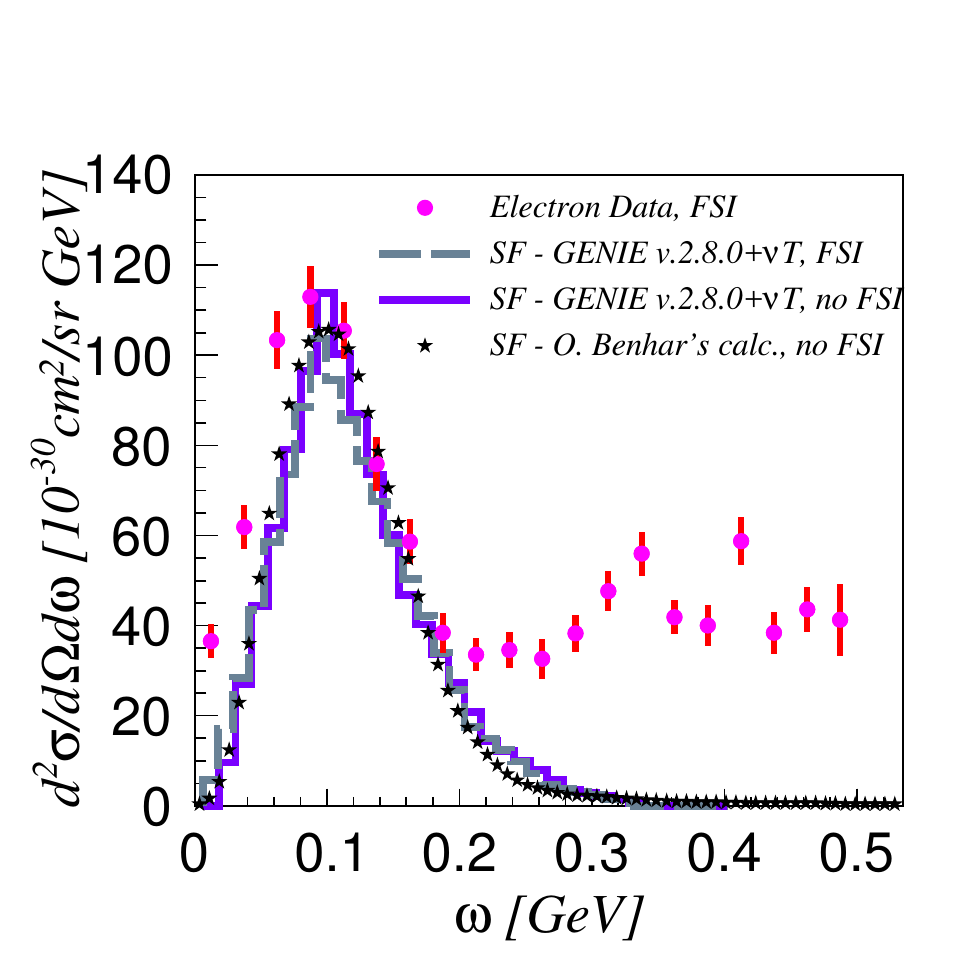}}
\caption{Double differential electron-nucleus cross section in the quasielastic channel. The curves labeled SF have been obtained using Eq.~\eqref{MC1} and the model spectral functions of  Refs.~\cite{LDA} (for carbon) and~\cite{Ankowski:2007uy} (for calcium and argon)  \cite{Jen:2014aja}. The data are taken from Refs.~\cite{12C2} (for carbon), \cite{Williamson:1997} (for calcium), and \cite{Anghinolfi:95n} (for argon).}\label{fig:F11}
\end{center}
\end{figure*}

As an example of the GENIE $2.8.0+\nu T$ results, in Fig.~\ref{fig:F11}
we show the electron scattering cross sections obtained using the LDA carbon spectral function of Ref.~\cite{LDA} and the calcium and argon spectral functions obtained from the simplified approach of Ref.~\cite{Ankowski:2007uy}. In order to allow for a consistent comparison with the data---that were not corrected to remove the effects of the FSI---the results of the simulations are presented with and without inclusion of FSI~\cite{Dytman:2014}.

The agreement appears to be quite satisfactory, the differences being largely ascribable to numerical accuracy. The effects of FSI discussed in Section~\ref{fsi}\textemdash that is, a shift of the 
energy loss distribution and a redistribution of the strength from the quasi elastic peak to the tails\textemdash can be clearly observed in the GENIE 2.8.0 + $\nu T$ results, and appear to be 
more pronounced in the case of heavier targets. 
Note that
the tail of the  inelastic contributions
extending into the region of the QE peak are quite small (see, e.g., Ref.~\cite{Benhar05}), and their inclusion does not appreciably affect the emerging scenario. 

A consistent implementation of the IA requires a careful consideration of the $Q^2$ selection, taking  into account the fact that, while the tensor $L^{\lambda \mu}$ of Eq.~\eqref{leptensor} is determined from lepton kinematical variables only, the nucleon tensor depends on the initial nucleon momentum, ${\bf k}$, and ${\widetilde q} \equiv (\widetilde \omega,{\bf q})$, which in turn depends on the removal energy $E$ through its time component ${\widetilde \omega}$,
defined by Eq.~\eqref{def:omegatilde}.  The energy transfer at the elementary interaction vergex is $\widetilde \omega < \omega$, while the difference $\delta \omega = \omega -\widetilde \omega$ provides a measure of the energy transfer to the spectator system. The details of the $Q^2$-selection procedure can be found in the paper of Jen and collaborators \cite{Jen:2014aja}.

\begin{figure}[htb!]
\subfigure[Inclusive electron-carbon cross section at beam energy E$_e$=0.961~GeV and scattering angle $\theta_e$=37.5~deg.]
{\includegraphics[width=\columnwidth]{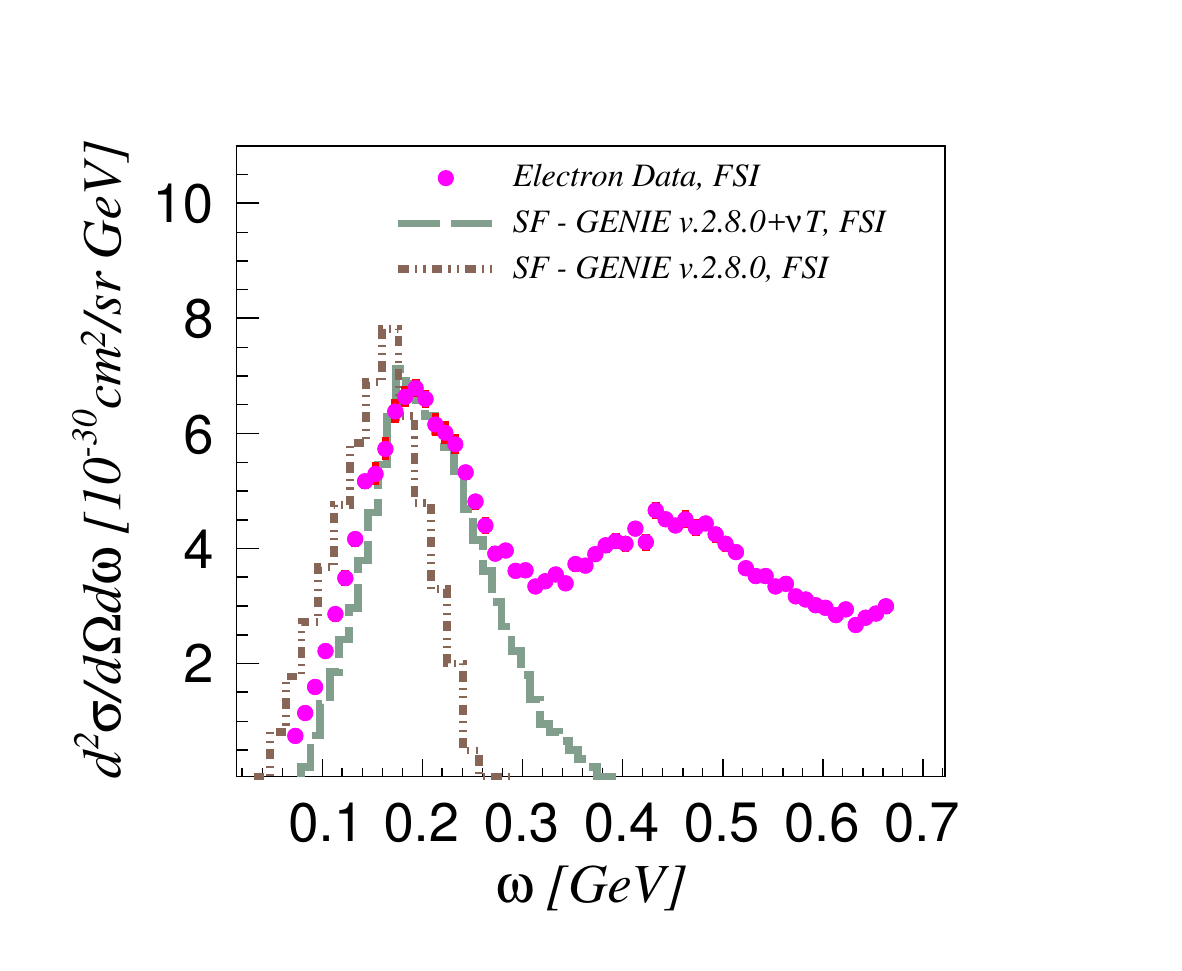}}
\subfigure[Outgoing muon energy distribution corresponding to muon neutrinos of 0.8~GeV interacting with oxygen. ]{\includegraphics[width=\columnwidth]{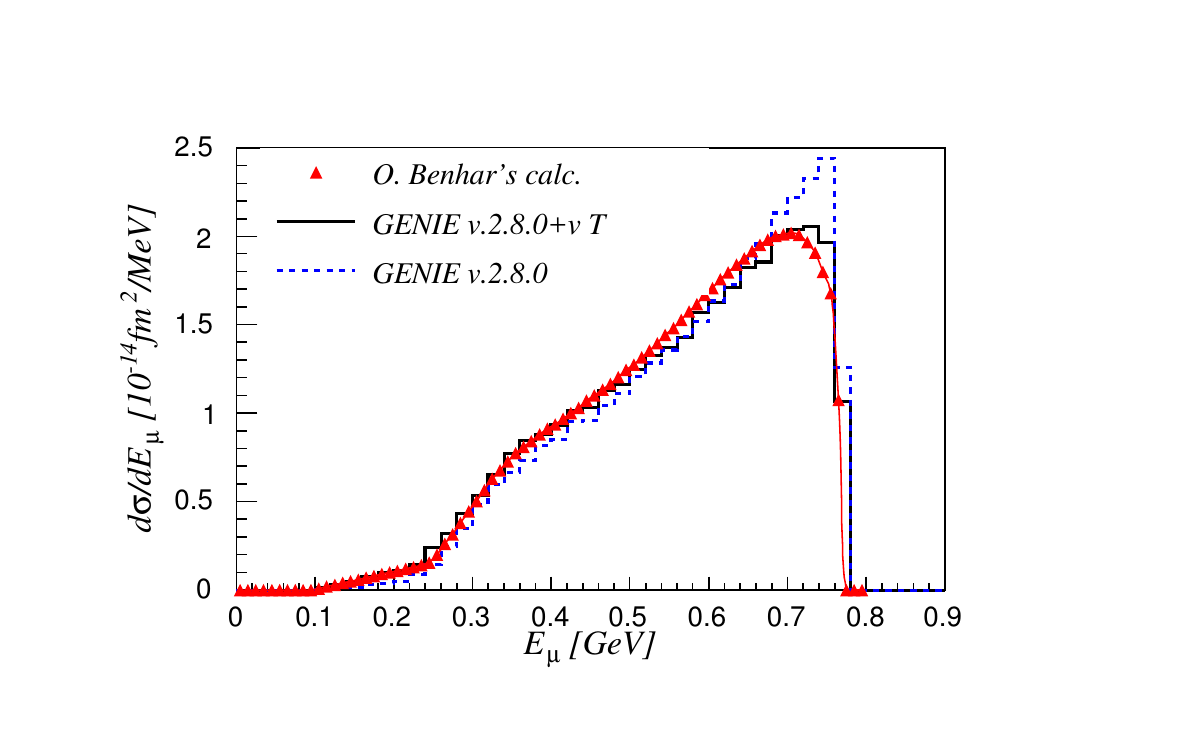}}
\caption{Comparison between different implementation of the spectral function model  \cite{Jen:2014aja}. 
In panel (a) the results obtained from GENIE $2.8.0$ and GENIE $2.8.0+\nu T$ are displayed
using dots and the dot-dash line, respectively, and compared to the data of Ref.~\cite{12C1}. In panel (b) the GENIE $2.8.0$ and GENIE $2.8.0+T$ results are represented by the dotted and solid line, while  
the theoretical results of Benhar and collaborators \cite{Benhar05} (labeled O. Benhar's calc.)
are shown by triangles. The effect of Pauli blocking is included in all three calculations, while FSI are not taken into account.}
\label{fig:electron_data_comparison}
\end{figure}

Unfortunately, the full description of the complex dynamics of nuclear scattering processes requires the use of more variables and observables than just the four-momenta of the target nucleon, the outgoing leptons and all the produced hadrons and
gammas. Hence, the comparison between the observed neutrino interactions and the predictions of nuclear models needs to be done with great care. Because neutrino event generators are usually tuned to existing data, the implementation of a
new model in an existing framework may result in making the comparisons more difficult. Individual nuclear 
models should be evaluated considering their limitations and the kinematical region in which  they are expected to be valid,
 and must be compared to a set of {\em external} data, when available. An example of this procedure is given in Fig.~\ref{fig:electron_data_comparison}, showing a comparison between two different implementations of the spectral function approach
 and the results of a  ($e,e^\prime$) measurement [panel (a)], as well as the corresponding predictions for the outgoing muon distribution in neutrino interactions [panel (b)]. As an external data set, the electron scattering cross sections turn out to
 provide an excellent validation tool.
 
\section{Dependence of oscillation parameters on the description of nuclear effects}
\label{dop}

The discovery of neutrino oscillation has been one of the major
results of particle physics in the past two decades. As a
consequence, we are witnessing a massive global experimental effort, 
in the form of ongoing and planned long- and short-baseline oscillation
experiments.  At short baselines, of order 1\,km or less, the goal is
to conclusively test the hypothesis of an eV-scale sterile neutrino (see, 
e.g., Ref.~\cite{Abazajian:2012ys}), whereas at long baselines, of
several 100\,km or more, the goal is to measure the leptonic CP phase, 
and to ultimately test the validity of the three neutrino oscillation
paradigm. All of these experiments will use detectors made of nuclei
with masses in the range A=12-56. Therefore, a precise
understanding of the electro-weak response of these nuclei to
both neutrino and antineutrino interactions is necessary. The relevant
neutrino energies range from a few hundreds MeV up to approximately
10\,GeV, thus fully covering the quasi-elastics region, but also
largely extending into the domain of resonance production and DIS. 
As explained in the previous Sections, this is a considerable challenge
for theory. From an experimental view point, the problem is severely
compounded by the limitations of currently used neutrino sources and
the fact that in the detector different underlying events yield
identical signatures.

In this Section, we establish a connection between the oscillation
physics measurements and the required level of precision in the
understanding of neutrino-nucleus interactions. We also demonstrate,
using a number of specific examples which have been studied in detail
in the literature, the impact the current level of uncertainty
surrounding neutrino-nucleus cross sections has on the ability to
correctly interpret experimental results. Regretfully, despite these efforts
important special cases\textemdash notably CP violation searches using liquid argon
detectors\textemdash have not yet been studied in sufficient detail to make
reliable, quantitative statements.

Before embarking on detailed case studies, we first describe the basic
problem based on a number of approximations and give some examples
based on these approximations. We analyze in detail the appearance
$\nu_\mu \to \nu_e$ and disappearance $\nu_\mu \to \nu_\mu$ channels
in the T2K experiment, to show that such uncertainties are relevant also in the
extraction of the atmospheric parameters. This analysis is performed comparing the
RFGM, the approach based on a a realistic spectral function and two different
implementations of the RPA scheme, as well as different
event generators for the disappearance analysis. Here, we limit
ourselves to the quasi-elastic regime, where the neutrino cross
sections can be evaluated with less uncertainties.

It is worthwhile to point out that many of the problems arise from the
need to reconstruct the neutrino energy precisely, since it is the
neutrino energy which enters the oscillation probability. At the rate
level many quantitative studies have been performed. The uncertainties
in the energy reconstruction resulting from our lack of an
understanding of the underlying micro-physics are very hard to analyze
quantitatively for two reasons: First, this relies on the ability to
model exclusive cross sections, since the energy distribution of
secondary particles produced in the neutrino event has to be
quantified. There is very little data to support this modeling and the
sparse data we have does not fit with any existing model. It is
therefore, very difficult to establish the range of plausible
variations of energy dependencies. Second, neutrino energy
reconstruction also depends on fine details of detector performance and
in particular, for liquid argon detectors, there is not yet a good
quantitative understanding of detector performance. As a corollary,
there clearly is awareness that energy reconstruction is profoundly
affected by exclusive cross sections and their uncertainties, but
there a few quantitative studies and those which exist are highly
simplified, we discuss some examples in section~\ref{sec:energy}.

Finally, we point out necessary steps towards a full quantitative
understanding of the relation between cross section uncertainties and
oscillation physics measurements for the next generation of
long-baseline experiments.


\subsection{The problem at rate-level}

Neutrino physics is on its way to become precision science, and this
brings new challenges for future experiments. The next 10--20 years
will be centered on a long-baseline neutrino oscillation program using
conventional neutrino beams obtained from pion decay-in-flight. This
technique of making neutrino beams has been used for many decades: an
intense proton beam impacts on a thick target producing mostly pions,
but also other mesons. The pions are focused using a magnetic horn,
and the polarity of the magnetic field allows to select predominantly
one charge-sign of the pion, which implies the ability to make
neutrino and antineutrino beams. It is important to note, that the
resulting fluxes and purities of neutrino and antineutrino beams are
not related in a meaningful way. Therefore, the neutrino and
antineutrino beams derived from the same target and horn configuration
are essentially independent experiments. The next generation of beams
will exceed the 1\,MW level of power on target, which represent a
major advance in engineering and accelerator physics. Hence, these
beams are dubbed superbeams. It is a very difficult task to determine
the resulting neutrino flux, energy spectrum and flavor composition
purely from data on meson production in thick targets, beam parameters
and horn configuration. Currently, the state of the art in controlling
beam systematic uncertainties is represented by the MINOS~\cite{Adamson:2009ju} and 
 MINERvA~\cite{Higuera:2014azj} experiments. Much of the information used, in both
cases, comes from the ability to change the position of the target
with respect to the horn system. This ability most likely will not
exist for beams with target powers in excess of 1\,MW, due to the
resulting very harsh operating conditions in close proximity to the
target. Therefore, it appears reasonable to assume that the 
understanding of the beam at the level of
roughly 5\%, demonstrated by MINOS and MINERvA, represents
the best case for future experiments as well. More precise neutrino beams require a different
technological approach. For instance, muon decay offers the possibility
to obtain high-intensity $\nu_\mu$ and $\nu_e$ beams with beams
systematics well below 1\%. This concept is know as neutrino
factory~\cite{Choubey:2011zzq}, and a low-energy entry-level version is
know as $\nu$STORM~\cite{Adey:2013pio}, which would allow, among other
applications, a very precise and accurate measurement of neutrino
cross sections.

As explained in the previous Sections, the current understanding of cross
sections---where understanding implies the ability to describe actual
experimental data---is generally at the 10\% level, and neutrino
beams are known at the 5\% level. Thus, the question is:
what level of accuracy is needed for future neutrino oscillation
measurements? One of the main goals of the future neutrino program is
a measurement of the leptonic CP phase, $\delta$. It also happens
that this measurement puts the most stringent demands on the overall
accuracy, since it involves both neutrinos and antineutrinos and the
relevant oscillation probability $P(\nu_\mu\rightarrow\nu_\mu)$
depends on all three mixing angles and both $\delta m^2$ in leading
order, see {\it e.g.} Ref.~\cite{Freund:2001pn}. The resulting
requirement on systematic uncertainties is closely tied to the
required level of accuracy in the determination of the CP phase. There
is no a priori physics arguments which would argue for an error of $x$
degrees in the measurement of $\delta$, like for instance in the case
of QED, where new effects clearly appear as powers of the fine
structure constant. Arguments can be made based on certain neutrino
flavor models, like sum rules~\cite{Antusch:2007rk}, but these
arguments remain model-dependent. Another line of argumentation, is
based on the recognition that CP violation is not well understood in
the Standard Model, in the sense that the QCD Lagrangian would allow
for CP violation but is CP conserving to a very high degree of
accuracy, the so-called strong CP problem---for a brief introduction
see Ref.~\cite{Peccei:2006as}\textemdash whereas mixing in the quark sector
shows large CP violation. Framed in this way, the question of how
large CP violation in the lepton sector is, if it exists at all,
becomes very relevant. The ability of an experiment to discover CP
violation depends on the one hand on the existence of CP violation, in
this case $\delta \neq 0,\pi$, and on the ability to distinguish the
measured value for $\delta$ from the CP conserving cases $0$ and
$\pi$. A reasonable goal for an experiment could be the ability to
discover CP violation at 3\,$\sigma$ confidence level for 75\% of all
CP phases. This goal was for instance adopted recently in the
U.S.~\cite{P5}.

\begin{figure}
\vspace*{.1in}
\begin{center}
\includegraphics[width=0.95\columnwidth]{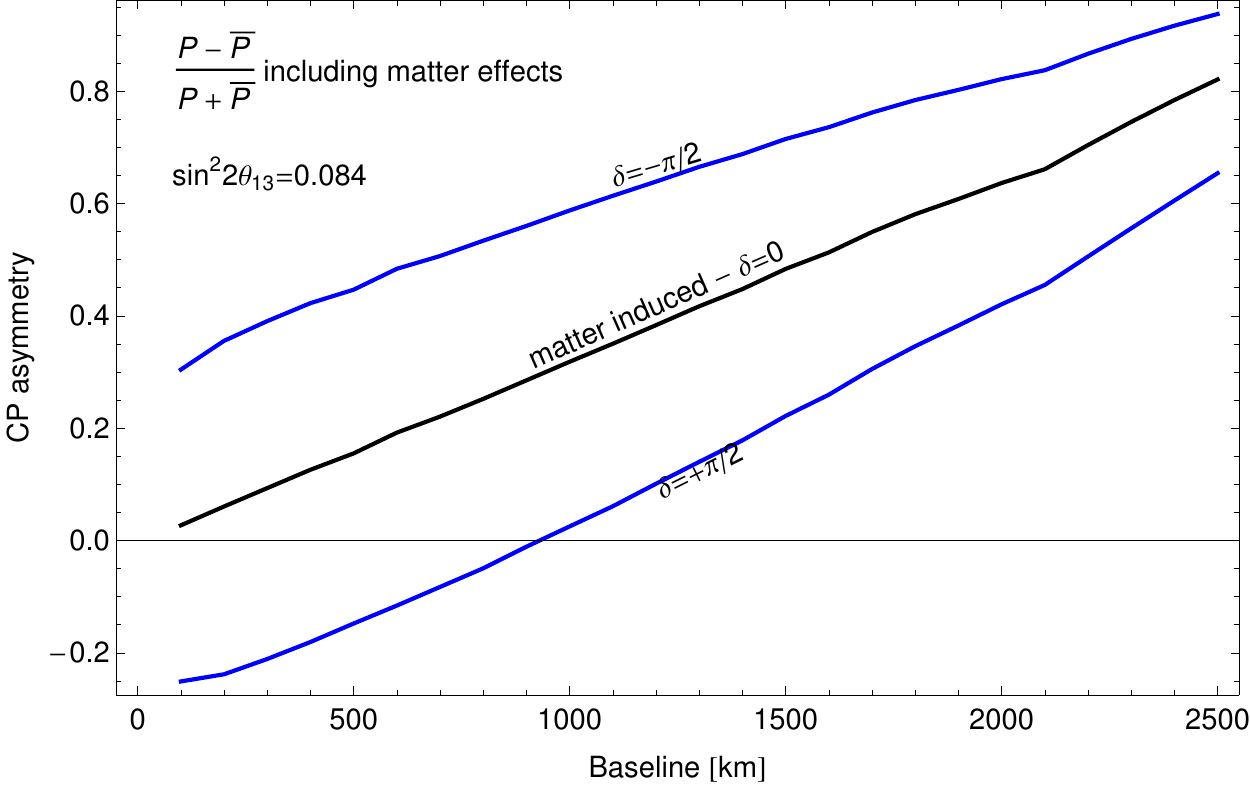}
\end{center}
\vspace*{-.2in}
\caption{\label{fig:cpa} Value of the CP asymmetry $A$, for different
choices of $\delta$, as a function of the baseline.}
\end{figure}

A good proxy for the ability to measure the CP phase is given by the CP
asymmetry, $A$, defined as
\begin{equation}
\label{eq:cpa}
A=\frac{\langle P\rangle-\langle \bar P\rangle}{\langle P\rangle+\langle \bar P\rangle}\,,
\end{equation}
where $\langle P\rangle$ is the energy averaged oscillation
probability for $\nu_\mu\rightarrow\nu_e$ and $\bar P$ is the
corresponding quantity for antineutrinos. The energy average is taken
over the range defined by having one half of the peak probability
around the first oscillation maximum. In vacuo, $A$ is proportional to
$\sin\delta$ and thus the errors on $A$ and $\delta$ are very similar
for $\delta=0,\pi$. $A$ also will receive contributions from matter
effects, which manifestly break CP invariance, as shown by the line
labeled $\delta=0$ in Fig.~\ref{fig:cpa}. Here, the value of the
asymmetry
as a function of the baseline is shown for oscillation parameters from
a recent global fit~\cite{GonzalezGarcia:2012sz}. From this figure it
is evident, that the maximal CP induced asymmetry is $\pm25\%$ for
baselines of about 1,500\,km on top a of a similar contribution from
matter effects. For 75\% of all CP phases the genuine CP asymmetry can
become as small as 5\%, which at 3\,$\sigma$ translates into about
1.5\% required accuracy. Further assuming, that the contributions from
statistics and systematics should be about equal, this translates into
a systematics requirement of 1\%. Obviously, for different CP
violation discovery goals, this value will change correspondingly.

In practice experiments do not measure oscillation probabilities but
event rate distributions, $R(E_\mathrm{vis})$, as a function of the
visible energy, $E_\mathrm{vis}$,
\begin{eqnarray}
\label{eq:rate}
&&R^\alpha_\beta(E_\mathrm{vis})=\\
&&N\int dE\,\Phi_\alpha(E)\,\sigma_\beta(E,E_\mathrm{vis})\,\epsilon_\beta(E)\,P(\nu_\alpha\rightarrow\nu_\beta,E) \,,\nonumber
\end{eqnarray}

where $N$ is a normalization factor, $\Phi_\alpha(E)$ is the neutrino
flux as a function of the neutrino energy, $E$ and
$P(\nu_\alpha\rightarrow\nu_\beta,E)$ is the oscillation probability
as function of the neutrino energy. The differential cross section $\sigma_\beta(E,E_\mathrm{vis})$
describes the probability that a neutrino of
energy $E$ produces a distribution of visible energies $E_\mathrm{vis}$ in the
detector. Finally, $\epsilon_\beta(E)$ is the detection efficiency, and
since it appears always in combination with $\sigma$ one can define
the effective cross section,
$\tilde\sigma_\beta:=\sigma_\beta\epsilon_\beta$. Note, that for this
qualitative discussion we neglect any effects from the detector energy
response and reconstruction efficiencies, which in general will add another level of complexity in
terms of the relation between visible and reconstructed
energy. Neglecting all energy dependencies of the flux, $\Phi_\alpha$, the
effective cross section $\tilde\sigma$, Eq.~\eqref{eq:rate} describes
how the total event rate depends on the average oscillation
probability. Even in this simplifying limit, there is no reason to
assume that any of the quantities on the right hand side of
Eq.~\eqref{eq:rate} will be known better than 5--10\%. Also, ratios for
neutrino and antineutrinos as well as flavor ratios are not a priori
constrained to a better level of accuracy. At quark level lepton
universality prevails, but since we deal with nuclei and do not
resolve the $Q^2$ of the interaction, non-trivial flavor effects are
found especially at low energies~\cite{Day:2012gb}. Finally, the ratio
of detection efficiencies $\epsilon_e/\epsilon_\mu$ has to be
determined experimentally.

The problem of performing accurate measurements of oscillation
parameters in the presence of significant cross section and/or flux
uncertainties has been encountered before. A practical solution in
many cases has been to use a near detector to measure the unoscillated
event rate and exploit the fact that in the comparison of near and far detector data many
uncertainties cancel. This method was used with great success in the
Daya Bay experiment to measure $\theta_{13}$~\cite{An:2012eh}. For
this cancellation to occur efficiently it is essential that near and
far detectors have as close as possible an identical response to the
neutrino signal. Any differences have to be understood with great
precision. There are many potential sources for different near and far
detector response functions, {\it e.g.} geometric acceptance or
different background levels. Using the simplifying assumption that
near and far detectors have \emph{identical} response, for the total
event rate ratio one obtains
\begin{eqnarray}
\label{eq:cancellation}
\frac{R^\alpha_\alpha(\mathrm{far})L^2}{R^\alpha_\alpha(\mathrm{near})}
&=&\frac{N_\mathrm{far}\Phi_\alpha\,\tilde{\sigma}_\alpha\,
  P(\nu_\alpha\rightarrow\nu_\alpha)}{N_\mathrm{near}
  \Phi_\alpha\,\tilde{\sigma}_\alpha 1 }\nonumber\\
&=&\frac{N_\mathrm{far}}{N_\mathrm{near}}\,P(\nu_\alpha\rightarrow\nu_\alpha)\,.
\end{eqnarray}
In Daya Bay the conditions for this cancellation to occur were, by
design, nearly ideal: the near and far detectors have the same size,
they are made from the same materials, the reactors appear as point
sources to both, the inverse beta-decay cross section is independently
known, and the initial and final flavor is the same.

In extrapolating from the Daya Bay experience to future, long-baseline
experiments a number of factors should be considered. First, to make the
neutrino source, which in reality is the whole length of the decay
pipe, point-like in the near detector, a not-so-near near detector is
required. For baselines longer than 1,000\,km the required tunnelling is likely to be
prohibitively expensive. Second, the enormous size of the far
detector renders an equally sized near detector unfeasible, and thus
the detectors cannot have identical response. Third, the beam energy
spread in a neutrino beam is large enough that a wide variety of
interaction mechanisms will contribute to the signal, and thus the
energy dependence of the near/far ratio can no longer be neglected.
For a disappearance measurement, MINOS can serve as a benchmark of how
well a near/far comparison does reduce systematic
errors~\cite{Michael:2006rx}.

In an appearance measurement final and initial neutrino flavors are
different, which will lead to an additional term in
Eq.~\eqref{eq:cancellation} of the form
$\tilde\sigma_\beta/\tilde\sigma_\alpha$. Measuring
$\tilde\sigma_\beta$ in a beam of purely flavor $\alpha$ is
impossible. The small component of $\nu_e$ present in the beam is
overall even less well known than the primary beam flux and the
relative smallness of the $\nu_e$ component will result in reduced
statistics in the near detector. Recently, the T2K Collaboration
presented a result on the $\nu_e$ cross section in a predominantly
$\nu_\mu$ beam~\cite{Abe:2014agb}. The total systematic error is about
16\%, mostly originating from the beam flux uncertainty and the
detector response. There are no obvious methods to improve this
situation and it stands to reason that for antineutrinos the situation
will be worse. On general grounds, the ability to measure a cross
section with much better accuracy than the accuracy at which the beam
flux is known appears doubtful.

\begin{figure}
\vspace*{.1in}
\begin{center}
\includegraphics[width=0.95\columnwidth]{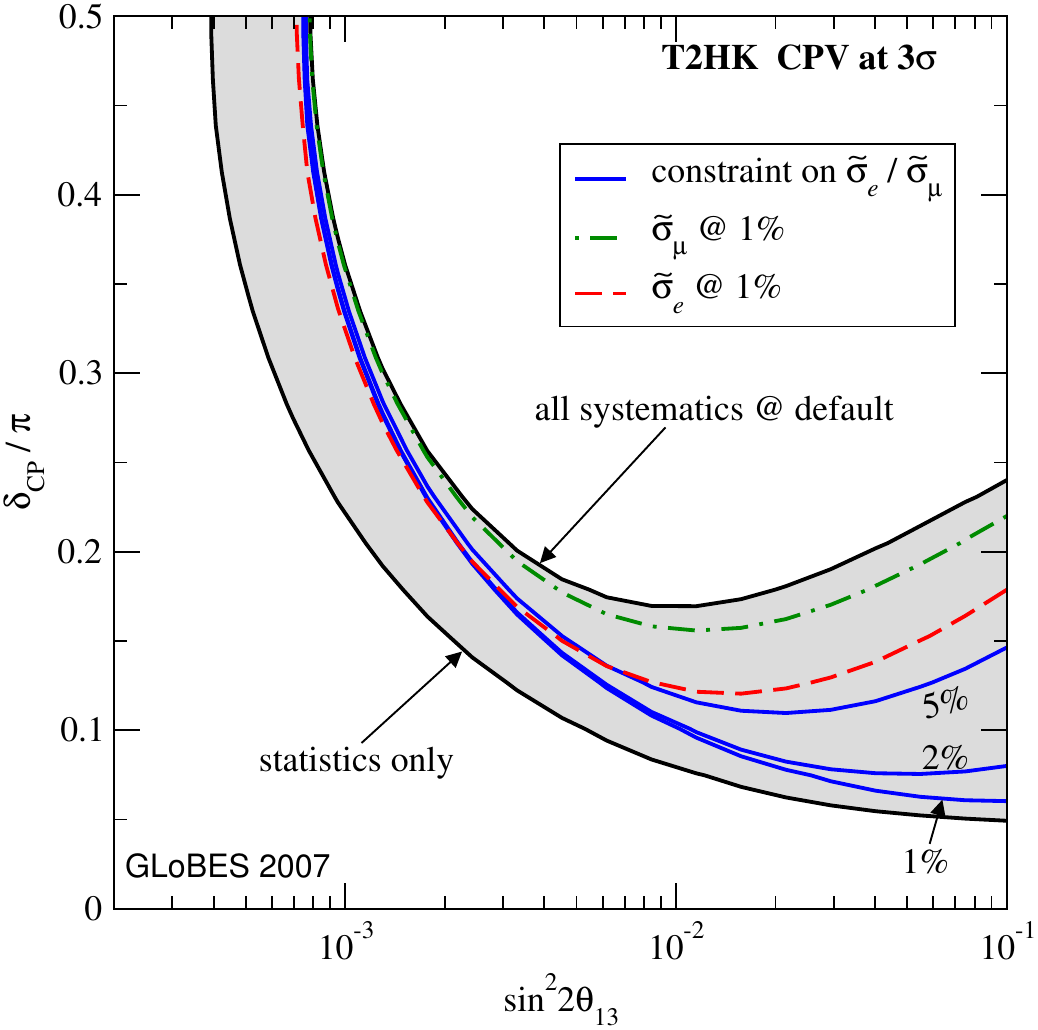}
\end{center}
\vspace*{-.15in}
\caption{\label{fig:t2hk} CP violation sensitivity at 3$\,\sigma$ level for
  a certain choice of systematical errors  and for
  statistical errors only (curves delimiting the shaded region). We
  show also the sensitivity if certain constraints on the product of
  cross sections times efficiencies $\tilde\sigma$ are available: 1\%
  accuracies on $\tilde\sigma_{\mu}$ and $\tilde\sigma_{e}$ for
  neutrinos and antineutrinos, and 5\%, 2\%, 1\% accuracies on the
  ratios $\tilde\sigma_\mu/\tilde\sigma_e$ for neutrinos and
  antineutrinos. Figure and caption adapted from Ref.~\cite{Huber:2007em}.}
\end{figure}

A quantitative analysis along these lines, based on only
energy-independent implementations of uncertainties has been presented
in Ref.~\cite{Huber:2007em}, where a T2HK-like experiment was used to
illustrate the impact of more than 20 potential sources of
systematical uncertainties. This analysis, for the first time,
explicitly included an idealized near detector. The main result for
the sensitivity to measure CP violation---defined as the
ability to exclude the CP conserving values of $\delta$ at the given
confidence level---is shown in
Fig.~\ref{fig:t2hk}. It is apparent, that a tight constraint of 1\% on
the ratio of $\tilde\sigma_e/\tilde\sigma_\mu$ is required to restore
the statistics-only result at large $\sin^22\theta_{13}$. A corollary
from this work is that even a perfect near detector is not a panacea.

The analysis presented by the authors of Ref.~\cite{Huber:2007em} relied on a significant
number of simplifying assumptions and is, in the present context, not so
much important for its quantitative results, but it conceptually helps
to frame the problem. It has been extended to a wide range of
different experiments~\cite{Coloma:2012ji}, without however improving
on the underlying assumptions. In comparison of various different
experiments, it turns out that experiments which rely on a relatively
narrow beam spectrum and operate at energies below 1\,GeV, like T2HK,
are particularly sensitive to uncertainties on flavor ratios. On the
other hand, experiments which employ a wide beam spectrum at multi-GeV
energies, like LBNE, are much less affected by these rate-only
uncertainties. The implementation of cross section uncertainties of
Refs.~\cite{Huber:2007em} and \cite{Coloma:2012ji} is naive at best, and the remainder
of this section is devoted to more sophisticated case studies
based either  on specific cross section models or event generators. The spirit
of these examples, is that in order to estimate an unknown (and
uncomputable) theory error, an evaluation of the spread between
different theory calculation is performed, and this spread somehow is
indicative of the associated theory uncertainty. As will become
obvious the situation is complex and each experiment faces very
specific challenges, and while the challenges are specific the
solutions likely are not.


\subsection{The impact on the mixing angle measurement at  T2K}

Many of the techniques discussed in this subsection have been developed for
applications to the so-called $\beta$-beams~\cite{Zucchelli:2002sa}, and
can also be applied to real data, with the intent to estimate the
systematic effects introduced in the analysis from the non perfect
knowledge of the neutrino-nucleus cross
section~\cite{FernandezMartinez:2010dm}. A tentative step along this
line has been undertaken in the work of Ref.~\cite{Meloni:2012fq}, where a three-flavor
fit to the recent $\nu_\mu \to \nu_e$ and $\nu_\mu \to \nu_\mu$ T2K
oscillation data with different models for the neutrino-nucleus cross
section has been discussed.  It was shown that, even with a limited
statistics, the allowed regions and best fit points in the
$(\theta_{13},\delta_{CP})$ and $(\theta_{23},\Delta m^2_{atm})$
planes are affected if, instead of using the RFGM to
describe the QE cross section, a model including
multi-nucleon emission processes was employed.

The sample of analyzed data comprises the $\nu_\mu\to\nu_e$
appearance~\cite{Abe:2013hdq} and $\nu_\mu\to\nu_\mu$
disappearance~\cite{Abe:2014ugx} modes; in the first case, 28 events
passed all the selection criteria, implying (for a normal ordering
case):
\bea
\sin^2(2\theta_{13})_{T2K}=0.14 \,,
\eea
with the CP phase $\delta_{CP}$ undetermined.  In the disappearance
channel, the 120 events collected by T2K were fitted with:
\begin{align}
(\sin^2 \theta_{23})_{T2K} & = 0.51  \ , \\  
\nonumber
 |\Delta m^2_{atm}|_{T2K} & = 2.51 \cdot 10^{-3} \,\text{eV}^2 \ .
\end{align}
In this case different models were considered, involving not
only QE interactions but also pion production and inclusive cross
sections. On the one hand, it was chosen a model as similar as possible to
the one used by the T2K Collaboration. The T2K Collaboration simulates
neutrino-nucleus interactions using the NEUT Monte Carlo
Generator~\cite{Hayato:2002sd}.  Even if the details of NEUT\textemdash as well as the effects of
the latest tunings performed by the T2K Collaboration to take into account
the recent measurements of K2K \cite{Gran:2006jn,Rodriguez:2008aa},
MiniBooNE~\cite{AguilarArevalo:2009eb,AguilarArevalo:2010zc} and
SciBooNE~\cite{Kurimoto:2009wq,Nakajima:2010fp}\textemdash are not known,
exclusive channels were treated using the same models implemented in NEUT.  As a
consequence, the RFGM~\cite{Smith:1972xh} was used for the
QE channel and the Rein and Sehgal model \cite{Rein:1980wg}
was used for pion production.  The second model considered in the analysis was
the one developed by the authors of Ref.~\cite{Martini:2009uj}, that in the 
following will be referred to as RPA-2p2h, (see Section ~\ref{eA:xsec}).

In the following, cross sections obtained using the two
different approaches described above will be used.
Those cross sections were computed in several exclusive channels
(quasi-elastic and pion production), as well as in the inclusive one,
for both charged current (CC) and neutral current (NC) interactions on
carbon and oxygen (the targets used in near and far T2K detectors,
respectively) and for two different neutrino flavors $\nu_\mu$ and
$\nu_e$. Although all exclusive channels are involved in the analysis,
the first model will be referred to as RFGM
and the second approach as  RPA-2p2h model.

In order to perform the comparisons among the above-mentioned models,
the RFGM was firstly normalized to the T2K event
rates, at both near (ND) and far (FD) detectors using the following
algorithm:
\begin{enumerate}
\item Normalization of the cross section with the  $\nu_\mu$ inclusive CC at the ND; according to 
Ref.~\cite{Abe:2014ugx},
in order to reproduce $\sim 1.8 \times 10^4$ $\nu_\mu$ inclusive events, collected using
$\sim 6.4 \times 10^{20}$ POT, in the energy range $[0-3]$ GeV, with an
active detector mass of 1,529\,kg at a distance of 280\,m from the $\nu$
source and half a year of data taking (Run 1).  Since only the muon
neutrino cross sections can be correctly normalized, it was assumed that
the same normalization also applies for the $\nu_e$ cross section,
although they could differ at the $\mu$ production threshold (in any
case away from the peak of the neutrino flux);
\item Calculation of the expected events (and energy distributions) at the far detector in the appropriate two-parameter plane ($(\sin^2 2\theta_{13},\delta_{CP})$
for appearance and $(\theta_{23},\Delta m^2_{atm})$ for
disappearance);
\item Normalization to the T2K spectral distributions.
\end{enumerate}
Step 3 is needed to get rid of the experimental efficiencies to the
signal and background events. This means that the bin contents of the
simulated distributions (obtained at step 2) are corrected by
coefficients, generally of ${\cal O}(1)$, that were considered as a
detector property, and then not further modified.  For a different
model, step 1 was first redone, and then step 2 was repeated, using the same
normalization coefficients extracted in step 3 with the RFGM.
GLoBES \cite{Huber:2004ka,Huber:2007ji} and
MonteCUBES \cite{Blennow:2009pk} were the softwares used for the computation of event
rates (and related $\chi^2$ functions) expected at the T2K ND and FD.  The
fluxes of $\nu_\mu$, $\nu_e$ and their CP-conjugate
counterparts predicted at the FD in absence of oscillations were
extracted directly from Fig.~1 of Ref.~\cite{Abe:2011sj}, whereas the
$\nu_\mu$ flux at the ND was obtained from the authors of Ref.~\cite{giganti}.

\paragraph*{The appearance channel}
The $\nu_\mu \to \nu_e$ transition probability is particularly
suited for extracting information on $\theta_{13}$ and
$\delta_{CP}$; at the T2K energies ($E_\nu$) and baseline (L), the full 3-flavor probability
can be expanded up to second order in the small
parameters $\tetaot, \Delta_{12} / \Delta_{13}$ and $\Delta_{12} L$,
with $\Delta_{ij}=\Delta m^2_{ij}/4 E_\nu$ \cite{Cervera:2000kp}. The resulting expression is

\begin{align}
& P_{\nu_\mu \to \nu_e} = s_{23}^2 \, \sin^2 2 \tetaot \, \sin^2 \left (\delot \, L \right )  \\
\nonumber & \ \ \ \ + c_{23}^2 \, \sin^2 2 \theta_{12} \, \sin^2 \left( \Delta_{sol} \, L \right ) \\
\nonumber
& \ \ \ \ + \tilde J \, \cos \left (\delta_{CP} + \delot \, L \right ) \;
(\Delta_{sol} \, L)\, \sin \left ( 2\, \delot \, L \right ) \, , \nn
\label{vacexpand} 
\end{align}
where
\be
\tilde J \equiv c_{13} \, \sin 2 \theta_{12} \sin 2 \tetatt \sin 2 \tetaot \ \ , \ \ s_{23} = \sin \theta_{23} \ .
\ee
We clearly see that CP violating effects are encoded in the
interference term proportional to the product of the solar mass
splitting and the baseline, implying a weak dependence of this
facility on $\delta_{CP}$ when only the $\nu_\mu\to \nu_e$ channel
(and the current luminosity) is considered.

\paragraph*{Extracting the T2K data}
Events in the far detector are $\nu_e$ CC QE from $\nu_\mu \to \nu_e$ oscillation, with the main
backgrounds given by $\nu_e$ contamination in the beam and neutral
current events with a misidentified $\pi^0$.  The experimental data
have been grouped in 25 reconstructed-energy bins, from 0 to 1.25 GeV.  The expectations for signal
and backgrounds have been computed by the T2K Collaboration from Monte
Carlo simulations, for the following fixed values of the oscillation parameters:
\begin{align}
\nonumber
\sin^2 \theta_{12}=0.306 \  , \  \sin^2 2\theta_{13}=0.14  \ ,  \  \sin^2 2\theta_{23}=1 \ ,  
\end{align}
and
\begin{align}
\nonumber
\Delta m^2_{sol}=7.6\times 10^{-5} eV^2 \ , \ \Delta m^2_{atm}=+2.4\times 10^{-3} eV^2 \ .
\end{align}
 In order to normalize the event
rates to the T2K Monte Carlo expectations, few numerical value were extracted
from Table I of Ref.~\cite{Abe:2013hdq}.
For the sake of simplicity, the central 
value was used as the reference value for the neutrino energy in a given bin; this could be
different from the reconstructed neutrino energies used by the T2K Collaboration. To mimic
possible uncertainties associated with the neutrino energy reconstruction,
an energy smearing function was used to distribute the event rates in the various energy bins.
The ratios among the computation presented in Ref.~\cite{Meloni:2012fq} and the T2K data
define a sort of {\it energy dependent} efficiencies, $\varepsilon$,  which, for the $\nu_\mu \to \nu_e$ signal turn out to be
$\varepsilon\sim 0.4$. This procedure (corresponding to step 3 of the previous paragraph) allows
to take into account all the detection efficiencies to different
neutrino flavors in the Super Kamiokande detector. Once computed,
these corrective factors are used in the simulations done with a
different cross section model, since it was assumed that those are due to detector features
and not to the neutrino interactions.

\paragraph*{Fit to the data}
Using these results, the authors of Ref.~\cite{Meloni:2012fq} performed a $\chi^2$ analysis to
reproduce the allowed regions of the $(\sin^2 2\theta_{13},\delta_{CP})$-plane as shown in Fig. 5 of
Ref.~\cite{Abe:2013hdq}, based on a complete three-neutrino
analysis of the experimental data shown in Fig. 4 of the same paper, marginalising over all parameters
not shown in the confidence regions. As external input errors, the following list of parameters was used:
3\% on $\theta_{12}$ and $\Delta m^2_{sol}$, 8\% on $\theta_{23}$ and
6\% on $\Delta m^2_{atm}$.  In addition, a constant energy resolution
function $\sigma(E_\nu)=0.085$ was used and, for simplicity, a 7\%
normalization error for the signal and 30\% for the backgrounds were adopted. The energy calibration errors were
fixed to $10^{-4}$ for the signal and $5\cdot 10^{-2}$ for the backgrounds; normalization and energy
calibration errors were taken into account in evaluating the impact of systematic errors
in the $\chi^2$ computation.\\ Assuming a normal hierarchy spectrum,
the best-fit point from the fit procedure is:
\bea
\sin^2(2\theta_{13})=0.126  \ \  ,  \  \ \delta_{CP} = 0.45
\eea
with $\chi^2_{min} = 19.8$.
Compared to the official release, the best fit points are in quite good
agreement.

The same procedure was then applied to determine $\theta_{13}$ using the
RPA-2p2h cross sections described in Ref.~\cite{Martini:2009uj}.  In doing
that, the cross sections were normalized to the ND events and then
the number of oscillated events (and related backgrounds) were computed, and
compared with the experimental T2K data. Assuming that the
energy dependent efficiencies computed in the previous section are exactly the same, since
they are a property of the SK detectors, and therefore independent of the
cross section model, and considering that the CC RPA-2p2h cross section is
a bit larger than the RFGM cross section,
a larger bin-to-bin rate was obtained, for a total of 33 events (signal plus backgrounds).

\begin{figure}[!h]
\centering
\includegraphics[width=0.45\textwidth]{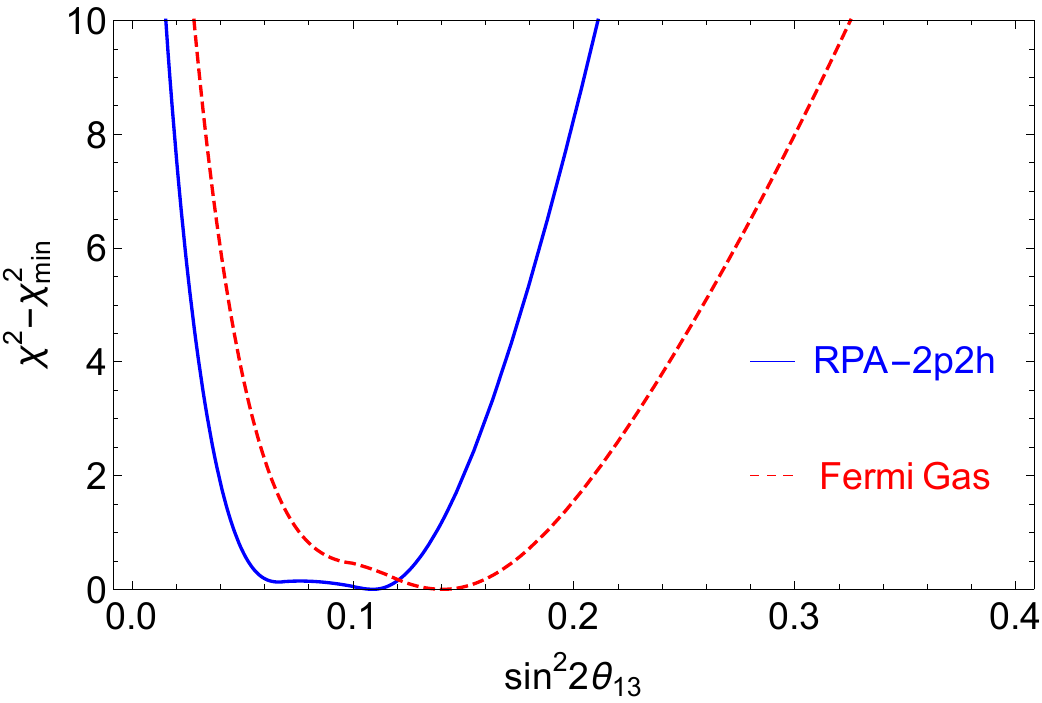}
\caption{Behavior of the function $\chi^2 - \chi^2_{min}$ (see text) as a function of $\theta_{13}$, for the RPA-2p2h model (solid line) and the RFGM (dashed line).}
\label{chi2}
\end{figure}

It is clear that larger rates need smaller $\theta_{13}$ to reproduce
the data (the effect of the CP phase $\delta$ is negligible with such
a statistics). The best fit point is:
\bea
\sin^2(2\theta_{13})=0.08 \ \ \ , \ \ \ \delta_{CP} = 0\,,
\eea
with $\chi^2_{min} = 19.2$. 
To make a more direct comparison on
$\theta_{13}$ between RFGM and RPA-2p2h results, in Fig.~\ref{chi2}
we show the  function $\chi^2-\chi^2_{min}$, computed marginalizing
over all other oscillation parameters (including $\delta_{CP}$). At
1$\sigma$ level, the following result was obtained:
\begin{align}
\nonumber
\sin^2 2\theta_{13}^{RPA-2p2h} = 0.11 {^{+0.03}_{-0.06}}  \ \ \ , \ \ \ 
\sin^2 2\theta_{13}^{FG} &= 0.14 {^{+0.05}_{-0.06}}  \ .
\end{align}
The results are clearly compatible although, as expected, $\theta_{13}^{RPA-2p2h}  < \theta_{13}^{FG} $.

\paragraph*{The disappearance channel}\label{par:disappearance}

\begin{figure}[!h]
\vspace*{.1in}
\centering
\includegraphics[width=0.430\textwidth]{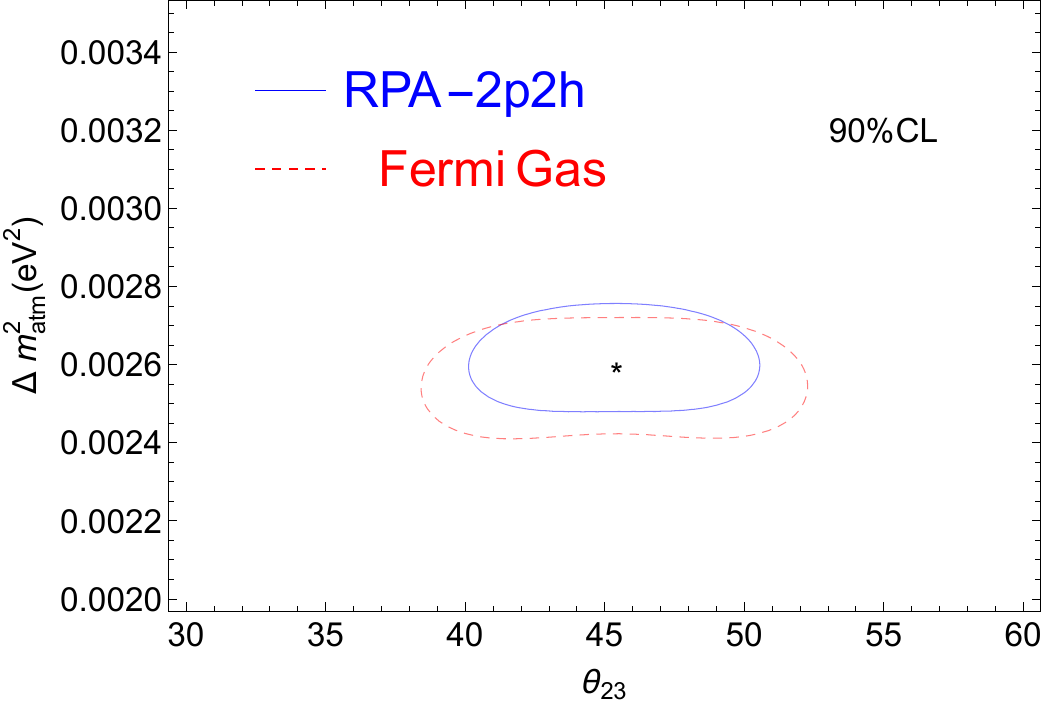}
\caption{ 90\% contour levels for the RPA-2p2h model (solid line) and the RFFM (dashed line),
in the $(\theta_{23},\Delta m^2_{atm})$ plane. Star indicates the best fit obtained in the  RPA-2p2h model.}
\label{distort}
\end{figure}

The previous analysis was then extended to include the disappearance
$\nu_\mu \to \nu_\mu$ data \cite{Abe:2014ugx}.  In the two-flavor limit,
(the one where both $\theta_{13}$ and $\Delta m^2_{sol}$ are
vanishing) the $\nu_\mu \to \nu_\mu$ probability
reads \cite{Donini:2005db}:
\bea
P(\nu_\mu \to \nu_\mu) & = & 1-   \sin^2 2 \theta_{23} \, \sin^2 \left(\Delta_{atm} L\right )
 \label{eq:probdismu}\,.
\eea
Effects related to $\theta_{13}$ are clearly sub-dominant, so that
this channel is particularly useful to extract information on the
atmospheric parameters.
The T2K Collaboration collected 120 data events, grouped in 30 energy
bins, as one can see from Fig.~2 of \cite{Abe:2014ugx}. The sample extends
above 5~GeV and is mainly given by $\nu_\mu$CC QE, $\nu_\mu$CC
non-QE, $\nu_e$ CC and NC.
The RFGM cross section was normalized to the rates shown in Fig. 2 of~\cite{Abe:2014ugx}.
In the fit procedure a conservative 15\%
normalization error and an energy calibration error at the level of
$10^{-3}$ for both the signal and the background were adopted .
 The results of the analysis are shown in Fig.~\ref{distort}, where it can be found the 90\% CL limit for the RFGM (dashed
 line) and the RPA-2p2h model (solid line), in the case of normal
 hierarchy together with the 2 degrees of freedom (dof) confidence levels in
 the $(\theta_{23},\Delta m^2_{atm})$-plane.  Again, the plots
 have been obtained marginalizing over the  parameters that are  not shown (a full
 three-flavor analysis). The plot in Fig.~\ref{distort} was obtained considering a 50\% error on $\sin ^2
 2\theta_{13}$ (with best fit at $\sin ^2 2\theta_{13}=0.14$) and
 $\delta_{CP}$ undetermined.

In summary the following results were obtained:
\begin{align}
\nn
& RFG: \\ 
\nn
         & 38 <  \theta_{23} <   52 \ {\rm deg}  \ , \\
        \nn
         & 2.40 \cdot 10^{-3}<\Delta m^2_{atm} \,(\text{eV}^2)< 2.70 \cdot 10^{-3}\ , \\
\nn         
& \\
\nn
& RPA-2p2h: \\
\nn
       &  40 <  \theta_{23} <   50 \ {\rm deg}    \ , \\
        &   2.50 \cdot 10^{-3}<\Delta m^2_{atm}\, (\text{eV}^2)< 2.80 \cdot 10^{-3}  \ , \nn 
\end{align}
with best fit points:
\begin{align}
&RFG: \nn \\
&\theta_{23} =47.9 \ {\rm deg} \ \ , \ \   \Delta m^2_{atm} = 2.56 \cdot 10^{-3} \,\text{eV}^2\nn  \\ \nn
& \\
& RPA-2p2h:\nn \\ 
& \theta_{23} =45.4 \ {\rm deg} \ \ , \ \     \Delta m^2_{atm} = 2.62 \cdot 10^{-3}\, \text{eV}^2\nn \ .
\end{align}
Here, some comments are in order. First of all, we should observe that, for both
models, the best fit point is different from the T2K one, which corresponds to
\begin{align}
(\theta_{23})_{T2K} =45.8 \ {\rm deg}    \ \ , \ \     |\Delta m^2_{atm}|_{T2K} = 2.51 \cdot 10^{-3} \,\text{eV}^2 \nn  \ .
\end{align}
This is somehow obvious since we normalized our events to the MC
predictions obtained for a different set of atmospheric
parameters. The RPA-2p2h cross section gives a better determination of
both $\theta_{23}$ and $\Delta m^2_{atm}$, mainly due to the larger
statistics with respect to the RFGM; at the same time, the disappearance
probability in Eq.(\ref{eq:probdismu}), for negligible solar mass
difference and reactor angle, is smaller if the atmospheric mass
difference is larger, for fixed $\sin^2 2\theta_{23}$. This is what
happens here, where a smaller $P(\nu_\mu \to \nu_\mu)$ (and then a
larger $\Delta m^2_{atm}$) is needed in the RPA-2p2h model to
partially compensate for the larger cross section.

For the sake of completeness, the same computations were repeated
as above under the hypothesis that the neutrino mass spectrum is of
inverted type (IH). With the current T2K statistics, one does not find
significant differences in the results obtained using the two
different models for the cross section.

\subsection{Reconstruction of neutrino energy}

\begin{figure*}
\begin{center}
\includegraphics[width=0.475\textwidth]{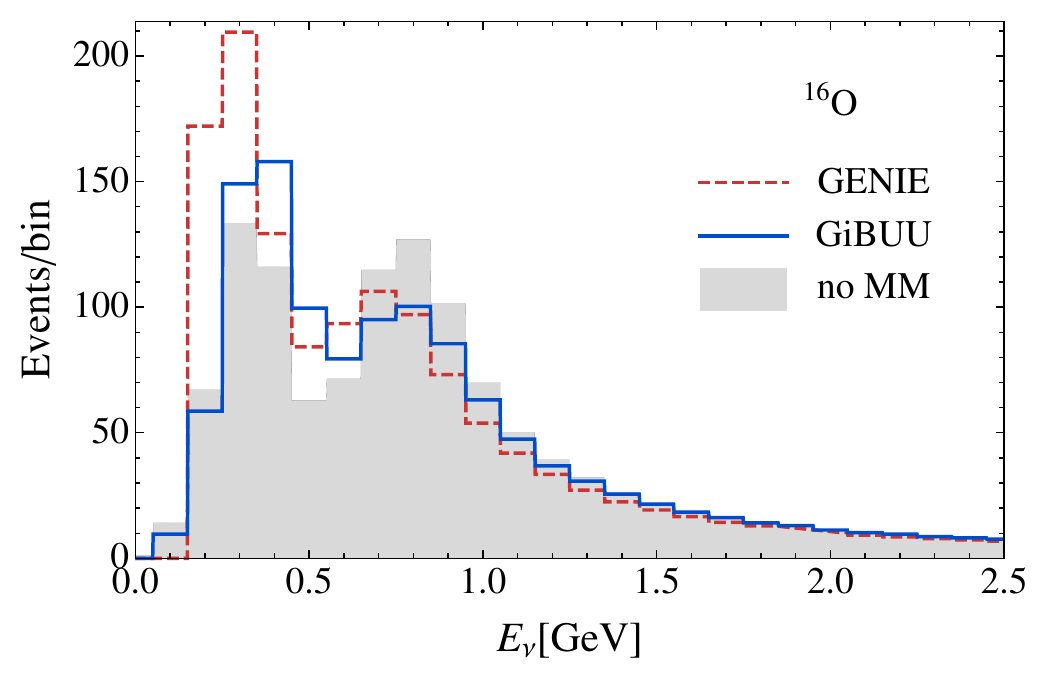}%
\includegraphics[width=0.475\textwidth]{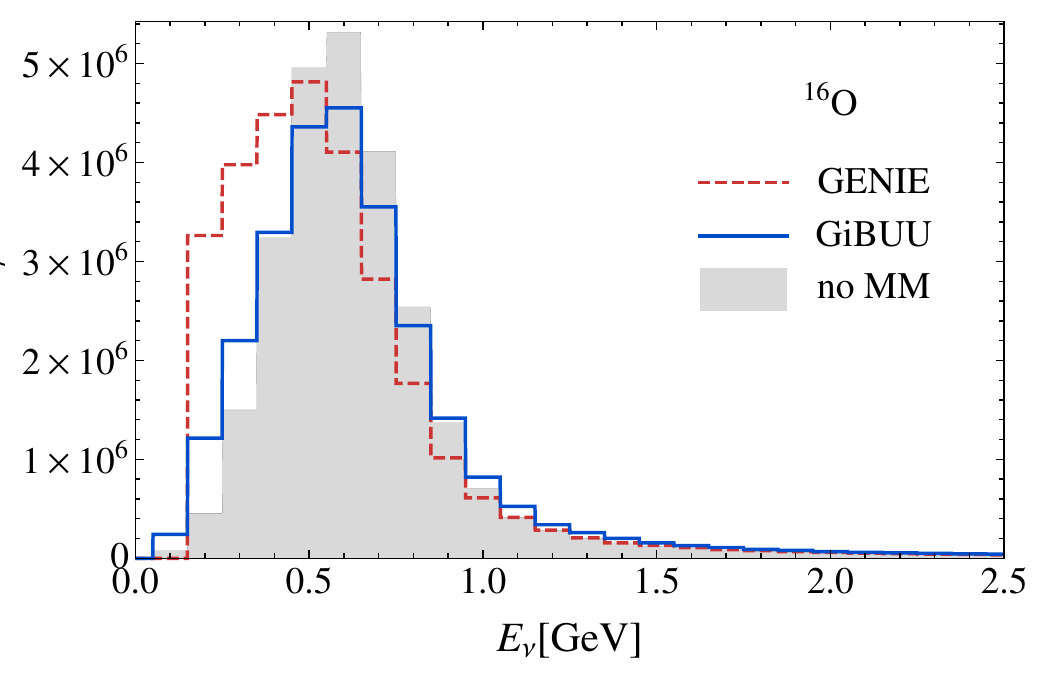}
\end{center}
\vspace*{-0.25in}
\caption{\label{fig:qev} Binned QE-like event rates as a function of
  the reconstructed neutrino energy in GeV. The solid (dashed) line shows the event rates obtained after migration using the
  GiBUU (GENIE) event generator. The shaded areas show the expected
  event rates coming from the QE-like event sample computed using the
  GiBUU cross-section for $^{16}$O, as for the solid lines, but
  without including any migration matrices. For the shaded areas, a
  Gaussian energy resolution function with a constant standard
  deviation of 85\,MeV is added to account for the finite resolution of
  the detector. The left and right panels show the event rates at the near
  and far detectors, respectively. Figure and caption
  adapted from Ref.~\cite{Coloma:2013tba}.}
\end{figure*}
Future long-baseline neutrino experiments will rely on their ability
to map the energy dependence of the oscillation probability to measure
oscillation mixing parameters and test the validity of the
three-flavor oscillation framework.

The oscillation probability is a
non-trivial function of the \emph{true} neutrino energy, and thus the
problem of reconstructing the neutrino energy arises. Also, the fact that some
experiments seem to be less affected by rate-only systematics is
largely due to their ability to exploit the differences in energy
dependence of the various contributions to the error budget to control
systematic uncertainties. There are a number of
works~\cite{Lalakulich:2012hs,Martini:2012fa,Nieves:2012yz,Mosel:2013fxa}
where it is shown that event rate distributions in reconstructed
neutrino energy will change significantly based on the underlying
interaction model. A conceptual laboratory is provided by quasi-elastic
scattering, which, due its relative simplicity and amenability to
theoretical calculations, has been also the focus of many published
studies. In a true quasi-elastic scattering event involving a neutron
at rest there is a one-to-one correspondence between the charged lepton
momentum and emission angle and the incoming true neutrino energy, given by
\begin{equation}
\label{eq:qe-energy}
E_\nu= \frac{ 2(m_n - \epsilon) E_\ell - ( \epsilon^2-2m_n  \epsilon + m^2_\ell+\Delta m^2)}
{2(m_n - \epsilon - E_\ell+|{\bf k}_\ell | \cos \theta_\ell)} \ . 
\end{equation}
In the above equation,  ${\bf k}_\ell$ and $E_\ell$ are the momentum and energy of the outgoing charged lepton,
$\theta_\ell$ is the scattering angle in the laboratory frame,
$\Delta m^2=m_n^2-m_p^2$ is the neutron-proton squared mass difference and $\epsilon$ denotes the average binding energy
of the neutron.

The problem, now, is that in any experiment there will be events which
are not quasi-elastic but exhibit all the same \emph{experimental}
signatures of a true quasi-elastic event, {\it e.g.} the so-called
stuck-pion events where in addition to the charged lepton a pion is
produced at the vertex, but this pion is then re-absorbed within the
nucleus. As a result, any real QE event sample will contain non-QE
events as well, and for those non-QE events the simple kinematic
relation in Eq.~\eqref{eq:qe-energy} will not be valid.

In a water
Cherenkov detector the selection criterion for QE events is that only
one charged particle is above Cherenkov threshold, resulting in a
single ring of light. Taking the output of an event generator and
selecting events using this criterion it is possible to construct the
appropriate migration matrix between true and reconstructed energy,
which, if the generator were perfect, would completely describe these
effects. In practice, different generators lead to very different
migration matrices and, as a result, to very different reconstructed
energy distributions in both near and far detectors, as shown in
Fig.~\ref{fig:qev}. Interestingly, an offset in the distributions
between GENIE and GiBUU is observed and the overall effect is to
change the position and depth of the oscillation dip. Similar effects
have been observed previously by several
authors~\cite{Lalakulich:2012hs,Martini:2012fa,Nieves:2012yz,Mosel:2013fxa}. The
next question to address is what will the impact on the extraction of
oscillation parameters be, and whether the inclusion of the near
detector will solve the problem.

A first step in this direction was
taken by the authors of Ref.~~\cite{Coloma:2013rqa}, where a comparison between ideal
energy reconstruction and a fit performed using migration matrices derived from
an event generator\textemdash in this case GiBUU\textemdash  was made. Specifically, the
ability to measure the atmospheric mixing parameters in
$\nu_\mu\rightarrow\nu_\mu$ disappearance was studied. The main
finding is that the use of a near detector leads to a high $\chi^2$-value
per degree of freedom, that is a bad fit, but does not prevent
significant bias in the parameter determination, which can potentially be
as large as several times the statistical error. Both the
$\chi^2$-value and bias can be reduced if the energy scale of the
experiment is allowed to shift by as much as 5\%. The key to this
behavior is the fact, that the beam flux is not known a priori to
better than 5--10\% and hence the near detector cannot determine the
right energy migration matrix \emph{and} the beam flux simultaneously.
In essence, there are fewer observables than unknowns.

\begin{figure*}
\includegraphics[width=0.5\textwidth]{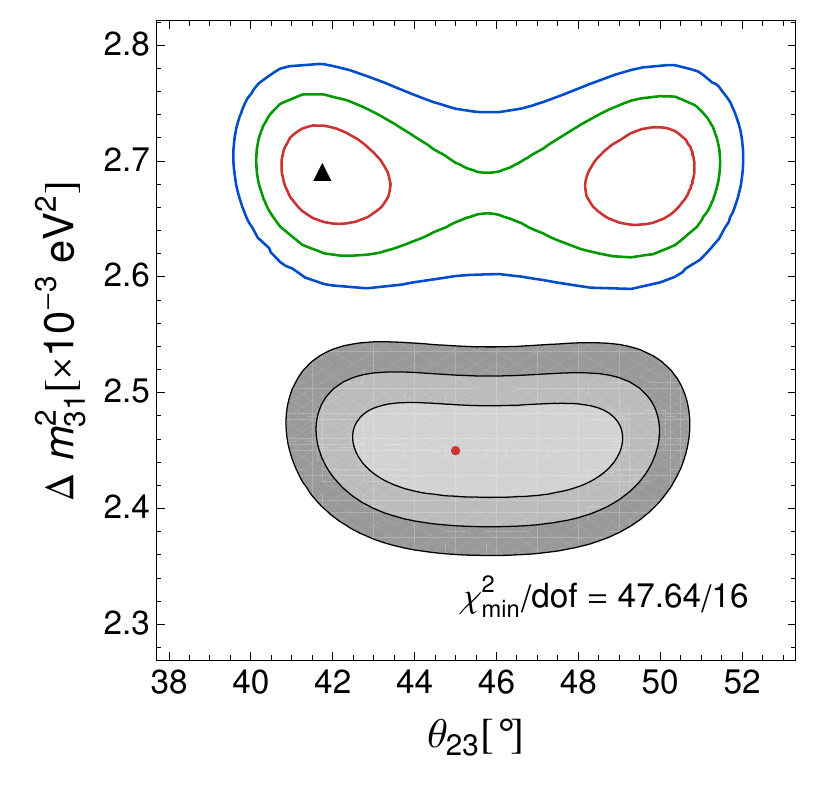}%
\includegraphics[width=0.5\textwidth]{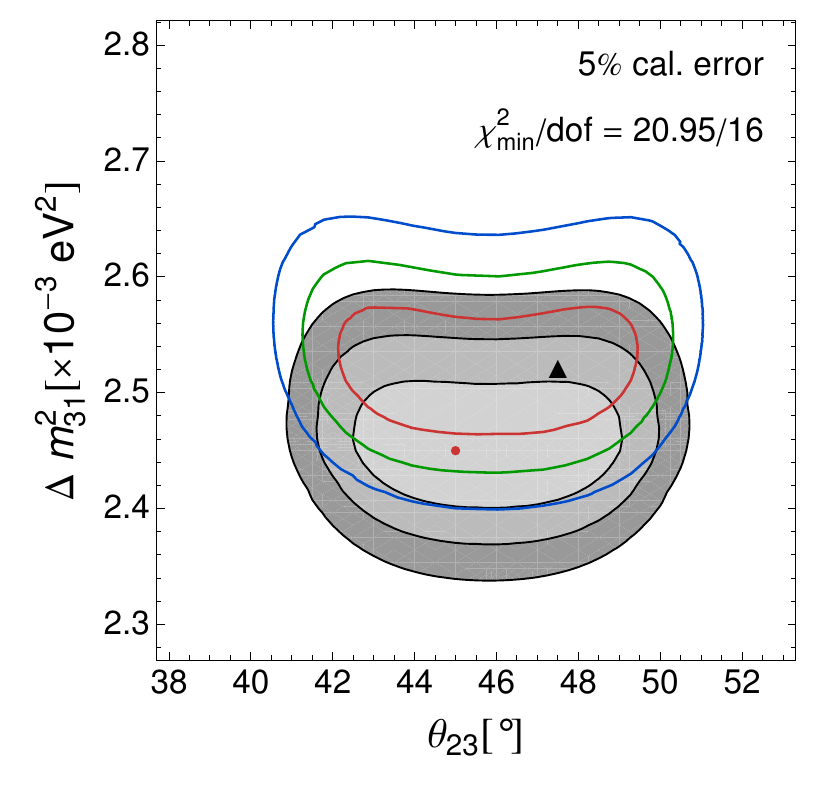}
\vspace*{-.25in}
\caption{Impact on the results if a different generator is used to
  compute the true and fitted rates in the analysis. The shaded areas
  show the confidence regions at 1, 2 and 3$\sigma$ that are
  obtained in the $\theta_{23}-\Delta m_{31}^2$ plane if the true and
  fitted rates are generated using the same set of migration matrices
  (obtained from GiBUU, with oxygen as the target nucleus). The
  solid lines show the same confidence regions if the true rates are
  generated using matrices produced with GiBUU, but the fitted rates
  are computed using matrices produced with GENIE. Both sets of
  matrices are generated using oxygen as the target nucleus. The
  dot indicates the true input value, while the triangle shows
  the location of the best fit point. The value of the $\chi^2$ at the
  best fit is also shown, together with the number of degrees of
  freedom. In the left panel no energy scale uncertainty is
  considered, while for the right panel an energy scale uncertainty of
  $5\%$ is assumed, see text for details. Figure and caption adapted
  from Ref.~\cite{Coloma:2013tba} \label{fig:generators}}
\end{figure*}

A somewhat more sophisticated analysis has been presented
in Ref.~\cite{Coloma:2013tba}, where a comparison between two event
generators, GiBUU and GENIE, was performed. One of the
generators was used to compute virtual data, and this virtual data was in
turn fitted using the other event generator. The specific choice
of generators for this comparison was guided by convenience and
availability and does not imply that one of them is more accurate than
the other, or that any of them is more accurate than some other
generator. The results are somewhat sobering, as can be seen from
Fig.~\ref{fig:generators}, clearly indicating that a large bias with
acceptable $\chi^2$-values could occur. The closed contours are
obtained from fitting data generated with GENIE with GENIE, whereas the
open contours are obtained from fitting data generated with GENIE with
GiBUU. In the left hand panel the energy scale is fixed, while in the right
panel a 5\% energy scale shift is allowed.

The authors of Ref.~\cite{Mosel:2013fxa} pointed out that an event sample which
is the combination of 0-pion events (traditionally selected as QE),
1-pion and N-neutron events has a much more benign behavior in terms
of shifting the oscillation peak. It is clear, that such a
selection would require a liquid argon detector. Moreover, the
statistics is strongly reduced with respect to the full event sample
in an experiment in the multi-GeV region. There is some risk that an
improved systematic error is bought at the price of a greatly enlarged
statistical error. On the other hand, having a reliable sub-sample may be
sufficient to also tie down the energy scale of DIS events. This issue
awaits further detailed study along the lines
of Ref.~\cite{Coloma:2013tba}.

The results of Ref.~\cite{Mosel:2013fxa} indicate that a more detailed
treatment of the hadronic energy deposition, which can be measured to
a large extent in liquid argon detectors, will likely improve the
results. At the same time, missing energy will    become a crucially
important problem. Neutral secondary particles like $\pi^0$ and
neutrons have to either decay or interact further in the detector to
leave a signature. The amount of these neutral secondaries will be
very different in neutrino and antineutrino interactions, as will their energy
distributions. This can be easily seen from the very different
$y$-distributions in DIS for neutrinos and antineutrinos. The
contained fraction of neutral particles will be a sensitive function
of detector size and surface to volume ratio. Therefore near and far
detectors are guaranteed to behave very differently in this
respect. Even if neutral particles are contained in the detector,
associating their signature with the (correct) primary vertex will be
complicated, in particular in the near detector, which may see more
than one event per beam spill.

A further real-world issue will be energy thresholds, implying that a proton 
has to exceed a certain value of kinetic energy in order to be detected. The relatively poor energy
resolution for hadronic energy deposition, compared to the one for
leptons, will  impose further limitations.  A quantitative study of the
impact of hadronic calorimetry with a special emphasis on the missing
energy from neutral secondaries is urgently needed. This also implies
the need for reliable theoretical models\textemdash most notably of  neutral
particle production in neutrino-nucleus interactions\textemdash capable to 
correctly predict multiplicity and momentum distributions,
since they affect the fraction of contained events and their
signature. This information is required to correctly account for the
aforementioned thresholds and the hadronic energy resolution, even if
detailed test beam data on the detector response exist. Furthermore,
it is obvious that the energy scales and systematic bias for electron-
and muon-type events will be quite different, which adds another level
of complexity, an will cause practical difficulties for attempts to use
the disappearance data, typically based on
$\nu_\mu\rightarrow\nu_\mu$, to ``calibrate'' the energy scale of the
appearance data set, based on $\nu_\mu\rightarrow\nu_e$. It is also worth
noting that this approach effectively assumes the correctness of the
three flavor oscillation framework,  thus defeating one the major
reasons for pursuing long-baseline experiments:  \emph{tesing} the
validity of the three-flavor description.


\subsection{Detector effects impact on disappearance and appearance results}
\label{sec:energy}

As we have seen in the previous sections in long/short baseline
neutrino oscillation experiments neutrino energy is the key
element. The neutrino energy is reconstructed looking at the
kinematics of particles produced in the neutrino interaction with the
detector. There are two reconstruction techniques:
one, already described in Eq.~\eqref{eq:qe-energy}, is based on the QE
assumption and only uses the  information on muon momentum. We will call this
method kinematical. A different technique to reconstruct the energy, that we
will call calorimetric, consists in collecting all the energy of the
particles deposited in the detector, and from that infer the neutrino
energy. In both methods the reconstructed neutrino energy will depend
upon detection capabilities, such as energy resolutions,
reconstruction efficiencies and detector energy thresholds for the
different particles.  

The impact of detector capabilities on the oscillation analysisIn has been recently investigated in 
Ref.~\cite{Ankowski:2015jya,Ankowski:2015kya}. The
authors studied how uncertainties related to the detector effects
influence the oscillation analysis by using event distributions simulated
in the far detector which include detector effects and analyzing them
neglecting, or partially neglecting, detector effects in the final
oscillation analysis.

In Ref.~\cite{Ankowski:2015jya,Ankowski:2015kya} the kinematical observables have been 
smeared according to a normal distributions centered
at their true values. For muons, the smearing was applied to both 
the momentum and emission angle, using the following realistic
parameters~\cite{Aliaga:2013uqz}:

\begin{equation}
\sigma(\n{k_\mu})=0.02\n{k_\mu}\quad\textrm{ and }\quad\sigma(\theta)=0.7^\circ.
\label{eq:muonResolutions}
\end{equation}
The energy resolutions of for $\pi^0$'s producing electromagnetic showers and other hadrons have been set to
\begin{align}
\label{eq:hadResolutions}
 \frac{\sigma(E_{\pi^0})}{E_{\pi^0}}& =\max\left\{\frac{0.107}{\sqrt{E_{\pi^0}}},\, \frac{0.02}{E_{\pi^0}}\right\}  \ ,  \\ 
 \notag
\frac{\sigma(E_h)}{E_h}& =\max\left\{\frac{0.145}{\sqrt{E_h}},\,0.067\right\} \ , 
\end{align}
respectively, with detection thresholds corresponding to a measured kinetic energy of 20~MeV for mesons and 40~MeV for protons. 
The efficiencies, on the other hand,  have bee considered as energy independent and set to 60\% for $\pi^0$'s, 80\% for other mesons, and 50\% for protons. 
Neutrons were assumed to always escape detections.

In the context of the $\nu_\mu$ disappearance analysis~\cite{Ankowski:2015jya}, the authors analyzed the T2K-like setup described in Section~\ref{par:disappearance}, with a number of expected unoscillated events of $\sim$4900.

Instead of changing individual parameters related to the detector performances, the authors used a linear combination of migration matrices computed with and without detector effects.

\begin{figure*}
\includegraphics[width=0.4\textwidth,angle=-90]{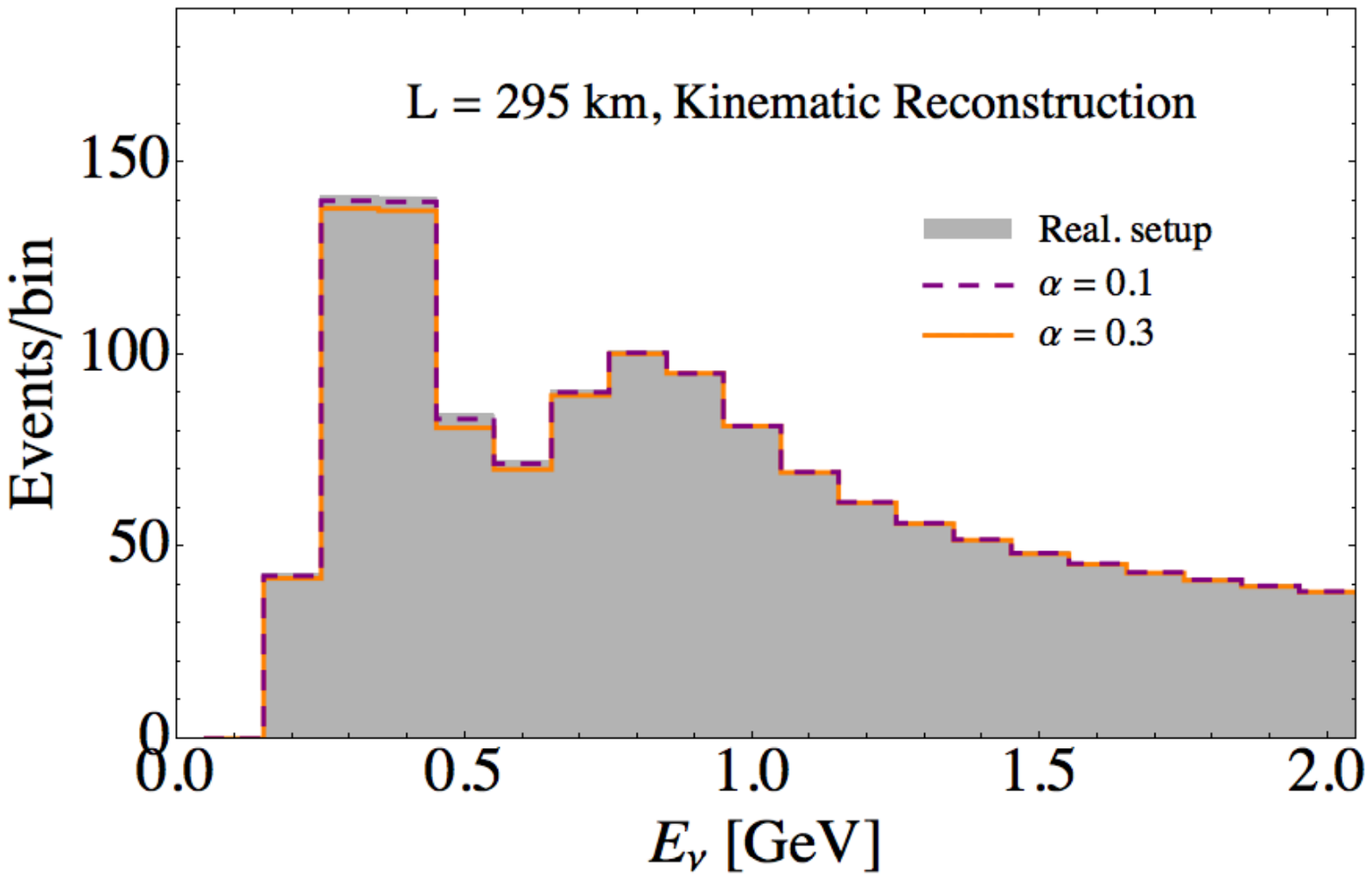}
\includegraphics[width=0.4\textwidth,angle=-90]{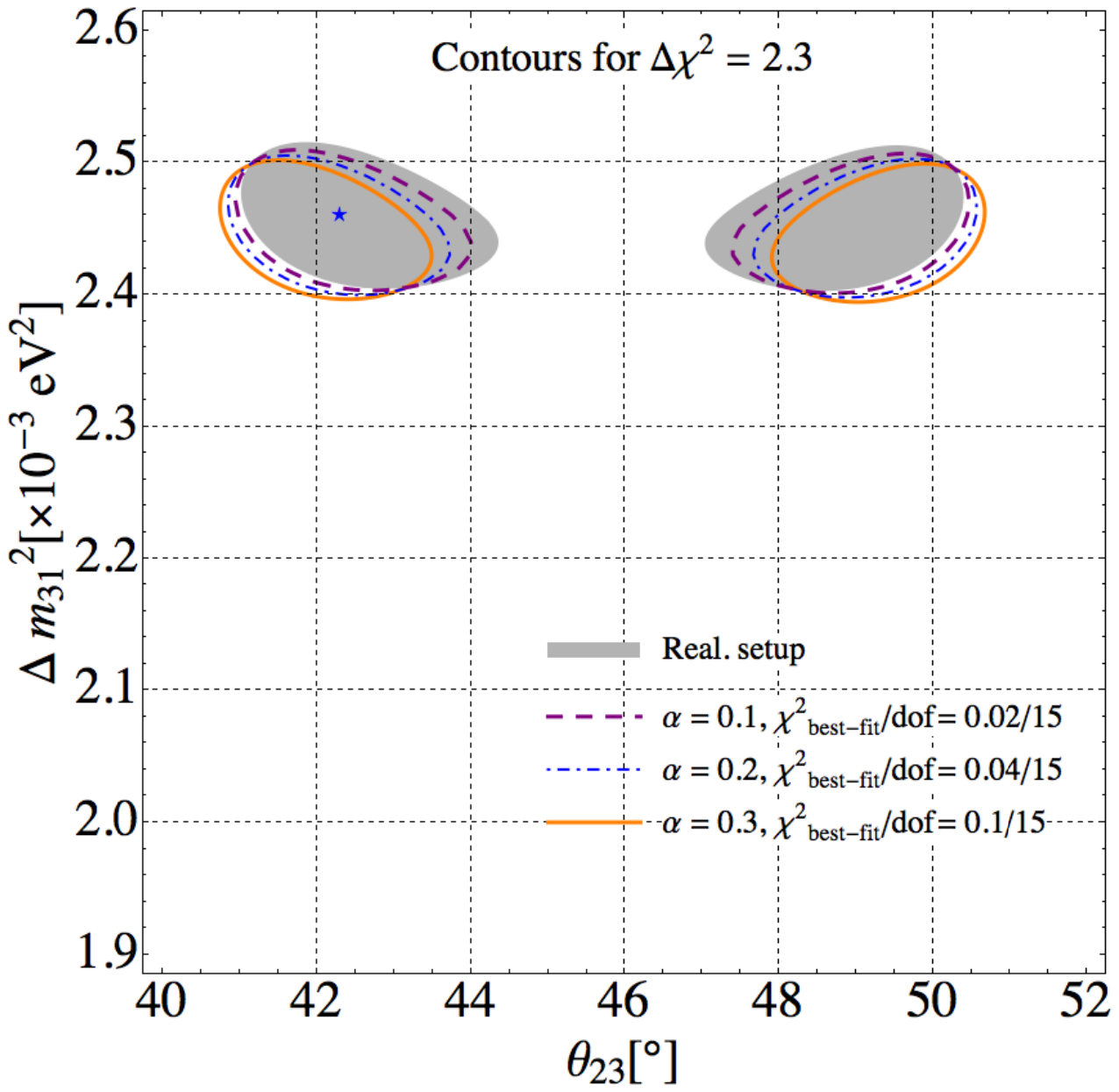}
\caption{Effect of detector-related uncertainties on the oscillation analysis performed using kinematic energy reconstruction. Left: comparison between the event distributions obtained 
overestimating the detector performance by 10\% ($\alpha = 0.1$, dashed line) and by 30\% ($\alpha = 0.3$, solid line). The accurate estimate of detector effects is represented in the shaded histogram. Right: $1\sigma$ confidence regions in the $(\theta_{23},\,\Delta m_{31}^2)$ plane obtained using data simulated including detector effects and oscillation parameters extracted using migration matrices with 100\% detector effects (shaded area) and at 10\%, 20\%, 30\% level only (lines). }
\label{fig:detEff_kin}
\end{figure*}

As shown in Fig.~\ref{fig:detEff_kin}, the kinematic method turns out to be mostly independent of detector effects, and uncertainties at the level of 30\% (corresponding to $\alpha= 0.3$), do not significantly affect the results of the oscillation analysis, due to the good resolution in the determination of the muon kinematics.

\begin{figure*}
\includegraphics[width=0.4\textwidth,angle=-90]{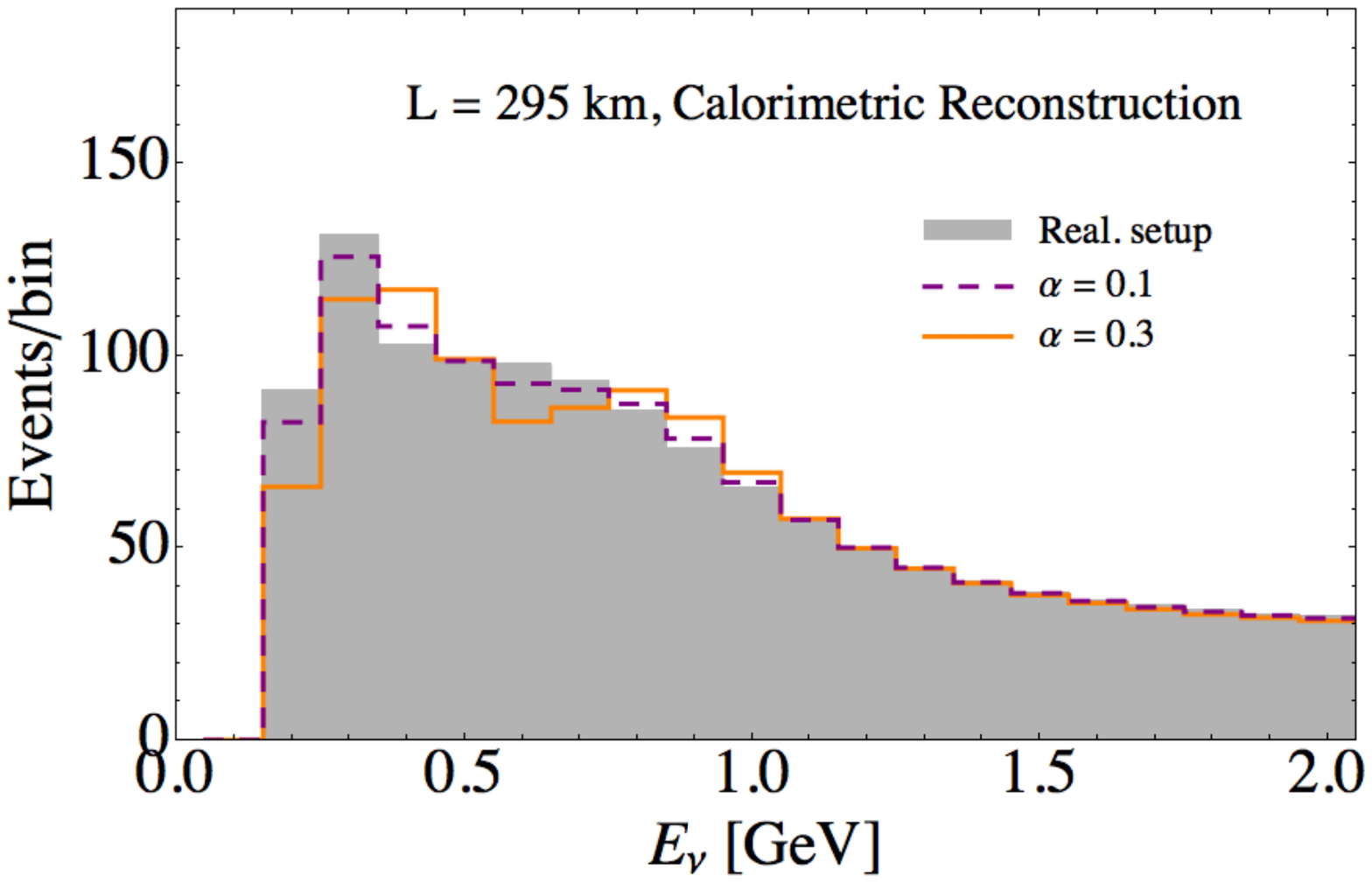}
\includegraphics[width=0.4\textwidth,angle=-90]{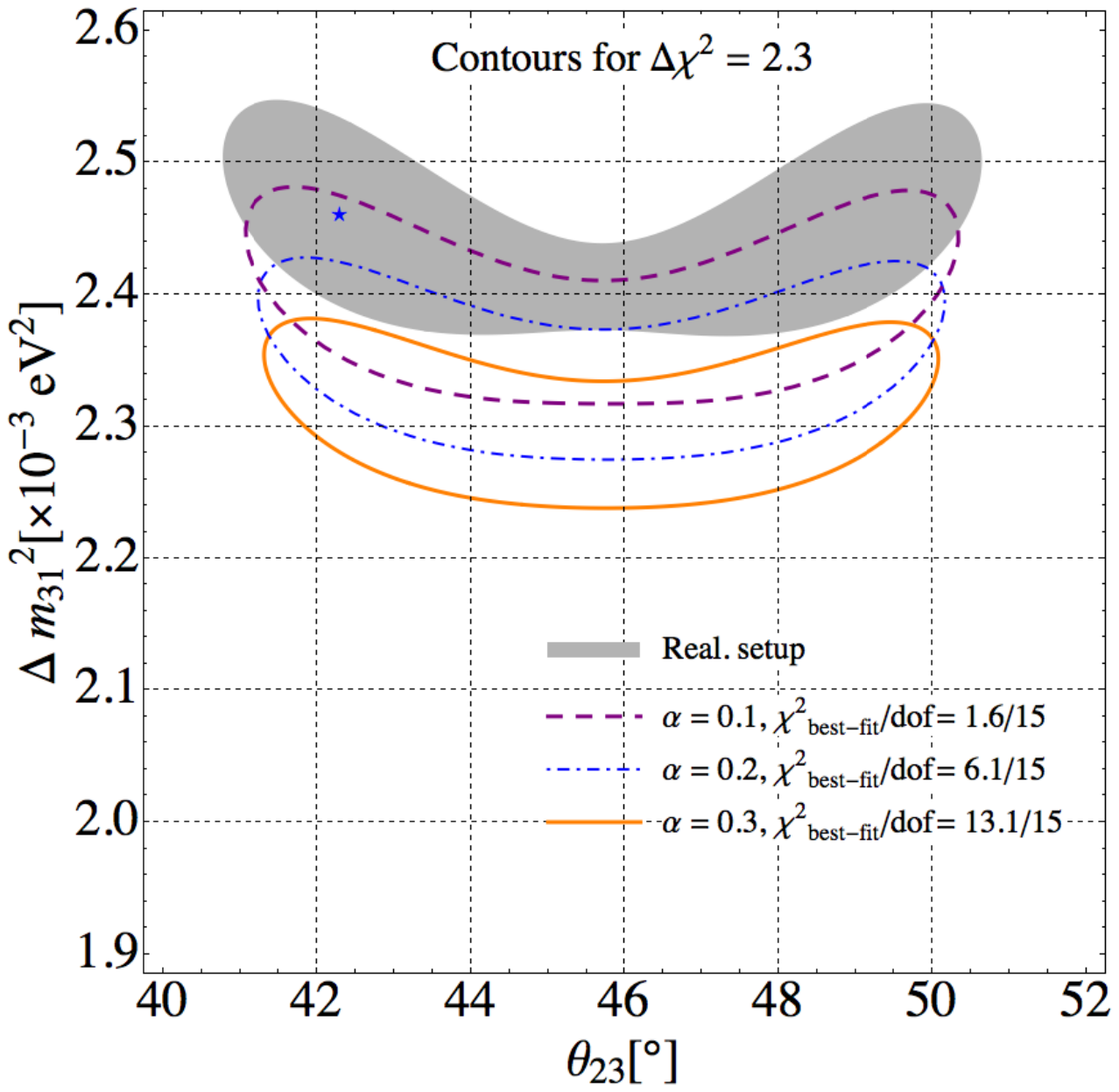}
\caption{Same as figure \ref{fig:detEff_kin} but considering the calorimetric energy reconstruction.}
\label{fig:detEff_cal}
\end{figure*}

In the case of calorimetric energy reconstruction, on the other hand, the authors of Ref.~~\cite{Ankowski:2015jya} see a large dependence of the neutrino energy on detector effects. Figure \ref{fig:detEff_cal} shows that  the detector response has to be determined with high precision\textemdash at least 10\%\textemdash to avoid large biases in the determination of the oscillation parameters. This behavior is mainly due to the large uncertainties in the determination of hadron energies , see Eq.~\eqref{eq:hadResolutions}, associated with the calorimetric technique.

Although the results of Figs.~\ref{fig:detEff_kin} and \ref{fig:detEff_cal} have been obtained for beam peaked at $\sim$0.6~GeV~\cite{Huber:2009cw}, the above conclusions can be extended to 
the case of a wide-band beam peaked at $\sim$1.6 GeV, as discussed in Ref.~\cite{Ankowski:2015jya}.

The same effects have been studied in the case of oscillation experiments aimed at accurately measuring $\delta_{CP}$. In this instance the oscillation probability is more complicated, due to the presence of different neutrino/anti-neutrino species (like electron (anti)neutrinos from oscillations of muon (anti)neutrinos). The authors of Ref.~\cite{Ankowski:2015kya} analyzed the role of missing energy on the 
$\delta_{CP}$ sensitivity for an experiment similar to DUNE~\cite{Acciarri:2016crz}.

In the DUNE-like setup, a wide-band neutrino beam has been analyzed with a
far detector of 40~kton (fiducial mass), located 1300~km from the
neutrino beam. The authors assumed 6 years of running, 3 years in
neutrino mode and 3 years in antineutrino mode. For the signal, 2\%
uncertainties of normalization (bin-to-bin correlated) and shape
(bin-to-bin uncorrelated) were considered, and in the case of the
background only a global normalization uncertainty of 5\% was taken
into account~\cite{Ankowski:2015kya}.

The event distributions have been simulated including all detector
effects---resolutions, efficiencies, and energy thresholds. The
oscillation parameters were extracted using linear combinations of
migration matrices calculated with and without missing energy
shifts.

\begin{figure*}
\includegraphics[width=0.5\textwidth]{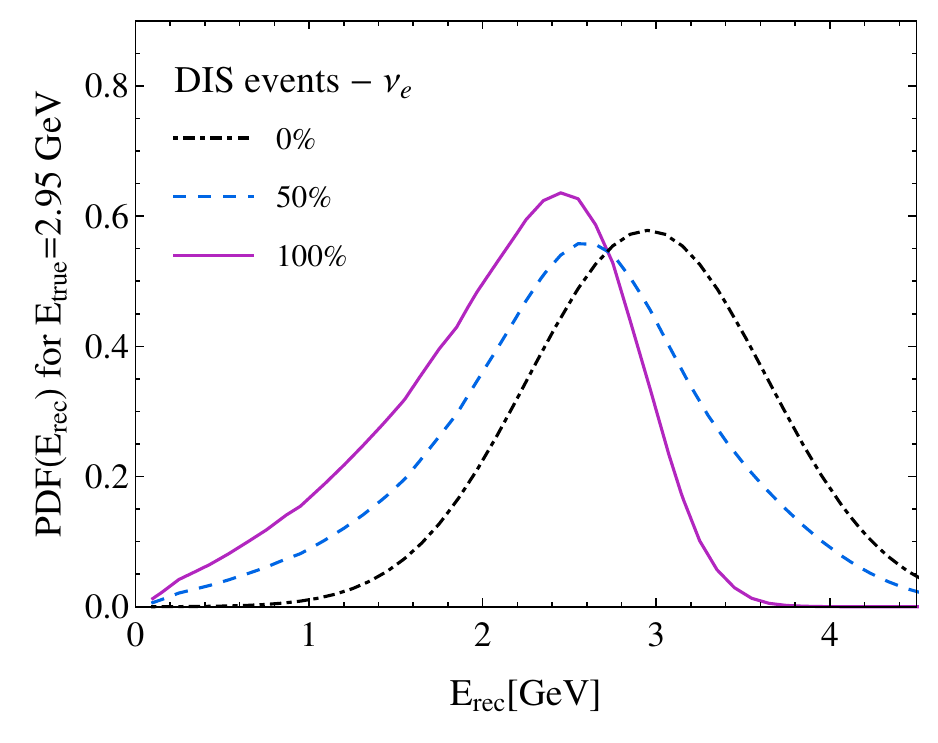}
\includegraphics[width=0.5\textwidth]{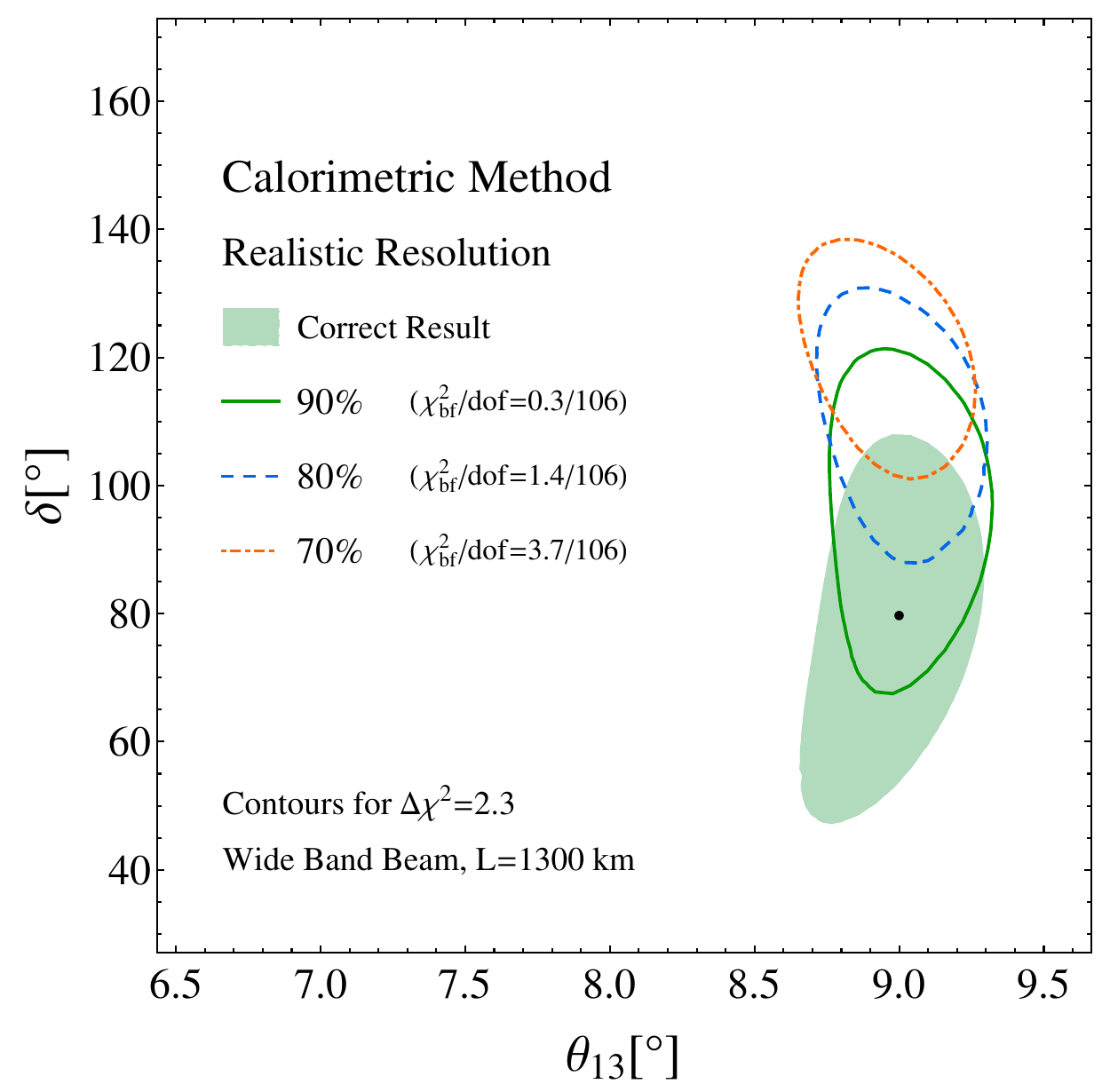}
\caption{Impact of the missing energy on the oscillation analysis for a DUNE like setup using the calorimetric energy reconstruction. Left: Reconstructed energy distributions for DIS of electron neutrino at energy 2.95 GeV (true), obtained including a 0\%, 50\% and 100\% shift of the reconstructed energy due to the missing energy. Right: $1\sigma$ confidence regions in $(\theta_{13},\,\delta_{CP})$ plane obtained using simulated data fitted using the migration matrices accounting for 90\%, 80\% and 70\% (lines) of the missing energy and all the missing energy (shaded area). The dot shows the assumed true values of the oscillation parameters.}
\label{fig:missE}
\end{figure*}

From the left panel of Fig.~\ref{fig:missE} it is evident that
the maximum of the distribution in energy is shifted to values lower
than the true neutrino energy, due to the missing energy. As the
nature of the particles in the final state is different for neutrinos and
antineutrinos, this shift will be different, since it depends on the
interaction channel, the momentum transfer and the nature of the
neutral secondary particles which give rise to missing energy.  As
shown in the right panel of Fig.~\ref{fig:missE}, a 20\%
underestimation of the missing energy introduces a large bias in the
extracted $\delta_{CP}$ value. If, instead, the missing energy will be
underestimated by 30\%, the oscillation analysis would exclude the
true value of $\delta_{CP}$ at a confidence level between 2 and
3$\sigma$. This result illustrates the importance of an accurate
determination of detector response in test-beam exposures and the
relevance of a realistic simulation of nuclear effects in neutrino
interactions, including intranuclear cascade.

In summary, a significant improvement of our theoretical understanding
of neutrino-nucleus cross section is required, since the currently
existing neutrino beams do no allow for measurements at a sufficient
level of precision. This, in turn, raises the question of how to validate
a theoretical model to a precision \emph{better} than the available
neutrino scattering data. One part of the answer, clearly can come
from electron scattering data, since\textemdash irrespective of the underlying
theoretical framework\textemdash  any model able to predict the electro-weak
nuclear response necessarily \emph{must} also describe the
electro-magnetic response, which can be accurately measured by
electron scattering experiments. This is a necessary condition, but whether it is
sufficient is difficult to judge. We are faced by a vicious cycle: the lack of
high quality data necessitates theory to be trusted, but to trust theory
it needs to be validated against data. It should be kept in mind that
if everything else fails, $\nu$STORM~\cite{Adey:2013pio} provides a way
out of this vicious cycle by ``simply'' providing very high quality
data, for neutrino and antineutrino as well as for $\nu_\mu$ and
$\nu_e$ interactions. This data, will likely be good enough to allow
extrapolation to any experimental situation of interest. In the case
extrapolation is insufficient, $\nu$STORM data would provide the cornerstone 
to decisively test the theoretical understanding of
neutrino-nucleus interactions.

At a phenomenological level, existing studies, of which a few
examples were shown in this section, have merely started to scratch
the surface of the issue and, at this stage, it is not clear, whether
many results are not just a mere consequence of the assumptions put
into the analysis. One example is,  for instance, the relative
robustness of LBNE-like experiments against cross section systematics
found by the authirs of Ref.~\cite{Coloma:2012ji}: in light of later studies, this may be
entirely due to using systematical uncertainties which only affect the
rate but not the shape of the signal. The conceptually expedient
simplification to focus on one interaction type, like QE,
severely restricts practical applicability for future experiments in
the multi-GeV energy range, where a multitude of interaction mechanisms
contributes.  Similarly, comparing different event generators may
create a false sense of the magnitude of the problem, in particular
since none of the existing generators is known to \emph{correctly}
describe neutrino scattering over a wide kinematic range and different
interaction modes.


\section{Summary and outlook}
\label{sec:concl}
The surge of activities aimed at improving the description of neutrino-nucleus interactions, critical
for the interpretation of oscillation signals, is now over a decade old, and still growing.
Beginning with the
first Workshop of the NUINT (Neutrino-Nucleus Interactions in the Few GeV Region) series---that
marked the dawning of the age of collaboration between the communities of nuclear theory and neutrino physics back
in 2001---a number of experimental and theoretical developments
contributed to steadily advance the field.

On the experimental side, the MiniBooNE Collaboration performed the first measurement of the double differential nuclear cross section
in the QE sector \cite{AguilarArevalo:2010zc}, thus providing an unprecedented opportunity to test the available
theoretical models and compare their predictions of the flux-integrated neutrino cross section. Additional 
information has been 
provided by the NOMAD \cite{Wu:2008,Lyubushkin:2008pe} and T2K \cite{NEUT,T2K:xsec} Collaborations, as well as 
by the SciBooNE \cite{SciBooNE} and Miner$\nu$a \cite{minerva_2014} experiments, specifically designed to
study neutrino-nucleus interactions in different kinematical regions and using different nuclear targets.

Theoretical studies, carried out using highly advanced models of nuclear structure and dynamics, shed new light on the complex reaction mechanisms
contributing to the flux integrated cross sections,  the understanding of which is needed to reduce the uncertainties associated  with
event identification and neutrino energy reconstruction. In this context, processes involving two-nucleon correlations and meson-exchange currents appear 
to play a significant role.

Thanks to the availability of ever more powerful computers, and to the steady
evolution of Monte Carlo computational algorithms, accurate {\em ab initio} calculations
of scattering observables, based on the formalism of nuclear many-body theory and realistic
nuclear Hamiltonians,  can presently be carried out for nuclei as large as carbon.
Pioneering calculations have been also carried out within the variational Monte Carlo approach using
a relativistic Hamiltonian, although these studies are limited to the ground state
energies of  $^3$H and $^4$He  \cite{PhysRevC.47.484}.
For the foreseeable future, applications of the GFMC computational technique to the calculation
of the nuclear response will be unavoidably limited
to the region of low to moderate momentum transfer, in which the non relativistic approximation
is expected to be workable. However, GFMC results will provide most valuable benchmarks to 
test the accuracy of more approximate approaches in the non relativistic limit.

Nuclear interactions at large momentum transfer are effectively described using the
formalism based on the  the factorisation {\em ansatz}.  Within this
scheme, the interaction vertex is treated using the full relativistic expression of the nuclear current, while
the initial state---which is obviously independent of momentum transfer---is described in terms of non relativistic
spectral functions. The development of improved models of the target spectral function, needed to study
neutrino interactions in liquid argon detectors, will require both the experimental information coming from
electron scattering experiments \cite{argon} and the use of Monte Carlo techniques to carry out
accurate calculations of the relevant nuclear amplitudes.

The factorisation {\em ansatz} provides a fully consistent framework to treat processes involving both 
one- and two-nucleon currents. Moreover, to the extent to which the matrix elements of the current can be 
parametrized exploiting the available proton and deuteron data, it can be applied to all interaction channels: quasi elastic scattering, 
resonance production and deep inelastic scattering. This scheme is 
ideally suited for implementation in simulation codes, but disregards the effects
of final state interactions between the particles produced at the interaction vertex and the spectator nucleons, which, depending
on kinematics, may be quite significant.  Extensive theoretical work on electron-nucleus scattering suggests that the effects of final state 
interactions can be systematically
included within the spectral function formalism using the eikonal approximation. The existing applications of this approach, 
which allows for a consistent treatment of initial- and final-state correlations, are limited to  the quasi elastic sector. 
However, the underlying conceptual scheme, based on the assumptions that (i) the struck nucleon
travels along a straight trajectory with constant velocity, and (ii) the spectator nucleons can be seen as a collection
of fixed scattering centers, has a much broader range of applications. Cascade Monte Carlo simulations, designed 
to provide an event-by-event description
of the complex final states occurring in neutrino interactions at energies between
few hundreds MeV and few GeV, are in fact largely based on the same assumptions.
The development of simulation codes whose basic inputs---coordinate-space
distribution of the spectator nucleons and medium-modified hadronic cross sections---are
consistent with the theoretical description of the target initial state based on
many-body theory appears to feasible, and needs to be thoroughly investigated.

The picture emerging from this review suggests that there are routes worth exploring to achieve  a
better \emph{quantitative} understanding of what needs---and
needs not---to be known about neutrino-nucleus interactions for
long-baseline oscillation experiments.

First, comparison between
event generators will remain indicative of potential issues,  and will help
to pin down where there needs to be more theoretical work, or where existing
theoretical results need to be implemented into generators. However,
there is an exponential number of possibilities to combine various
generators and their options. Therefore, care should be used to pick
illuminating cases for comparison, instead of attempting to completely
exhaust the possibilities (and likely the audiences as well). In this
process, agreement between results from different generators should
not be misinterpreted as a sign of correctness, since there is a distinct
possibility of several generators being wrong about the same physics
in the same or very similar way. For example, most generators rely on
the RFGM as a description of the initial state.

It is worth mentioning that the above considerations also apply to the 
cross section models, which often predict very similar results in spite of being 
based on totally different, and sometimes incompatible, physics 
assumptions~\cite{Benhar:NUINT15}.   

Second, completeness should come before accuracy. A complete cross
section model which can cover all the relevant regions of kinematics
and interaction modes at a coarse level of approximation is, at this
stage, preferable to a microphysical accurate description of a narrow
kinematic range or a single interaction channel. The devil is the in
interaction of the various pieces, and many clever schemes to solve one
particular problem eventually fail because of all the other moving
parts. The same applies for the subsequent phenomenological
analysis. For instance, flux uncertainties clearly limit what a near
detector can do in terms of eliminating
systematics~\cite{Huber:2007em}.

Third, detector effects need to be
included, even if only approximately. Thresholds, energy resolution and
particle identification are all interrelated and typically rely on the
underlying event generator.

Fourth, the three flavor oscillation should not be assumed since testing this
framework is a major objective for the next generation of long-baseline experiments.

Obviously,  achieving all the above goals in a combined analysis will require a strong synergy
between experimentalists, theorists and developers of simulation codes.


\begin{thebibliography}{189}
\providecommand{\natexlab}[1]{#1}
\providecommand{\url}[1]{\texttt{#1}}
\providecommand{\urlprefix}{URL }
\expandafter\ifx\csname urlstyle\endcsname\relax
  \providecommand{\doi}[1]{doi:\discretionary{}{}{}#1}\else
  \providecommand{\doi}[1]{doi:\discretionary{}{}{}\begingroup
  \urlstyle{rm}\url{#1}\endgroup}\fi
\providecommand{\bibinfo}[2]{#2}

\bibitem[{{Q.R. Ahmad, {\em et al.} (SNO Collaboration)}(2002)}]{SNO}
\bibinfo{author}{{Q.R. Ahmad, {\em et al.} (SNO Collaboration)}},
  \bibinfo{journal}{Phys. Rev. Lett.} \bibinfo{volume}{89}
  (\bibinfo{year}{2002}) \bibinfo{pages}{011301}.

\bibitem[{{T. Araki, {\em et al.} (KamLAND Collaboration)}(2005)}]{KAMLAND}
\bibinfo{author}{{T. Araki, {\em et al.} (KamLAND Collaboration)}},
  \bibinfo{journal}{Phys. Rev. Lett.} \bibinfo{volume}{94}
  (\bibinfo{year}{2005}) \bibinfo{pages}{081801}.

\bibitem[{{Y. Ashie, {\em et al.} (Super-Kamiokande
  Collaboration)}(2004)}]{SUPERK}
\bibinfo{author}{{Y. Ashie, {\em et al.} (Super-Kamiokande Collaboration)}},
  \bibinfo{journal}{Phys. Rev. Lett.} \bibinfo{volume}{93}
  (\bibinfo{year}{2004}) \bibinfo{pages}{101801}.

\bibitem[{{E. Aliu, {\em et al.} (K2K Collaboration)}(2005)}]{Aliu:2004sq}
\bibinfo{author}{{E. Aliu, {\em et al.} (K2K Collaboration)}},
  \bibinfo{journal}{Phys. Rev. Lett.} \bibinfo{volume}{94}
  (\bibinfo{year}{2005}) \bibinfo{pages}{081802}.

\bibitem[{{K. Abe, {\em et al.} (T2K
  Collaboration)}(2014{\natexlab{a}})}]{Abe:2014ugx}
\bibinfo{author}{{K. Abe, {\em et al.} (T2K Collaboration)}},
  \bibinfo{journal}{Phys. Rev. Lett.} \bibinfo{volume}{112}
  (\bibinfo{year}{2014}{\natexlab{a}}) \bibinfo{pages}{181801}.

\bibitem[{{P. Adamson, {\em et al.} (MINOS
  Collaboration)}(2011)}]{Adamson:2011qu}
\bibinfo{author}{{P. Adamson, {\em et al.} (MINOS Collaboration)}},
  \bibinfo{journal}{Phys. Rev. Lett.} \bibinfo{volume}{107}
  (\bibinfo{year}{2011}) \bibinfo{pages}{181802}.

\bibitem[{{F.P. An, {\em et al.} (Daya Bay Collaboration)}(2012)}]{An:2012eh}
\bibinfo{author}{{F.P. An, {\em et al.} (Daya Bay Collaboration)}},
  \bibinfo{journal}{Phys. Rev. Lett.} \bibinfo{volume}{108}
  (\bibinfo{year}{2012}) \bibinfo{pages}{171803}.

\bibitem[{{J.K. Ahn, {\em et al.} (RENO Collaboration)}(2012)}]{Ahn:2012nd}
\bibinfo{author}{{J.K. Ahn, {\em et al.} (RENO Collaboration)}},
  \bibinfo{journal}{Phys. Rev. Lett.} \bibinfo{volume}{108}
  (\bibinfo{year}{2012}) \bibinfo{pages}{191802}.

\bibitem[{{K. Abe, {\em et al.} (T2K
  Collaboration)}(2014{\natexlab{b}})}]{Abe:2013hdq}
\bibinfo{author}{{K. Abe, {\em et al.} (T2K Collaboration)}},
  \bibinfo{journal}{Phys. Rev. Lett.} \bibinfo{volume}{112}
  (\bibinfo{year}{2014}{\natexlab{b}}) \bibinfo{pages}{061802}.

\bibitem[{{R. Acciari, {\em et al.} (DUNE Collaboration)}(2015)}]{DUNE}
\bibinfo{author}{{R. Acciari, {\em et al.} (DUNE Collaboration)}},
  \bibinfo{journal}{arXiv:1512.06148 [physics.ins-det]} .

\bibitem[{{K. Abe, {\em et al.} (T2K Collaboration)}(2013)}]{Abe:2013xua}
\bibinfo{author}{{K. Abe, {\em et al.} (T2K Collaboration)}},
  \bibinfo{journal}{Phys. Rev. D} \bibinfo{volume}{88} (\bibinfo{year}{2013})
  \bibinfo{pages}{032002}.

\bibitem[{{A.A. Aguilar-Arevalo, {\em et al.} (MiniBooNE
  Collaboration)}(2010)}]{AguilarArevalo:2010zc}
\bibinfo{author}{{A.A. Aguilar-Arevalo, {\em et al.} (MiniBooNE
  Collaboration)}}, \bibinfo{journal}{Phys. Rev. D} \bibinfo{volume}{81}
  (\bibinfo{year}{2010}) \bibinfo{pages}{092005}.

\bibitem[{{A.A. Aguilar-Arevalo, {\em et al.} (MiniBooNE
  Collaboration)}(2008)}]{miniboone_ccqe_2}
\bibinfo{author}{{A.A. Aguilar-Arevalo, {\em et al.} (MiniBooNE
  Collaboration)}}, \bibinfo{journal}{Phys. Rev. Lett.} \bibinfo{volume}{108}
  (\bibinfo{year}{2008}) \bibinfo{pages}{191802}.

\bibitem[{{A.A. Aguilar-Arevalo, {\em et al.} (MiniBooNE
  Collaboration)}(2013)}]{miniboone_antinu}
\bibinfo{author}{{A.A. Aguilar-Arevalo, {\em et al.} (MiniBooNE
  Collaboration)}}, \bibinfo{journal}{Phys. Rev. D} \bibinfo{volume}{88}
  (\bibinfo{year}{2013}) \bibinfo{pages}{032001}.

\bibitem[{{B.G. Tice, {\em et al.} (MINER$\nu$A
  Collaboration)}(2014)}]{minerva_2014}
\bibinfo{author}{{B.G. Tice, {\em et al.} (MINER$\nu$A Collaboration)}},
  \bibinfo{journal}{Phys. Rev. Lett.} \bibinfo{volume}{112}
  (\bibinfo{year}{2014}) \bibinfo{pages}{231801}.

\bibitem[{{O. Benhar {\em et al.} (JLab E12-014012
  Collaboration)}(2014)}]{argon}
\bibinfo{author}{{O. Benhar {\em et al.} (JLab E12-014012 Collaboration)}},
  \bibinfo{journal}{arXiv:1406.4080 [nucl-ex]} .

\bibitem[{Benhar et~al.(2008)Benhar, Day, and Sick}]{Benhar:2006wy}
\bibinfo{author}{O.~Benhar}, \bibinfo{author}{D.~Day},
  \bibinfo{author}{I.~Sick}, \bibinfo{journal}{Rev. Mod. Phys.}
  \bibinfo{volume}{80} (\bibinfo{year}{2008}) \bibinfo{pages}{189}.

\bibitem[{Benhar et~al.(1993{\natexlab{a}})Benhar, Pieper, and
  Pandharipande}]{RMP_old}
\bibinfo{author}{O.~Benhar}, \bibinfo{author}{S.~C. Pieper},
  \bibinfo{author}{V.~R. Pandharipande}, \bibinfo{journal}{Rev. Mod. Phys.}
  \bibinfo{volume}{65} (\bibinfo{year}{1993}{\natexlab{a}})
  \bibinfo{pages}{817}.

\bibitem[{Benhar(2016{\natexlab{a}})}]{Benhar:NPN}
\bibinfo{author}{O.~Benhar}, \bibinfo{journal}{Nucl. Phys. News}
  \bibinfo{volume}{26} (\bibinfo{year}{2016}{\natexlab{a}})
  \bibinfo{pages}{15}.

\bibitem[{{C. Marchand, {\em et al.}}(1988)}]{Marchand}
\bibinfo{author}{{C. Marchand, {\em et al.}}}, \bibinfo{journal}{Phys. Rev.
  Lett} \bibinfo{volume}{60} (\bibinfo{year}{1988}) \bibinfo{pages}{1703}.

\bibitem[{{D. Rohe, {\em et al} (JLab E97-006 Collaboration)}(2004)}]{daniela}
\bibinfo{author}{{D. Rohe, {\em et al} (JLab E97-006 Collaboration)}},
  \bibinfo{journal}{Phys. Rev. Lett.} \bibinfo{volume}{93}
  (\bibinfo{year}{2004}) \bibinfo{pages}{182501}.

\bibitem[{{D. Rohe (JLab E97-006 Collaboration)}(2006)}]{Rohe2006152}
\bibinfo{author}{{D. Rohe (JLab E97-006 Collaboration)}},
  \bibinfo{journal}{Nucl. Phys. B, Proc. Suppl.} \bibinfo{volume}{159}
  (\bibinfo{year}{2006}) \bibinfo{pages}{152}.

\bibitem[{{O. Benhar, {\em et al.}}(2005)}]{Benhar05}
\bibinfo{author}{{O. Benhar, {\em et al.}}}, \bibinfo{journal}{Phys. Rev. D}
  \bibinfo{volume}{72} (\bibinfo{year}{2005}) \bibinfo{pages}{053005}.

\bibitem[{Ankowski et~al.(2015)Ankowski, Benhar, and Sakuda}]{Ankowski2013}
\bibinfo{author}{A.~M. Ankowski}, \bibinfo{author}{O.~Benhar},
  \bibinfo{author}{M.~Sakuda}, \bibinfo{journal}{Phys. Rev. D}
  \bibinfo{volume}{91} (\bibinfo{year}{2015}) \bibinfo{pages}{054616}.

\bibitem[{Wiringa et~al.(1995)Wiringa, Stoks, and Schiavilla}]{Wiringa95}
\bibinfo{author}{R.~B. Wiringa}, \bibinfo{author}{V.~G.~J. Stoks},
  \bibinfo{author}{R.~Schiavilla}, \bibinfo{journal}{Phys. Rev. C}
  \bibinfo{volume}{51} (\bibinfo{year}{1995}) \bibinfo{pages}{38}.

\bibitem[{{B.S. Pudliner, {\em et al.}}(1995)}]{Pudliner95b}
\bibinfo{author}{{B.S. Pudliner, {\em et al.}}}, \bibinfo{journal}{Phys. Rev.
  C} \bibinfo{volume}{56} (\bibinfo{year}{1995}) \bibinfo{pages}{1720}.

\bibitem[{Epelbaum et~al.(2009)Epelbaum, Hammer, and Mei{\ss}ner}]{Epelbaum}
\bibinfo{author}{E.~Epelbaum}, \bibinfo{author}{H.~Hammer},
  \bibinfo{author}{U.~Mei{\ss}ner}, \bibinfo{journal}{Rev. Mod. Phys.}
  \bibinfo{volume}{81} (\bibinfo{year}{2009}) \bibinfo{pages}{1773}.

\bibitem[{Machleidt and Entemb(2011)}]{chiral:review}
\bibinfo{author}{R.~Machleidt}, \bibinfo{author}{D.~R. Entemb},
  \bibinfo{journal}{Phys. Rep.} \bibinfo{volume}{503} (\bibinfo{year}{2011})
  \bibinfo{pages}{1}.

\bibitem[{Weinberg(2005)}]{WeinBook1}
\bibinfo{author}{S.~Weinberg}, \bibinfo{title}{The Quantum Theory of Field},
  \bibinfo{publisher}{Cambridge University Press}, \bibinfo{year}{2005}.

\bibitem[{Riska(1989)}]{Riska89}
\bibinfo{author}{D.~O. Riska}, \bibinfo{journal}{Phys. Rep.}
  \bibinfo{volume}{181} (\bibinfo{year}{1989}) \bibinfo{pages}{207}.

\bibitem[{{J.S. O'Connell, {\em et al}}(1987)}]{12C1}
\bibinfo{author}{{J.S. O'Connell, {\em et al}}}, \bibinfo{journal}{Phys. Rev.
  C} \bibinfo{volume}{35} (\bibinfo{year}{1987}) \bibinfo{pages}{1063}.

\bibitem[{Sealock et~al.(1989)}]{12C2}
\bibinfo{author}{R.~Sealock}, et~al., \bibinfo{journal}{Phys. Rev. Lett.}
  \bibinfo{volume}{62} (\bibinfo{year}{1989}) \bibinfo{pages}{1350}.

\bibitem[{Benhar(2013{\natexlab{a}})}]{Benhar:NUFACT11}
\bibinfo{author}{O.~Benhar}, \bibinfo{journal}{J. Phys. Conf. Ser.}
  \bibinfo{volume}{408} (\bibinfo{year}{2013}{\natexlab{a}})
  \bibinfo{pages}{012042}.

\bibitem[{Itzykson and Zuber(1980)}]{Itzykson80}
\bibinfo{author}{C.~Itzykson}, \bibinfo{author}{J.~Zuber},
  \bibinfo{title}{Quantum Field Theory}, \bibinfo{publisher}{McGraw-Hill, New
  York}, \bibinfo{year}{1980}.

\bibitem[{{J. Golak, {\em et al.}}(1995)}]{Golak}
\bibinfo{author}{{J. Golak, {\em et al.}}}, \bibinfo{journal}{Phys. Rev. C}
  \bibinfo{volume}{52} (\bibinfo{year}{1995}) \bibinfo{pages}{1216}.

\bibitem[{Efros et~al.(1994)Efros, Leidemann, and Orlandini}]{Efros94}
\bibinfo{author}{V.~D. Efros}, \bibinfo{author}{W.~Leidemann},
  \bibinfo{author}{G.~Orlandini}, \bibinfo{journal}{Phys. Lett. B}
  \bibinfo{volume}{338} (\bibinfo{year}{1994}) \bibinfo{pages}{130}.

\bibitem[{Efros et~al.(1997)Efros, Leidemann, and Orlandini}]{Efros97}
\bibinfo{author}{V.~D. Efros}, \bibinfo{author}{W.~Leidemann},
  \bibinfo{author}{G.~Orlandini}, \bibinfo{journal}{Phys. Rev. Lett.}
  \bibinfo{volume}{78} (\bibinfo{year}{1997}) \bibinfo{pages}{432}.

\bibitem[{Efros et~al.(2004)Efros, Leidemann, and Orlandini}]{Efros04}
\bibinfo{author}{V.~D. Efros}, \bibinfo{author}{W.~Leidemann},
  \bibinfo{author}{G.~Orlandini}, \bibinfo{journal}{Phys. Rev. C}
  \bibinfo{volume}{69} (\bibinfo{year}{2004}) \bibinfo{pages}{044001}.

\bibitem[{Carlson and Schiavilla(1992)}]{CS92}
\bibinfo{author}{J.~Carlson}, \bibinfo{author}{R.~Schiavilla},
  \bibinfo{journal}{Phys. Rev. Lett.} \bibinfo{volume}{68}
  (\bibinfo{year}{1992}) \bibinfo{pages}{3682}.

\bibitem[{Carlson and Schiavilla(1998)}]{Carlson98}
\bibinfo{author}{J.~Carlson}, \bibinfo{author}{R.~Schiavilla},
  \bibinfo{journal}{Rev. Mod. Phys.} \bibinfo{volume}{70}
  (\bibinfo{year}{1998}) \bibinfo{pages}{743}.

\bibitem[{Benhar et~al.(1989)Benhar, Fabrocini, and Fantoni}]{PKE}
\bibinfo{author}{O.~Benhar}, \bibinfo{author}{A.~Fabrocini},
  \bibinfo{author}{S.~Fantoni}, \bibinfo{journal}{Nucl. Phys. A}
  \bibinfo{volume}{505} (\bibinfo{year}{1989}) \bibinfo{pages}{267}.

\bibitem[{Benhar et~al.(1994)Benhar, Fabrocini, Fantoni, and Sick}]{LDA}
\bibinfo{author}{O.~Benhar}, \bibinfo{author}{A.~Fabrocini},
  \bibinfo{author}{S.~Fantoni}, \bibinfo{author}{I.~Sick},
  \bibinfo{journal}{Nucl. Phys. A} \bibinfo{volume}{579} (\bibinfo{year}{1994})
  \bibinfo{pages}{493 -- 517}.

\bibitem[{de~Forest~Jr.(1983)}]{Forest83}
\bibinfo{author}{T.~de~Forest~Jr.}, \bibinfo{journal}{Nucl. Phys.}
  \bibinfo{volume}{A392} (\bibinfo{year}{1983}) \bibinfo{pages}{232}.

\bibitem[{Dieperink et~al.(1976)Dieperink, de~Forest, Sick, and
  Brandenburg}]{Dieperink76}
\bibinfo{author}{A.~E. Dieperink}, \bibinfo{author}{T.~de~Forest},
  \bibinfo{author}{I.~Sick}, \bibinfo{author}{R.~A. Brandenburg},
  \bibinfo{journal}{Phys. Lett. B} \bibinfo{volume}{63} (\bibinfo{year}{1976})
  \bibinfo{pages}{261}.

\bibitem[{Frullani and Mougey(1984)}]{eep1}
\bibinfo{author}{S.~Frullani}, \bibinfo{author}{J.~Mougey},
  \bibinfo{journal}{Adv. Nucl. Phys.} \bibinfo{volume}{14}
  (\bibinfo{year}{1984}) \bibinfo{pages}{1}.

\bibitem[{Dieperink and de~Witt~Huberts(1990)}]{eep2}
\bibinfo{author}{A.~E.~L. Dieperink}, \bibinfo{author}{P.~K. de~Witt~Huberts},
  \bibinfo{journal}{Ann. Rev. Nucl. Part. Sci.} \bibinfo{volume}{40}
  (\bibinfo{year}{1990}) \bibinfo{pages}{239}.

\bibitem[{{E.J. Moniz, {\em et al.}}(1971)}]{Moniz}
\bibinfo{author}{{E.J. Moniz, {\em et al.}}}, \bibinfo{journal}{Phys. Rev.
  Lett.} \bibinfo{volume}{26} (\bibinfo{year}{1971}) \bibinfo{pages}{445}.

\bibitem[{Schiavilla et~al.(1986)Schiavilla, Pandharipande, and
  Wiringa}]{Schiavilla86}
\bibinfo{author}{R.~Schiavilla}, \bibinfo{author}{V.~Pandharipande},
  \bibinfo{author}{R.~Wiringa}, \bibinfo{journal}{Nucl. Phys. A}
  \bibinfo{volume}{449} (\bibinfo{year}{1986}) \bibinfo{pages}{219}.

\bibitem[{Benhar et~al.(1990)Benhar, Fabrocini, and Fantoni}]{Benhar:KL}
\bibinfo{author}{O.~Benhar}, \bibinfo{author}{A.~Fabrocini},
  \bibinfo{author}{S.~Fantoni}, \bibinfo{journal}{Phys. Rev. C}
  \bibinfo{volume}{41} (\bibinfo{year}{1990}) \bibinfo{pages}{R24(R)}.

\bibitem[{Benhar et~al.(2015)Benhar, Lovato, and Rocco}]{Benhar2015a}
\bibinfo{author}{O.~Benhar}, \bibinfo{author}{A.~Lovato},
  \bibinfo{author}{N.~Rocco}, \bibinfo{journal}{Phys. Rev. C}
  \bibinfo{volume}{92} (\bibinfo{year}{2015}) \bibinfo{pages}{024602}.

\bibitem[{West(1975)}]{West75}
\bibinfo{author}{G.~B. West}, \bibinfo{journal}{Phys. Rep.}
  \bibinfo{volume}{18} (\bibinfo{year}{1975}) \bibinfo{pages}{263}.

\bibitem[{Sick et~al.(1980)Sick, Day, and McCarthy}]{Sick80}
\bibinfo{author}{I.~Sick}, \bibinfo{author}{D.~Day}, \bibinfo{author}{J.~S.
  McCarthy}, \bibinfo{journal}{Phys. Rev. Lett.} \bibinfo{volume}{45}
  (\bibinfo{year}{1980}) \bibinfo{pages}{871}.

\bibitem[{{J. Arrington, {\em et al}}(????)}]{E89-008_1}
\bibinfo{author}{{J. Arrington, {\em et al}}}, \bibinfo{journal}{Phys. Rev.
  Lett.} \bibinfo{volume}{82} (????) \bibinfo{pages}{2056}.

\bibitem[{Donnelly and Sick(1999)}]{Donnelly_superscaling}
\bibinfo{author}{T.~W. Donnelly}, \bibinfo{author}{I.~Sick},
  \bibinfo{journal}{Phys. Rev. C} \bibinfo{volume}{60} (\bibinfo{year}{1999})
  \bibinfo{pages}{065502}.

\bibitem[{Day et~al.(1987)}]{SLAC}
\bibinfo{author}{D.~B. Day}, et~al., \bibinfo{journal}{Phys. Rev. Lett.}
  \bibinfo{volume}{59} (\bibinfo{year}{1987}) \bibinfo{pages}{427}.

\bibitem[{{C. Maieron, {\em et al.}}(2009)}]{Maieron}
\bibinfo{author}{{C. Maieron, {\em et al.}}}, \bibinfo{journal}{Phys. Rev. C}
  \bibinfo{volume}{80} (\bibinfo{year}{2009}) \bibinfo{pages}{035504}.

\bibitem[{Amaro et~al.(2007)Amaro, Barbaro, Caballero, and Donnelly}]{Amaro}
\bibinfo{author}{J.~E. Amaro}, \bibinfo{author}{M.~B. Barbaro},
  \bibinfo{author}{J.~A. Caballero}, \bibinfo{author}{T.~W. Donnelly},
  \bibinfo{journal}{Phys. Rev. Lett.} \bibinfo{volume}{98}
  (\bibinfo{year}{2007}) \bibinfo{pages}{242501}.

\bibitem[{Mart{\'i}nez et~al.(2008)Mart{\'i}nez, Caballero, Donnelly, and
  Ud{\'i}as}]{Martinez}
\bibinfo{author}{M.~C. Mart{\'i}nez}, \bibinfo{author}{J.~A. Caballero},
  \bibinfo{author}{T.~W. Donnelly}, \bibinfo{author}{J.~M. Ud{\'i}as},
  \bibinfo{journal}{Phys. Rev. Lett.} \bibinfo{volume}{100}
  (\bibinfo{year}{2008}) \bibinfo{pages}{052502}.

\bibitem[{{G. Garino, {\em et al}}(1992)}]{Garino92}
\bibinfo{author}{{G. Garino, {\em et al}}}, \bibinfo{journal}{Phys. Rev. C}
  \bibinfo{volume}{45} (\bibinfo{year}{1992}) \bibinfo{pages}{780}.

\bibitem[{{T. G. O'Neill, {\em et al}}(1995)}]{O'Neill95}
\bibinfo{author}{{T. G. O'Neill, {\em et al}}}, \bibinfo{journal}{Phys. Lett.
  B} \bibinfo{volume}{351} (\bibinfo{year}{1995}) \bibinfo{pages}{87}.

\bibitem[{{D. Abbott, {\em et al}}(1998)}]{Abbott98}
\bibinfo{author}{{D. Abbott, {\em et al}}}, \bibinfo{journal}{Phys. Rev. Lett.}
  \bibinfo{volume}{80} (\bibinfo{year}{1998}) \bibinfo{pages}{5072}.

\bibitem[{{K. Garrow, {\em et al}}(2002)}]{Garrow02}
\bibinfo{author}{{K. Garrow, {\em et al}}}, \bibinfo{journal}{Phys. Rev. C}
  \bibinfo{volume}{66} (\bibinfo{year}{2002}) \bibinfo{pages}{044613}.

\bibitem[{{D. Rohe, {\em et al} (JLab E97-006 Collaboration)}(2005)}]{Rohe05}
\bibinfo{author}{{D. Rohe, {\em et al} (JLab E97-006 Collaboration)}},
  \bibinfo{journal}{Phys. Rev. C} \bibinfo{volume}{72} (\bibinfo{year}{2005})
  \bibinfo{pages}{054602}.

\bibitem[{Benhar(1999)}]{BenharPRL99}
\bibinfo{author}{O.~Benhar}, \bibinfo{journal}{Phys. Rev. Lett.}
  \bibinfo{volume}{83} (\bibinfo{year}{1999}) \bibinfo{pages}{3130}.

\bibitem[{Boffi et~al.(1992)Boffi, Giusti, and Pacati}]{BoffiPR}
\bibinfo{author}{S.~Boffi}, \bibinfo{author}{C.~Giusti}, \bibinfo{author}{F.~D.
  Pacati}, \bibinfo{journal}{Phys. Rep.} \bibinfo{volume}{69}
  (\bibinfo{year}{1992}) \bibinfo{pages}{881}.

\bibitem[{{J.M. Ud{\'i}as, {\em et al.} }(1993)}]{Udias}
\bibinfo{author}{{J.M. Ud{\'i}as, {\em et al.} }}, \bibinfo{journal}{Phys. Rev.
  C} \bibinfo{volume}{48} (\bibinfo{year}{1993}) \bibinfo{pages}{2731}.

\bibitem[{Meucci et~al.(2001)Meucci, Giusti, and Pacati}]{Meucci_etal}
\bibinfo{author}{A.~Meucci}, \bibinfo{author}{C.~Giusti},
  \bibinfo{author}{F.~D. Pacati}, \bibinfo{journal}{Phys. Rev. C}
  \bibinfo{volume}{64} (\bibinfo{year}{2001}) \bibinfo{pages}{014604}.

\bibitem[{Meucci(2002)}]{Meucci}
\bibinfo{author}{A.~Meucci}, \bibinfo{journal}{Phys. Rev. C}
  \bibinfo{volume}{65} (\bibinfo{year}{2002}) \bibinfo{pages}{044601}.

\bibitem[{Giusti et~al.(2011)Giusti, Meucci, Pacati, Co', and
  De~Donno}]{Giusti_etal}
\bibinfo{author}{C.~Giusti}, \bibinfo{author}{A.~Meucci},
  \bibinfo{author}{F.~D. Pacati}, \bibinfo{author}{G.~Co'},
  \bibinfo{author}{V.~De~Donno}, \bibinfo{journal}{Phys. Rev. C}
  \bibinfo{volume}{84} (\bibinfo{year}{2011}) \bibinfo{pages}{024615}.

\bibitem[{Capuzzi et~al.(1991)Capuzzi, Giusti, and Pacati}]{Capuzzi1991681}
\bibinfo{author}{F.~Capuzzi}, \bibinfo{author}{C.~Giusti},
  \bibinfo{author}{F.~Pacati}, \bibinfo{journal}{Nucl. Phys. A}
  \bibinfo{volume}{524} (\bibinfo{year}{1991}) \bibinfo{pages}{681}.

\bibitem[{Sosnik et~al.(1991)Sosnik, Snow, Silver, and Sokol}]{Sosnik91}
\bibinfo{author}{T.~R. Sosnik}, \bibinfo{author}{W.~M. Snow},
  \bibinfo{author}{R.~N. Silver}, \bibinfo{author}{P.~E. Sokol},
  \bibinfo{journal}{Phys. Rev. B} \bibinfo{volume}{43} (\bibinfo{year}{1991})
  \bibinfo{pages}{216}.

\bibitem[{{O. Benhar, {\em et al.} }(1991)}]{Benhar91}
\bibinfo{author}{{O. Benhar, {\em et al.} }}, \bibinfo{journal}{Phys. Rev. C}
  \bibinfo{volume}{44} (\bibinfo{year}{1991}) \bibinfo{pages}{2328}.

\bibitem[{Benhar et~al.(1993{\natexlab{b}})Benhar, Fabrocini, Fantoni,
  Pandharipande, and Sick}]{BenharCT}
\bibinfo{author}{O.~Benhar}, \bibinfo{author}{A.~Fabrocini},
  \bibinfo{author}{S.~Fantoni}, \bibinfo{author}{V.~R. Pandharipande},
  \bibinfo{author}{I.~Sick}, \bibinfo{journal}{Phys. Rev. Lett.}
  \bibinfo{volume}{226} (\bibinfo{year}{1993}{\natexlab{b}})
  \bibinfo{pages}{1}.

\bibitem[{Benhar(2013{\natexlab{b}})}]{Benhar2013}
\bibinfo{author}{O.~Benhar}, \bibinfo{journal}{Phys. Rev. C}
  \bibinfo{volume}{87} (\bibinfo{year}{2013}{\natexlab{b}})
  \bibinfo{pages}{024606}.

\bibitem[{Pandharipande and Pieper(1992)}]{panpiep}
\bibinfo{author}{V.~R. Pandharipande}, \bibinfo{author}{S.~C. Pieper},
  \bibinfo{journal}{Phys. Rev. C} \bibinfo{volume}{45} (\bibinfo{year}{1992})
  \bibinfo{pages}{791}.

\bibitem[{Benhar et~al.(1995)Benhar, Fabrocini, Fantoni, Pandharipande, Pieper,
  and Sick}]{higherord}
\bibinfo{author}{O.~Benhar}, \bibinfo{author}{A.~Fabrocini},
  \bibinfo{author}{S.~Fantoni}, \bibinfo{author}{V.~R. Pandharipande},
  \bibinfo{author}{S.~C. Pieper}, \bibinfo{author}{I.~Sick},
  \bibinfo{journal}{Phys. Lett. B} \bibinfo{volume}{359} (\bibinfo{year}{1995})
  \bibinfo{pages}{8 -- 12}.

\bibitem[{Schiavilla et~al.(1990)Schiavilla, Pandharipande, and
  Riska}]{Schiavilla90}
\bibinfo{author}{R.~Schiavilla}, \bibinfo{author}{V.~R. Pandharipande},
  \bibinfo{author}{D.~O. Riska}, \bibinfo{journal}{Phys. Rev. C}
  \bibinfo{volume}{41} (\bibinfo{year}{1990}) \bibinfo{pages}{309}.

\bibitem[{Finn et~al.(1984)Finn, Lourie, and Cottmann}]{Finn84}
\bibinfo{author}{J.~Finn}, \bibinfo{author}{R.~W. Lourie},
  \bibinfo{author}{B.~H. Cottmann}, \bibinfo{journal}{Phys. Rev. C}
  \bibinfo{volume}{29} (\bibinfo{year}{1984}) \bibinfo{pages}{2230}.

\bibitem[{Barreau et~al.(1983)}]{Barreau83}
\bibinfo{author}{P.~Barreau}, et~al., \bibinfo{journal}{Nucl. Phys. A}
  \bibinfo{volume}{402} (\bibinfo{year}{1983}) \bibinfo{pages}{515}.

\bibitem[{Cenni et~al.(1989)Cenni, degli Atti, and Salm\`e}]{Cennietal}
\bibinfo{author}{R.~Cenni}, \bibinfo{author}{C.~C. degli Atti},
  \bibinfo{author}{G.~Salm\`e}, \bibinfo{journal}{Phys. Rev. C}
  \bibinfo{volume}{39} (\bibinfo{year}{1989}) \bibinfo{pages}{1425--1437}.

\bibitem[{Carlson et~al.(2002)Carlson, Jourdan, Schiavilla, and
  Sick}]{Carlson02}
\bibinfo{author}{J.~Carlson}, \bibinfo{author}{J.~Jourdan},
  \bibinfo{author}{R.~Schiavilla}, \bibinfo{author}{I.~Sick},
  \bibinfo{journal}{Phys. Rev. C} \bibinfo{volume}{65} (\bibinfo{year}{2002})
  \bibinfo{pages}{024002--}.

\bibitem[{Lovato et~al.(2016)Lovato, Gandolfi, Carlson, Pieper, and
  Schiavilla}]{Inversion}
\bibinfo{author}{A.~Lovato}, \bibinfo{author}{S.~Gandolfi},
  \bibinfo{author}{J.~Carlson}, \bibinfo{author}{S.~C. Pieper},
  \bibinfo{author}{R.~Schiavilla}, \bibinfo{journal}{Phys. Rev. Lett.}
  \bibinfo{volume}{117} (\bibinfo{year}{2016}) \bibinfo{pages}{082501}.

\bibitem[{{De Pace} et~al.(2003){De Pace}, Nardi, Alberico, Donnelly, and
  Molinari}]{MEC_TO}
\bibinfo{author}{A.~{De Pace}}, \bibinfo{author}{M.~Nardi},
  \bibinfo{author}{W.~M. Alberico}, \bibinfo{author}{T.~W. Donnelly},
  \bibinfo{author}{A.~Molinari}, \bibinfo{journal}{Nucl. Phys. A}
  \bibinfo{volume}{726} (\bibinfo{year}{2003}) \bibinfo{pages}{303 -- 326}.

\bibitem[{Meucci et~al.(2002)Meucci, Giusti, and Pacati}]{MEC_PV}
\bibinfo{author}{A.~Meucci}, \bibinfo{author}{C.~Giusti},
  \bibinfo{author}{F.~D. Pacati}, \bibinfo{journal}{Phys. Rev. C}
  \bibinfo{volume}{66} (\bibinfo{year}{2002}) \bibinfo{pages}{034610}.

\bibitem[{Benhar and Rocco(2013)}]{Benhar:2013bwa}
\bibinfo{author}{O.~Benhar}, \bibinfo{author}{N.~Rocco}, \bibinfo{journal}{Adv.
  High Energy Phys.} \bibinfo{volume}{2013} (\bibinfo{year}{2013})
  \bibinfo{pages}{912702}.

\bibitem[{Benhar and Fabrocini(2000)}]{spec2}
\bibinfo{author}{O.~Benhar}, \bibinfo{author}{A.~Fabrocini},
  \bibinfo{journal}{Phys. Rev. C} \bibinfo{volume}{62} (\bibinfo{year}{2000})
  \bibinfo{pages}{034304}.

\bibitem[{{W.M. Alberico, M. Ericson, and A. Molinari}(1984)}]{Alberico}
\bibinfo{author}{{W.M. Alberico, M. Ericson, and A. Molinari}},
  \bibinfo{journal}{Ann. Phys.} \bibinfo{volume}{154} (\bibinfo{year}{1984})
  \bibinfo{pages}{356}.

\bibitem[{Gil et~al.(1997{\natexlab{a}})Gil, Nieves, and Oset}]{Gil1997543}
\bibinfo{author}{A.~Gil}, \bibinfo{author}{J.~Nieves},
  \bibinfo{author}{E.~Oset}, \bibinfo{journal}{Nucl. Phys. A}
  \bibinfo{volume}{627} (\bibinfo{year}{1997}{\natexlab{a}})
  \bibinfo{pages}{543 -- 598}.

\bibitem[{Cowell and Pandharipande(2003)}]{Cowell}
\bibinfo{author}{S.~Cowell}, \bibinfo{author}{V.~R. Pandharipande},
  \bibinfo{journal}{Phys. Rev. C} \bibinfo{volume}{67} (\bibinfo{year}{2003})
  \bibinfo{pages}{035504}.

\bibitem[{Benhar and Farina(2009)}]{Farina}
\bibinfo{author}{O.~Benhar}, \bibinfo{author}{N.~Farina},
  \bibinfo{journal}{Phys. Lett. B} \bibinfo{volume}{680} (\bibinfo{year}{2009})
  \bibinfo{pages}{305}.

\bibitem[{Lovato et~al.(2013{\natexlab{a}})Lovato, Losa, and Benhar}]{LLB}
\bibinfo{author}{A.~Lovato}, \bibinfo{author}{C.~Losa},
  \bibinfo{author}{O.~Benhar}, \bibinfo{journal}{Nucl. Phys. A}
  \bibinfo{volume}{901} (\bibinfo{year}{2013}{\natexlab{a}}) \bibinfo{pages}{22
  -- 50}.

\bibitem[{Lovato et~al.(2014)Lovato, Benhar, Gandolfi, and Losa}]{LBGL}
\bibinfo{author}{A.~Lovato}, \bibinfo{author}{O.~Benhar},
  \bibinfo{author}{S.~Gandolfi}, \bibinfo{author}{C.~Losa},
  \bibinfo{journal}{Phys. Rev. C} \bibinfo{volume}{89} (\bibinfo{year}{2014})
  \bibinfo{pages}{025804}.

\bibitem[{Lovato et~al.(2015)Lovato, Gandolfi, Carlson, Pieper, and
  Schiavilla}]{Euclidean}
\bibinfo{author}{A.~Lovato}, \bibinfo{author}{S.~Gandolfi},
  \bibinfo{author}{J.~Carlson}, \bibinfo{author}{S.~C. Pieper},
  \bibinfo{author}{R.~Schiavilla}, \bibinfo{journal}{Phys. Rev. C}
  \bibinfo{volume}{91} (\bibinfo{year}{2015}) \bibinfo{pages}{062501(R)}.

\bibitem[{Lovato et~al.(2013{\natexlab{b}})}]{LovatoSR}
\bibinfo{author}{A.~Lovato}, et~al., \bibinfo{journal}{Phys. Rev. Lett.}
  \bibinfo{volume}{111} (\bibinfo{year}{2013}{\natexlab{b}})
  \bibinfo{pages}{092501}.

\bibitem[{Jourdan(1996)}]{Jourdan96}
\bibinfo{author}{J.~Jourdan}, \bibinfo{journal}{Nucl. Phys. A}
  \bibinfo{volume}{603} (\bibinfo{year}{1996}) \bibinfo{pages}{117}.

\bibitem[{Baran et~al.(1988)}]{Baran88}
\bibinfo{author}{D.~T. Baran}, et~al., \bibinfo{journal}{Phys. Rev. Lett.}
  \bibinfo{volume}{61} (\bibinfo{year}{1988}) \bibinfo{pages}{400--403}.

\bibitem[{Whitney et~al.(1974)}]{Whitney74}
\bibinfo{author}{R.~R. Whitney}, et~al., \bibinfo{journal}{Phys. Rev. C}
  \bibinfo{volume}{9} (\bibinfo{year}{1974}) \bibinfo{pages}{2230--2235}.

\bibitem[{Rocco et~al.(2016)Rocco, Lovato, and Benhar}]{NoemiPRL}
\bibinfo{author}{N.~Rocco}, \bibinfo{author}{A.~Lovato},
  \bibinfo{author}{O.~Benhar}, \bibinfo{journal}{Phys. Rev. Lett.}
  \bibinfo{volume}{116} (\bibinfo{year}{2016}) \bibinfo{pages}{192501}.

\bibitem[{Meucci et~al.(2009)Meucci, Caballero, Giusti, Pacati, and
  Ud{\'i}as}]{Meucci_etal_2}
\bibinfo{author}{A.~Meucci}, \bibinfo{author}{J.~A. Caballero},
  \bibinfo{author}{C.~Giusti}, \bibinfo{author}{F.~D. Pacati},
  \bibinfo{author}{J.~M. Ud{\'i}as}, \bibinfo{journal}{Phys. Rev. C}
  \bibinfo{volume}{80} (\bibinfo{year}{2009}) \bibinfo{pages}{024605}.

\bibitem[{{R. Gonzal\'ez-Jimenez, G.D. Megias, M.B. Barbaro, J.A. Caballero,
  T.W. Donnelly}(2014)}]{SuSav2}
\bibinfo{author}{{R. Gonzal\'ez-Jimenez, G.D. Megias, M.B. Barbaro, J.A.
  Caballero, T.W. Donnelly}}, \bibinfo{journal}{Phys. Rev. C}
  \bibinfo{volume}{90} (\bibinfo{year}{2014}) \bibinfo{pages}{035501}.

\bibitem[{{G.D. Megias, J.E. Amaro, M.B. Barbaro, J.A. Caballero, T.W.
  Donnelly}(2016)}]{Megias2016}
\bibinfo{author}{{G.D. Megias, J.E. Amaro, M.B. Barbaro, J.A. Caballero, T.W.
  Donnelly}}, \bibinfo{journal}{Phys. Rev. D} \bibinfo{volume}{94}
  (\bibinfo{year}{2016}) \bibinfo{pages}{013012}.

\bibitem[{Gil et~al.(1997{\natexlab{b}})Gil, Nieves, and Oset}]{Gil1997599}
\bibinfo{author}{A.~Gil}, \bibinfo{author}{J.~Nieves},
  \bibinfo{author}{E.~Oset}, \bibinfo{journal}{Nucl. Phys. A}
  \bibinfo{volume}{627} (\bibinfo{year}{1997}{\natexlab{b}})
  \bibinfo{pages}{599 -- 619}.

\bibitem[{Benhar and Veneziano(2011)}]{veneziano}
\bibinfo{author}{O.~Benhar}, \bibinfo{author}{G.~Veneziano},
  \bibinfo{journal}{Phys. Lett. B} \bibinfo{volume}{702} (\bibinfo{year}{2011})
  \bibinfo{pages}{433}.

\bibitem[{Benhar and Meloni(2007)}]{Benhar:2006nr}
\bibinfo{author}{O.~Benhar}, \bibinfo{author}{D.~Meloni},
  \bibinfo{journal}{Nucl. Phys. A} \bibinfo{volume}{789} (\bibinfo{year}{2007})
  \bibinfo{pages}{379--402}.

\bibitem[{{T. Akiri, {\em et al.}, (LBNE Collaboration)}(2011)}]{Akiri:2011dv}
\bibinfo{author}{{T. Akiri, {\em et al.}, (LBNE Collaboration)}},
  \bibinfo{title}{{The 2010 Interim Report of the Long-Baseline Neutrino
  Experiment Collaboration Physics Working Groups}} .

\bibitem[{Mosel et~al.(2014{\natexlab{a}})Mosel, Lalakulich, and
  Gallmeister}]{PhysRevD.89.093003}
\bibinfo{author}{U.~Mosel}, \bibinfo{author}{O.~Lalakulich},
  \bibinfo{author}{K.~Gallmeister}, \bibinfo{title}{Reaction mechanisms at
  $\mathrm{MINER}\ensuremath{\nu}\mathrm{A}$}, \bibinfo{journal}{Phys. Rev. D}
  \bibinfo{volume}{89} (\bibinfo{year}{2014}{\natexlab{a}})
  \bibinfo{pages}{093003}.

\bibitem[{Buss et~al.(2012)Buss, Gaitanos, Gallmeister, van Hees, Kaskulov
  et~al.}]{Buss:2011mx}
\bibinfo{author}{O.~Buss}, \bibinfo{author}{T.~Gaitanos},
  \bibinfo{author}{K.~Gallmeister}, \bibinfo{author}{H.~van Hees},
  \bibinfo{author}{M.~Kaskulov}, et~al., \bibinfo{title}{{Transport-theoretical
  Description of Nuclear Reactions}}, \bibinfo{journal}{Phys.Rept.}
  \bibinfo{volume}{512} (\bibinfo{year}{2012}) \bibinfo{pages}{1--124}.

\bibitem[{Kelly(2004)}]{Kelly2004}
\bibinfo{author}{J.~J. Kelly}, \bibinfo{journal}{Phys. Rev. C}
  \bibinfo{volume}{70} (\bibinfo{year}{2004}) \bibinfo{pages}{068202}.

\bibitem[{Bradford et~al.(2006)Bradford, Bodek, Budd, and Arrington}]{BBBA}
\bibinfo{author}{R.~Bradford}, \bibinfo{author}{A.~Bodek},
  \bibinfo{author}{H.~Budd}, \bibinfo{author}{J.~Arrington},
  \bibinfo{journal}{Nucl. Phys. B Proc. Suppl.} \bibinfo{volume}{159}
  (\bibinfo{year}{2006}) \bibinfo{pages}{127}.

\bibitem[{Mund et~al.(2013)Mund, M\"arkisch, Deissenroth, Krempel, Schumann,
  Abele, Petoukhov, and Soldner}]{g_A}
\bibinfo{author}{D.~Mund}, \bibinfo{author}{B.~M\"arkisch},
  \bibinfo{author}{M.~Deissenroth}, \bibinfo{author}{J.~Krempel},
  \bibinfo{author}{M.~Schumann}, \bibinfo{author}{H.~Abele},
  \bibinfo{author}{A.~Petoukhov}, \bibinfo{author}{T.~Soldner},
  \bibinfo{journal}{Phys. Rev. Lett.} \bibinfo{volume}{110}
  (\bibinfo{year}{2013}) \bibinfo{pages}{172502}.

\bibitem[{Bernard et~al.(2002)}]{bernard}
\bibinfo{author}{V.~Bernard}, et~al., \bibinfo{journal}{J. Phys. G}
  \bibinfo{volume}{28} (\bibinfo{year}{2002}) \bibinfo{pages}{R1}.

\bibitem[{Budd et~al.(2003)Budd, Bodek, and Arrington}]{bodek2}
\bibinfo{author}{H.~Budd}, \bibinfo{author}{A.~Bodek},
  \bibinfo{author}{J.~Arrington}, \bibinfo{journal}{arXiv:hep-ex/0308005} .

\bibitem[{Lalakulich and Paschos(2005)}]{LP}
\bibinfo{author}{O.~Lalakulich}, \bibinfo{author}{E.~A. Paschos},
  \bibinfo{journal}{Phys. Rev. D} \bibinfo{volume}{71} (\bibinfo{year}{2005})
  \bibinfo{pages}{074003}.

\bibitem[{Athar et~al.(2010)Athar, Chauhan, and Singh}]{sajjad}
\bibinfo{author}{M.~S. Athar}, \bibinfo{author}{S.~Chauhan},
  \bibinfo{author}{S.~K. Singh}, \bibinfo{journal}{Eur. Phys. J. A}
  \bibinfo{volume}{43} (\bibinfo{year}{2010}) \bibinfo{pages}{209}.

\bibitem[{Leitner et~al.(2009)Leitner, Buss, Alvarez-Ruso, and
  Mosel}]{Leitner:2008ue}
\bibinfo{author}{T.~Leitner}, \bibinfo{author}{O.~Buss},
  \bibinfo{author}{L.~Alvarez-Ruso}, \bibinfo{author}{U.~Mosel},
  \bibinfo{title}{{Electron- and neutrino-nucleus scattering from the
  quasielastic to the resonance region}}, \bibinfo{journal}{Phys. Rev. C}
  \bibinfo{volume}{79} (\bibinfo{year}{2009}) \bibinfo{pages}{034601}.

\bibitem[{Berge et~al.(1991)}]{CDHS}
\bibinfo{author}{P.~Berge}, et~al., \bibinfo{journal}{Zeit. Phys. C}
  \bibinfo{volume}{49} (\bibinfo{year}{1991}) \bibinfo{pages}{187}.

\bibitem[{Roberts(1990)}]{roberts}
\bibinfo{author}{R.~G. Roberts}, \bibinfo{title}{The Structure of the Proton},
  \bibinfo{publisher}{Cambridge University Press, Cambridge},
  \bibinfo{year}{1990}.

\bibitem[{{M. Gl\"uck, E. Reya, A. Vogt}(1998)}]{GRV98}
\bibinfo{author}{{M. Gl\"uck, E. Reya, A. Vogt}}, \bibinfo{journal}{Eur. Phys.
  J. C} \bibinfo{volume}{5} (\bibinfo{year}{1998}) \bibinfo{pages}{461}.

\bibitem[{Kulagin and Petti(2007)}]{petti}
\bibinfo{author}{S.~A. Kulagin}, \bibinfo{author}{R.~Petti},
  \bibinfo{journal}{Phys. Rev. D} \bibinfo{volume}{76} (\bibinfo{year}{2007})
  \bibinfo{pages}{094023}.

\bibitem[{Haider et~al.(2011)Haider, Simo, Athar, and Vacas}]{haider}
\bibinfo{author}{H.~Haider}, \bibinfo{author}{I.~R. Simo},
  \bibinfo{author}{M.~S. Athar}, \bibinfo{author}{M.~J.~V. Vacas},
  \bibinfo{journal}{Phys. Rev. C} \bibinfo{volume}{84} (\bibinfo{year}{2011})
  \bibinfo{pages}{054610}.

\bibitem[{Hirai et~al.(2001)Hirai, Kumano, and Miyama}]{kumano}
\bibinfo{author}{M.~Hirai}, \bibinfo{author}{S.~Kumano},
  \bibinfo{author}{M.~Miyama}, \bibinfo{journal}{Phys. Rev. D}
  \bibinfo{volume}{64} (\bibinfo{year}{2001}) \bibinfo{pages}{034003}.

\bibitem[{{R. Gran, {\em et al.}, (K2K Collaboration)}(2006)}]{Gran:2006jn}
\bibinfo{author}{{R. Gran, {\em et al.}, (K2K Collaboration)}},
  \bibinfo{journal}{Phys. Rev. D} \bibinfo{volume}{74} (\bibinfo{year}{2006})
  \bibinfo{pages}{052002}.

\bibitem[{{ V. Lyubushkin {\em et al.} (NOMAD
  Collaboration)}(2009)}]{Lyubushkin:2008pe}
\bibinfo{author}{{ V. Lyubushkin {\em et al.} (NOMAD Collaboration)}},
  \bibinfo{journal}{Eur. Phys. J. C} \bibinfo{volume}{63}
  (\bibinfo{year}{2009}) \bibinfo{pages}{355--381}.

\bibitem[{Benhar et~al.(2010)Benhar, Coletti, and Meloni}]{Benhar:2010nx}
\bibinfo{author}{O.~Benhar}, \bibinfo{author}{P.~Coletti},
  \bibinfo{author}{D.~Meloni}, \bibinfo{journal}{Phys. Rev. Lett.}
  \bibinfo{volume}{105} (\bibinfo{year}{2010}) \bibinfo{pages}{132301}.

\bibitem[{Martini et~al.(2011)Martini, Ericson, and Chanfray}]{martini}
\bibinfo{author}{M.~Martini}, \bibinfo{author}{M.~Ericson},
  \bibinfo{author}{G.~Chanfray}, \bibinfo{journal}{Phys. Rev. C}
  \bibinfo{volume}{84} (\bibinfo{year}{2011}) \bibinfo{pages}{055502}.

\bibitem[{Nieves et~al.(2012{\natexlab{a}})Nieves, Ruiz~Simo, and
  Vicente~Vacas}]{Nieves:2011yp}
\bibinfo{author}{J.~Nieves}, \bibinfo{author}{I.~Ruiz~Simo},
  \bibinfo{author}{M.~Vicente~Vacas}, \bibinfo{journal}{Phys. Lett. B}
  \bibinfo{volume}{707} (\bibinfo{year}{2012}{\natexlab{a}})
  \bibinfo{pages}{72--75}.

\bibitem[{Gran et~al.(2013)Gran, Nieves, Sanchez, and
  Vicente~Vacas}]{Gran:2013kda}
\bibinfo{author}{R.~Gran}, \bibinfo{author}{J.~Nieves},
  \bibinfo{author}{F.~Sanchez}, \bibinfo{author}{M.~Vicente~Vacas},
  \bibinfo{journal}{Phys. Rev. D} \bibinfo{volume}{88} (\bibinfo{year}{2013})
  \bibinfo{pages}{113007}.

\bibitem[{Marteau et~al.(2000)Marteau, Delorme, and Ericson}]{marteau2}
\bibinfo{author}{J.~Marteau}, \bibinfo{author}{J.~Delorme},
  \bibinfo{author}{M.~Ericson}, \bibinfo{journal}{Nucl. Phys. A}
  \bibinfo{volume}{663 \& 664} (\bibinfo{year}{2000}) \bibinfo{pages}{783c}.

\bibitem[{Delorme and Guichon(1991)}]{delorme}
\bibinfo{author}{J.~Delorme}, \bibinfo{author}{P.~A.~M. Guichon},
  \bibinfo{journal}{Phys. Lett. B} \bibinfo{volume}{293} (\bibinfo{year}{1991})
  \bibinfo{pages}{157}.

\bibitem[{Marteau(1999)}]{marteau1}
\bibinfo{author}{J.~Marteau}, \bibinfo{journal}{Eur. Phys. J. A}
  \bibinfo{volume}{5} (\bibinfo{year}{1999}) \bibinfo{pages}{183}.

\bibitem[{{G.D. Megias, {\em et al.}}(2016)}]{Megias:nu}
\bibinfo{author}{{G.D. Megias, {\em et al.}}}, \bibinfo{journal}{Phys. Rev. D}
  \bibinfo{volume}{94} (\bibinfo{year}{2016}) \bibinfo{pages}{093004}.

\bibitem[{Ankowski and Sobczyk(2008)}]{Ankowski:2007uy}
\bibinfo{author}{A.~M. Ankowski}, \bibinfo{author}{J.~T. Sobczyk},
  \bibinfo{journal}{Phys. Rev. C} \bibinfo{volume}{77} (\bibinfo{year}{2008})
  \bibinfo{pages}{044311}.

\bibitem[{Andreopoulos et~al.(2010)Andreopoulos, Bell, Bhattacharya, Cavanna,
  Dobson et~al.}]{Andreopoulos:2009rq}
\bibinfo{author}{C.~Andreopoulos}, \bibinfo{author}{A.~Bell},
  \bibinfo{author}{D.~Bhattacharya}, \bibinfo{author}{F.~Cavanna},
  \bibinfo{author}{J.~Dobson}, et~al., \bibinfo{journal}{Nucl. Instrum. Meth.}
  \bibinfo{volume}{A614} (\bibinfo{year}{2010}) \bibinfo{pages}{87--104}.

\bibitem[{Jen et~al.(2014)Jen, Ankowski, Benhar, Furmanski, Kalousis
  et~al.}]{Jen:2014aja}
\bibinfo{author}{C.~M. Jen}, \bibinfo{author}{A.~Ankowski},
  \bibinfo{author}{O.~Benhar}, \bibinfo{author}{A.~P. Furmanski},
  \bibinfo{author}{L.~N. Kalousis}, et~al., \bibinfo{journal}{Phys. Rev. D}
  \bibinfo{volume}{90} (\bibinfo{year}{2014}) \bibinfo{pages}{093004}.

\bibitem[{Golan et~al.(2012)Golan, Juszczak, and Sobczyk}]{NUWRO}
\bibinfo{author}{T.~Golan}, \bibinfo{author}{C.~Juszczak},
  \bibinfo{author}{J.~T. Sobczyk}, \bibinfo{journal}{Phys. Rev. C}
  \bibinfo{volume}{86} (\bibinfo{year}{2012}) \bibinfo{pages}{015505}.

\bibitem[{{K. Abe, {\em et al.} (T2K
  Collaboration)}(2015{\natexlab{a}})}]{NEUT}
\bibinfo{author}{{K. Abe, {\em et al.} (T2K Collaboration)}},
  \bibinfo{journal}{Phys. Rev. D} \bibinfo{volume}{91}
  (\bibinfo{year}{2015}{\natexlab{a}}) \bibinfo{pages}{112002}.

\bibitem[{Dytman(2011)}]{Dytman:2011zza}
\bibinfo{author}{S.~Dytman}, \bibinfo{journal}{AIP Conf. Proc.}
  \bibinfo{volume}{1382} (\bibinfo{year}{2011}) \bibinfo{pages}{156--157}.

\bibitem[{Dytman and Meyer(2011)}]{Dytman:2014}
\bibinfo{author}{S.~A. Dytman}, \bibinfo{author}{A.~S. Meyer},
  \bibinfo{journal}{AIP Conference Proceedings} \bibinfo{volume}{1405}
  (\bibinfo{year}{2011}) \bibinfo{pages}{213--218}.

\bibitem[{Williamson et~al.(1997)}]{Williamson:1997}
\bibinfo{author}{C.~Williamson}, et~al., \bibinfo{journal}{Phys. Rev. C}
  \bibinfo{volume}{56} (\bibinfo{year}{1997}) \bibinfo{pages}{3152--3172}.

\bibitem[{Anghinolfi et~al.(1997)Anghinolfi, Ripani, Cenni, Corvisiero, Longhi,
  Mazzaschi, Mokeev, Ricco, Taiuti, Teglia, Zucchiatti, Bianchi, Fantoni,
  Muccifora, LeviSandri, Lucherini, Polli, Reolon, Rossi, and
  Simula}]{Anghinolfi:95n}
\bibinfo{author}{M.~Anghinolfi}, \bibinfo{author}{M.~Ripani},
  \bibinfo{author}{R.~Cenni}, \bibinfo{author}{P.~Corvisiero},
  \bibinfo{author}{A.~Longhi}, \bibinfo{author}{L.~Mazzaschi},
  \bibinfo{author}{V.~Mokeev}, \bibinfo{author}{G.~Ricco},
  \bibinfo{author}{M.~Taiuti}, \bibinfo{author}{A.~Teglia},
  \bibinfo{author}{A.~Zucchiatti}, \bibinfo{author}{N.~Bianchi},
  \bibinfo{author}{A.~Fantoni}, \bibinfo{author}{V.~Muccifora},
  \bibinfo{author}{P.~LeviSandri}, \bibinfo{author}{V.~Lucherini},
  \bibinfo{author}{E.~Polli}, \bibinfo{author}{A.~R. Reolon},
  \bibinfo{author}{P.~Rossi}, \bibinfo{author}{S.~Simula}, \bibinfo{journal}{J.
  Phys. G} \bibinfo{volume}{21} (\bibinfo{year}{1997}) \bibinfo{pages}{L9}.

\bibitem[{{K.N. Abazajian, {\em et al.}}(2012)}]{Abazajian:2012ys}
\bibinfo{author}{{K.N. Abazajian, {\em et al.}}},
  \bibinfo{journal}{arXiv:1204.5379 [hep-ph]} .

\bibitem[{{P. Adamson, {\em et al.} (MINOS
  Collaboration)}(2010)}]{Adamson:2009ju}
\bibinfo{author}{{P. Adamson, {\em et al.} (MINOS Collaboration)}},
  \bibinfo{journal}{Phys. Rev. D} \bibinfo{volume}{81} (\bibinfo{year}{2010})
  \bibinfo{pages}{072002}.

\bibitem[{{H. Higuera, {\em et al.} (MINER$\nu$A
  Collaboration)}(2014)}]{Higuera:2014azj}
\bibinfo{author}{{H. Higuera, {\em et al.} (MINER$\nu$A Collaboration)}},
  \bibinfo{journal}{arXiv:1409.3835 [hep-ex]} .

\bibitem[{{S. Choubey, {\em et al.} (IDS-NF
  Collaboration)}(2011)}]{Choubey:2011zzq}
\bibinfo{author}{{S. Choubey, {\em et al.} (IDS-NF Collaboration)}},
  \bibinfo{journal}{arXiv:1112.2853 [hep-ex]} .

\bibitem[{{D. Adey, {\et al.} ($\nu$STORM Collaboration)}(2013)}]{Adey:2013pio}
\bibinfo{author}{{D. Adey, {\et al.} ($\nu$STORM Collaboration)}},
  \bibinfo{journal}{arXiv:1308.6822 [physics.acc-ph]} .

\bibitem[{Freund(2001)}]{Freund:2001pn}
\bibinfo{author}{M.~Freund}, \bibinfo{journal}{Phys. Rev. D}
  \bibinfo{volume}{64} (\bibinfo{year}{2001}) \bibinfo{pages}{053003}.

\bibitem[{Antusch et~al.(2007)Antusch, Huber, King, and
  Schwetz}]{Antusch:2007rk}
\bibinfo{author}{S.~Antusch}, \bibinfo{author}{P.~Huber},
  \bibinfo{author}{S.~King}, \bibinfo{author}{T.~Schwetz},
  \bibinfo{journal}{JHEP} \bibinfo{volume}{0704} (\bibinfo{year}{2007})
  \bibinfo{pages}{060}.

\bibitem[{Peccei(2008)}]{Peccei:2006as}
\bibinfo{author}{R.~D. Peccei}, \bibinfo{journal}{Lect. Notes Phys.}
  \bibinfo{volume}{741} (\bibinfo{year}{2008}) \bibinfo{pages}{3}.

\bibitem[{{S. Ritz, {\et al.} (HEPAP Subcommittee)}(2014)}]{P5}
\bibinfo{author}{{S. Ritz, {\et al.} (HEPAP Subcommittee)}},
  \bibinfo{title}{{{\em Building for Discovery: Strategic Plan for U.S.
  Particle Physics in the Global Context}}},
  \bibinfo{journal}{\url{http://science.energy.gov/\textasciitilde/media/hep/hepap/pdf/May2014/FINAL\_P5\_Report\_053014.pdf}}
  .

\bibitem[{Gonzalez-Garcia et~al.(2012)Gonzalez-Garcia, Maltoni, Salvado, and
  Schwetz}]{GonzalezGarcia:2012sz}
\bibinfo{author}{M.~Gonzalez-Garcia}, \bibinfo{author}{M.~Maltoni},
  \bibinfo{author}{J.~Salvado}, \bibinfo{author}{T.~Schwetz},
  \bibinfo{journal}{JHEP} \bibinfo{volume}{1212} (\bibinfo{year}{2012})
  \bibinfo{pages}{123}.

\bibitem[{Day and McFarland(2012)}]{Day:2012gb}
\bibinfo{author}{M.~Day}, \bibinfo{author}{K.~S. McFarland},
  \bibinfo{journal}{Phys. Rev. D} \bibinfo{volume}{86} (\bibinfo{year}{2012})
  \bibinfo{pages}{053003}.

\bibitem[{{D.G. Michael, {\em et al.} (MINOS
  COllaboration)}(2006)}]{Michael:2006rx}
\bibinfo{author}{{D.G. Michael, {\em et al.} (MINOS COllaboration)}},
  \bibinfo{journal}{Phys. Rev. Lett.} \bibinfo{volume}{97}
  (\bibinfo{year}{2006}) \bibinfo{pages}{191801}.

\bibitem[{{K. Abe, {\em et al.} (T2K
  Collaboration)}(2014{\natexlab{c}})}]{Abe:2014agb}
\bibinfo{author}{{K. Abe, {\em et al.} (T2K Collaboration)}},
  \bibinfo{journal}{Phys. Rev. Lett.} \bibinfo{volume}{113}
  (\bibinfo{year}{2014}{\natexlab{c}}) \bibinfo{pages}{241803}.

\bibitem[{Huber et~al.(2008)Huber, Mezzetto, and Schwetz}]{Huber:2007em}
\bibinfo{author}{P.~Huber}, \bibinfo{author}{M.~Mezzetto},
  \bibinfo{author}{T.~Schwetz}, \bibinfo{journal}{JHEP} \bibinfo{volume}{0803}
  (\bibinfo{year}{2008}) \bibinfo{pages}{021}.

\bibitem[{Coloma et~al.(2013)Coloma, Huber, Kopp, and Winter}]{Coloma:2012ji}
\bibinfo{author}{P.~Coloma}, \bibinfo{author}{P.~Huber},
  \bibinfo{author}{J.~Kopp}, \bibinfo{author}{W.~Winter},
  \bibinfo{journal}{Phys. Rev. D} \bibinfo{volume}{87} (\bibinfo{year}{2013})
  \bibinfo{pages}{033004}.

\bibitem[{Zucchelli(2002)}]{Zucchelli:2002sa}
\bibinfo{author}{P.~Zucchelli}, \bibinfo{journal}{Phys. Lett. B}
  \bibinfo{volume}{532} (\bibinfo{year}{2002}) \bibinfo{pages}{166}.

\bibitem[{Fernandez-Martinez and Meloni(????)}]{FernandezMartinez:2010dm}
\bibinfo{author}{E.~Fernandez-Martinez}, \bibinfo{author}{D.~Meloni},
  \bibinfo{journal}{Phys. Lett. B} \bibinfo{volume}{697} (????)
  \bibinfo{pages}{477}.

\bibitem[{Meloni and Martini(2012)}]{Meloni:2012fq}
\bibinfo{author}{D.~Meloni}, \bibinfo{author}{M.~Martini},
  \bibinfo{journal}{Phys. Lett. B} \bibinfo{volume}{716} (\bibinfo{year}{2012})
  \bibinfo{pages}{186}.

\bibitem[{Hayato(2002)}]{Hayato:2002sd}
\bibinfo{author}{Y.~Hayato}, \bibinfo{journal}{Nucl. Phys. B Proc. Suppl.}
  \bibinfo{volume}{112} (\bibinfo{year}{2002}) \bibinfo{pages}{171}.

\bibitem[{{A. Rodriguez, {\em et al.} (K2K
  Collaboration)}(2008)}]{Rodriguez:2008aa}
\bibinfo{author}{{A. Rodriguez, {\em et al.} (K2K Collaboration)}},
  \bibinfo{journal}{Phys. Rev. D} \bibinfo{volume}{78} (\bibinfo{year}{2008})
  \bibinfo{pages}{032003}.

\bibitem[{{A.A. Aguilar-Arevalo, {\em et al.} (MiniBooNE
  Collaboration}(2009)}]{AguilarArevalo:2009eb}
\bibinfo{author}{{A.A. Aguilar-Arevalo, {\em et al.} (MiniBooNE
  Collaboration}}, \bibinfo{journal}{Phys. Rev. Lett.} \bibinfo{volume}{103}
  (\bibinfo{year}{2009}) \bibinfo{pages}{081801}.

\bibitem[{{Y. Kurimoto, {\em et al.} (SciBooNE
  Collaboration)}(2010)}]{Kurimoto:2009wq}
\bibinfo{author}{{Y. Kurimoto, {\em et al.} (SciBooNE Collaboration)}},
  \bibinfo{journal}{Phys. Rev. D} \bibinfo{volume}{81} (\bibinfo{year}{2010})
  \bibinfo{pages}{033004}.

\bibitem[{{Y. Nakajima, {\em et al.} (SciBooNE
  Collaboration)}(2011{\natexlab{a}})}]{Nakajima:2010fp}
\bibinfo{author}{{Y. Nakajima, {\em et al.} (SciBooNE Collaboration)}},
  \bibinfo{journal}{Phys. Rev. D} \bibinfo{volume}{83}
  (\bibinfo{year}{2011}{\natexlab{a}}) \bibinfo{pages}{012005}.

\bibitem[{Smith and Moniz(1972)}]{Smith:1972xh}
\bibinfo{author}{R.~A. Smith}, \bibinfo{author}{E.~J. Moniz},
  \bibinfo{journal}{Nucl. Phys. B} \bibinfo{volume}{43} (\bibinfo{year}{1972})
  \bibinfo{pages}{605}.

\bibitem[{Rein and Sehgal(1981)}]{Rein:1980wg}
\bibinfo{author}{D.~Rein}, \bibinfo{author}{L.~M. Sehgal},
  \bibinfo{journal}{Ann. Phys.} \bibinfo{volume}{133} (\bibinfo{year}{1981})
  \bibinfo{pages}{79}.

\bibitem[{Martini et~al.(2009)Martini, Ericson, Chanfray, and
  Marteau}]{Martini:2009uj}
\bibinfo{author}{M.~Martini}, \bibinfo{author}{M.~Ericson},
  \bibinfo{author}{G.~Chanfray}, \bibinfo{author}{J.~Marteau},
  \bibinfo{journal}{Phys. Rev. C} \bibinfo{volume}{80} (\bibinfo{year}{2009})
  \bibinfo{pages}{065501}.

\bibitem[{Huber et~al.(2005)Huber, Lindner, and Winter}]{Huber:2004ka}
\bibinfo{author}{P.~Huber}, \bibinfo{author}{M.~Lindner},
  \bibinfo{author}{W.~Winter}, \bibinfo{journal}{Comput. Phys. Commun.}
  \bibinfo{volume}{167} (\bibinfo{year}{2005}) \bibinfo{pages}{195}.

\bibitem[{Huber et~al.(2007)Huber, Kopp, Lindner, Rolinec, and
  Winter}]{Huber:2007ji}
\bibinfo{author}{P.~Huber}, \bibinfo{author}{J.~Kopp},
  \bibinfo{author}{M.~Lindner}, \bibinfo{author}{M.~Rolinec},
  \bibinfo{author}{W.~Winter}, \bibinfo{journal}{Comput. Phys. Commun.}
  \bibinfo{volume}{177} (\bibinfo{year}{2007}) \bibinfo{pages}{432}.

\bibitem[{Blennow and Fernandez-Martinez(2010)}]{Blennow:2009pk}
\bibinfo{author}{M.~Blennow}, \bibinfo{author}{E.~Fernandez-Martinez},
  \bibinfo{journal}{Comput. Phys. Commun.} \bibinfo{volume}{181}
  (\bibinfo{year}{2010}) \bibinfo{pages}{227--231}.

\bibitem[{{K. Abe, {\em et al.} (T2K Collaboration)}(2011)}]{Abe:2011sj}
\bibinfo{author}{{K. Abe, {\em et al.} (T2K Collaboration)}},
  \bibinfo{journal}{Phys. Rev. Lett.} \bibinfo{volume}{107}
  (\bibinfo{year}{2011}) \bibinfo{pages}{041801}.

\bibitem[{{K. Abe, {\em et al.} (T2K Collaboration)}(2012)}]{giganti}
\bibinfo{author}{{K. Abe, {\em et al.} (T2K Collaboration)}},
  \bibinfo{journal}{Phys. Rev. D} \bibinfo{volume}{85} (\bibinfo{year}{2012})
  \bibinfo{pages}{031103}.

\bibitem[{{A. Cervera, {\em et al.}}(2000)}]{Cervera:2000kp}
\bibinfo{author}{{A. Cervera, {\em et al.}}}, \bibinfo{journal}{Nucl. Phys. b}
  \bibinfo{volume}{579} (\bibinfo{year}{2000}) \bibinfo{pages}{17}.

\bibitem[{Donini et~al.(2006)Donini, Fernandez-Martinez, Meloni, and
  Rigolin}]{Donini:2005db}
\bibinfo{author}{A.~Donini}, \bibinfo{author}{E.~Fernandez-Martinez},
  \bibinfo{author}{D.~Meloni}, \bibinfo{author}{S.~Rigolin},
  \bibinfo{journal}{Nucl. Phys. B} \bibinfo{volume}{743} (\bibinfo{year}{2006})
  \bibinfo{pages}{41}.

\bibitem[{Coloma et~al.(2014)Coloma, Huber, Jen, and Mariani}]{Coloma:2013tba}
\bibinfo{author}{P.~Coloma}, \bibinfo{author}{P.~Huber}, \bibinfo{author}{C.-M.
  Jen}, \bibinfo{author}{C.~Mariani}, \bibinfo{journal}{Phys. Rev. D}
  \bibinfo{volume}{89} (\bibinfo{year}{2014}) \bibinfo{pages}{073015}.

\bibitem[{Lalakulich et~al.(2012)Lalakulich, Mosel, and
  Gallmeister}]{Lalakulich:2012hs}
\bibinfo{author}{O.~Lalakulich}, \bibinfo{author}{U.~Mosel},
  \bibinfo{author}{K.~Gallmeister}, \bibinfo{journal}{Phys. Rev. C}
  \bibinfo{volume}{86} (\bibinfo{year}{2012}) \bibinfo{pages}{054606}.

\bibitem[{Martini et~al.(2012)Martini, Ericson, and Chanfray}]{Martini:2012fa}
\bibinfo{author}{M.~Martini}, \bibinfo{author}{M.~Ericson},
  \bibinfo{author}{G.~Chanfray}, \bibinfo{journal}{Phys. Rev. D}
  \bibinfo{volume}{85} (\bibinfo{year}{2012}) \bibinfo{pages}{093012}.

\bibitem[{Nieves et~al.(2012{\natexlab{b}})Nieves, Sanchez, Ruiz~Simo, and
  Vicente~Vacas}]{Nieves:2012yz}
\bibinfo{author}{J.~Nieves}, \bibinfo{author}{F.~Sanchez},
  \bibinfo{author}{I.~Ruiz~Simo}, \bibinfo{author}{M.~Vicente~Vacas},
  \bibinfo{journal}{Phys. Rev. D} \bibinfo{volume}{85}
  (\bibinfo{year}{2012}{\natexlab{b}}) \bibinfo{pages}{113008}.

\bibitem[{Mosel et~al.(2014{\natexlab{b}})Mosel, Lalakulich, and
  Gallmeister}]{Mosel:2013fxa}
\bibinfo{author}{U.~Mosel}, \bibinfo{author}{O.~Lalakulich},
  \bibinfo{author}{K.~Gallmeister}, \bibinfo{journal}{Phys. Rev. Lett.}
  \bibinfo{volume}{112} (\bibinfo{year}{2014}{\natexlab{b}})
  \bibinfo{pages}{151802}.

\bibitem[{Coloma and Huber(2013)}]{Coloma:2013rqa}
\bibinfo{author}{P.~Coloma}, \bibinfo{author}{P.~Huber},
  \bibinfo{journal}{Phys. Rev. Lett.} \bibinfo{volume}{111}
  (\bibinfo{year}{2013}) \bibinfo{pages}{221802}.

\bibitem[{Ankowski(2015{\natexlab{a}})}]{Ankowski:2015jya}
\bibinfo{author}{A.~M. e.~a. Ankowski}, \bibinfo{title}{{Comparison of the
  calorimetric and kinematic methods of neutrino energy reconstruction in
  disappearance experiments}}, \bibinfo{journal}{Phys. Rev.}
  \bibinfo{volume}{D92}~(\bibinfo{number}{7})
  (\bibinfo{year}{2015}{\natexlab{a}}) \bibinfo{pages}{073014}.

\bibitem[{Ankowski(2015{\natexlab{b}})}]{Ankowski:2015kya}
\bibinfo{author}{A.~M. e.~a. Ankowski}, \bibinfo{title}{{Missing energy and the
  measurement of the CP-violating phase in neutrino oscillations}},
  \bibinfo{journal}{Phys. Rev.} \bibinfo{volume}{D92}~(\bibinfo{number}{9})
  (\bibinfo{year}{2015}{\natexlab{b}}) \bibinfo{pages}{091301}.

\bibitem[{et~al(2014)}]{Aliaga:2013uqz}
\bibinfo{author}{L.~A. et~al}, \bibinfo{title}{Design, calibration, and
  performance of the \{MINERvA\} detector}, \bibinfo{journal}{Nuclear
  Instruments and Methods in Physics Research Section A: Accelerators,
  Spectrometers, Detectors and Associated Equipment} \bibinfo{volume}{743}
  (\bibinfo{year}{2014}) \bibinfo{pages}{130 -- 159}, ISSN
  \bibinfo{issn}{0168-9002}.

\bibitem[{Huber et~al.(2009)Huber, Lindner, Schwetz, and Winter}]{Huber:2009cw}
\bibinfo{author}{P.~Huber}, \bibinfo{author}{M.~Lindner},
  \bibinfo{author}{T.~Schwetz}, \bibinfo{author}{W.~Winter},
  \bibinfo{title}{{First hint for CP violation in neutrino oscillations from
  upcoming superbeam and reactor experiments}}, \bibinfo{journal}{JHEP}
  \bibinfo{volume}{0911} (\bibinfo{year}{2009}) \bibinfo{pages}{044}.

\bibitem[{Acciarri et~al.(2016)}]{Acciarri:2016crz}
\bibinfo{author}{R.~Acciarri}, et~al., \bibinfo{title}{{Long-Baseline Neutrino
  Facility (LBNF) and Deep Underground Neutrino Experiment (DUNE)}} .

\bibitem[{{Q. Wu, {\em et al.} (NOMAD Collaboration)}(2008)}]{Wu:2008}
\bibinfo{author}{{Q. Wu, {\em et al.} (NOMAD Collaboration)}},
  \bibinfo{journal}{Phys. Lett. B} \bibinfo{volume}{660} (\bibinfo{year}{2008})
  \bibinfo{pages}{19}.

\bibitem[{{K. Abe, {\em et al.} (T2K
  Collaboration)}(2015{\natexlab{b}})}]{T2K:xsec}
\bibinfo{author}{{K. Abe, {\em et al.} (T2K Collaboration)}},
  \bibinfo{journal}{Phys. Rev. D} \bibinfo{volume}{92}
  (\bibinfo{year}{2015}{\natexlab{b}}) \bibinfo{pages}{112003}.

\bibitem[{{Y. Nakajima, {\em et al.} (SciBooNE
  Collaboration)}(2011{\natexlab{b}})}]{SciBooNE}
\bibinfo{author}{{Y. Nakajima, {\em et al.} (SciBooNE Collaboration)}},
  \bibinfo{journal}{Phys. Rev. D} \bibinfo{volume}{83}
  (\bibinfo{year}{2011}{\natexlab{b}}) \bibinfo{pages}{012005}.

\bibitem[{Carlson et~al.(1993)Carlson, Pandharipande, and
  Schiavilla}]{PhysRevC.47.484}
\bibinfo{author}{J.~Carlson}, \bibinfo{author}{V.~R. Pandharipande},
  \bibinfo{author}{R.~Schiavilla}, \bibinfo{journal}{Phys. Rev. C}
  \bibinfo{volume}{47} (\bibinfo{year}{1993}) \bibinfo{pages}{484}.

\bibitem[{Benhar(2016{\natexlab{b}})}]{Benhar:NUINT15}
\bibinfo{author}{O.~Benhar}, \bibinfo{journal}{JPS Conf. Proc.}
  \bibinfo{volume}{12} (\bibinfo{year}{2016}{\natexlab{b}})
  \bibinfo{pages}{010001}.

\end{thebibliography}

\addcontentsline{toc}{section}{References}





\end{document}